\documentclass{article}
\usepackage[utf8]{inputenc}
\usepackage[usenames,dvipsnames]{color}
\usepackage{latexsym}
\usepackage{amsmath}
\usepackage{amssymb}
\usepackage{bbm}
\usepackage{graphicx}
\usepackage{pictexwd,dcpic}
\usepackage{subfigure}
\usepackage{dcolumn} 
\usepackage{bm}
\usepackage{cite}
\usepackage{yhmath}

\usepackage{pdfpages}

\usepackage{authblk} 
\usepackage{mathtools}

\usepackage[]{hyperref}

\usepackage{cleveref}
\usepackage{youngtab}
\usepackage{tikz}
\usepackage{tikz-cd}
\usetikzlibrary{ decorations.markings}

\usepackage{geometry}

\usetikzlibrary{arrows}
\usepackage{fancyhdr}
\usepackage{xcolor}
\newsavebox\MBox

\usepackage{setspace}
\usepackage{a4wide}
\usepackage{amsthm,array,booktabs}
\usepackage{hyperref}
\usepackage{cancel}
\usepackage{tikz}
\usepackage{amscd}
\usepackage{latexsym,cite}
\usepackage{ytableau}
\Yboxdim{5pt}
\usepackage{scalerel}
\usepackage{calc}
\input xy
\xyoption{all}

\newcommand{\be}{\begin{equation}}
\newcommand{\ee}{\end{equation}}
\newcommand{\bea}{\begin{eqnarray}}
\newcommand{\eea}{\end{eqnarray}}
\def \ba {\begin{aligned}}
\def \ea {\end{aligned}}

\newcommand{\ZZ}{{\mathbb Z}}

\newcommand{\CC}{{\mathbb C}}
\newcommand{\PP}{{\mathbb P}}

\DeclareMathOperator{\Arg}{Arg}

\DeclareMathOperator{\Tot}{Tot}

\def\l:{\mathopen{:}\,}
\def\r:{\,\mathclose{:}}

\newcommand{\cS}{\mathcal{S}}

\newcommand{\cO}{{\mathcal{O}}}
\newcommand{\cL}{\mathcal{L}}

\title{\textbf{A GLSM realization of derived equivalence \\in $U(1) \times U(2)$ models}}
\author[1,2]{Jirui Guo\footnote{jrkwok@tongji.edu.cn}}
\author[2,3,4,5]{Ban Lin \footnote{banlin@kias.re.kr}}
\author[6,7]{Hao Zou\footnote{haozou@fudan.edu.cn}}
\affil[1]{School of Mathematical Sciences and Institute for Advanced Study,\protect\\ Tongji University, Shanghai 200092, China} 
\affil[2]{Yau Mathematical Sciences Center, Tsinghua University, Beijing 100084, China}
\affil[3]{Department of Mathematical Sciences, Tsinghua University, Beijing 100084, China}
\affil[4]{School of Mathematics, University of Birmingham, Birmingham B15 2TT, UK}
\affil[5]{Korea Institute for Advanced Study, Seoul 02455, Republic of Korea}
\affil[6]{Center for Mathematics and Interdisciplinary Sciences, \protect\\Fudan University, Shanghai 200433, China}
\affil[7]{Shanghai Institute for Mathematics and Interdisciplinary Sciences, Shanghai 200433, China}

\date{\today}

\begin{document}
\maketitle
\begin{abstract}

This paper studies the derived equivalence between Calabi--Yau mixed branches using the B-brane hemisphere partition function in anomalous gauged linear sigma models (GLSMs). For a family of anomalous $U(2)$ GLSMs, we study the infrared behavior of B-branes under RG flow and the variation of FI parameters in the quantum K\"ahler moduli space. As characterized by the big and small window categories of the GLSM, the RG flow effect is analyzed according to the properties of their B-brane central charges as hemisphere partition functions. The result generalizes the band restriction rules in anomalous abelian GLSMs (Clingempeel--Floch--Romo) to some $U(2)$ models. As an application, we study a family of $U(1)\times U(k)$ models for $k=1,2$, which realize a Fano UV sigma model and two IR phases as mixed branches accompanied by local Calabi--Yau Higgs components. By applying the restriction rules, we establish new derived equivalences between Calabi--Yau varieties from these anomalous models.

\end{abstract}
\newpage

\setcounter{tocdepth}{2}
\tableofcontents

\section{Introduction}\label{intro}

The gauged linear sigma model (GLSM) with $\mathcal{N}=(2,2)$ supersymmetry has been a thriving research area that has spawned many topics in both physics and mathematics. See \cite{Sharpe:2024dcd} for a review of recent developments. In particular, the Fayet-Iliopoulos (FI) parameter space of a GLSM is divided into several chambers, each of which flows to an IR phase, such that it unifies different theories by a single GLSM \cite{Witten:1993yc}. 

The GLSM boundary conditions that preserve the B-type supersymmetry form the category of B-branes. The objects of this triangulated category for a GLSM are the equivariant matrix factorizations of the superpotential \cite{Sharpe:1999qz, Douglas:2000gi,Hori:2000ic,Hellerman:2001bu,Hori:2004zd,Herbst:2008jq}. Moreover, to define a GLSM B-brane, one needs to further assign a boundary condition for the twisted chiral fields \cite{Herbst:2008jq,Hori:2013ika}. This will be the key ingredient in our analysis. The UV-to-IR projection is a categorical quotient that maps GLSM B-branes to a subcategory consisting of the IR B-branes. If the low energy behavior of a phase can be described by a nonlinear sigma model (NLSM) with target space $X$, then the B-brane category projects onto $D^bCoh(X)$, or simply $D^b(X)$, the derived category of coherent sheaves on $X$ \cite{Aspinwall:2004jr,Govindarajan:2005im,Brunner:2006tc,Herbst:2008jq}. However, there exist more complicated sigma models in the IR that are not described by smooth geometry. One way to describe such a space is to regard the B-brane category as the noncommutative crepant resolution (NCCR) of a singular stack \cite{van2004non} (For its applications in GLSM see \cite{Addington:2012zv,Guo:2021aqj,Guo:2025yed}). In the singular case, this resolution essentially defines the IR phase, instead of a smooth target geometry. By lifting IR branes to the UV, and then projecting to another IR phase, \textit{B-brane transport} provides a functor mapping between different IR B-brane categories \cite{Herbst:2008jq, Hori:2013ika, Clingempeel:2018iub}. (See also \cite{Brunner:2021ulc} for a description of GLSM interfaces.) In a GLSM, this is essentially realized by analytically continuing FI parameters across different chambers. In certain cases, this functor gives an equivalence between the categories of B-branes. In particular, if two phases are described by two NLSMs with different target spaces $X_+$ and $X_-$ and the functor induced by the brane transport is an equivalence, then we get a derived equivalence between $X_+$ and $X_-$, i.e. $D^b(X_+) \cong D^b(X_-)$\footnote{If $X_+$ or $X_-$ is singular, we still use $D^b(X_\pm)$ to denote the category of B-branes in the corresponding phase, which defines a noncommutative resolution of the singular space.}. Therefore, B-brane transport in GLSMs has been used to study a large class of manifolds and their derived equivalences. (See also \cite{Segal:2010cz,halpern2015derived,Ballard:2016ncw} for developments in the mathematical literature.)

The B-brane transport by moving FI parameters depends mainly on two criteria from the GLSM \cite{Herbst:2008jq,Hori:2013ika,Clingempeel:2018iub}: 
\begin{enumerate}
    \item the {\it renormalization group} (RG) flow of FI parameters according to the anomaly of $U(1)_R$, the axial R-symmetry group,
    \item the \textit{grade restriction rule} (GRR) from the convergence in the Mellin-Barnes integral representation of hemisphere partition function (B-brane central charge).
\end{enumerate}
In the non-anomalous case, {\it i.e.} the Calabi-Yau (CY) case, where the axial $U(1)$ R-symmetry is anomaly free, the FI parameters are marginal and they parametrize a \textit{bona fide} complexified K\"ahler moduli space of the CY target space in a geometric phase. There are codimension-one singular loci in the K\"ahler moduli space, where non-compact Coulomb or mixed branches appear \cite{Witten:1993yc,Aspinwall:2001zq,Hori:2006dk}. Typically, these singular loci lie between the common boundary of K\"ahler chambers, so the path along which the brane is transported must avoid these loci. Then given such a path, only the branes satisfying the grade restriction rule can be smoothly transported from one phase to another. The branes satisfying the grade restriction rule constitute a subcategory, called the {\it window category}, of the category of matrix factorizations of the GLSM, which is equivalent to the IR B-brane category \cite{Herbst:2008jq}. The grade restriction rule depends on a choice of the $\theta$-angle, thus there are infinitely many windows from shifting the $\theta$-angle by $2\pi\ZZ$. A homotopically different path that is separated by the singular loci corresponds to a different window category. For abelian GLSMs, every path is homotopic to a straight line. For non-abelian GLSMs, the situation is more complicated, there are paths that are not homotopic to a straight line.
Therefore, even when the $\theta$-angle is restricted to a fixed interval of width $2\pi$ along the paths, different homotopy classes of the paths can give rise to different window categories. For this development in both the physics and mathematics literature, see \cite{halpern2016autoequivalences,Donovan:2013gia,Gerhardus:2015sla,Hori:2016txh,Lin:2024fpz, Guo:2025yed, Hori:202x}. In this paper, we only consider the paths that are homotopic to straight lines.

In asymptotically-free anomalous GLSMs (Fano case), lowering the energy scale triggers a negative flow on renormalized FI parameters according to \cite{Witten:1993yc}. At a UV gauge decoupling limit, the GLSM is a NLSM of a Fano variety. In the IR, the vacuum solutions are rich, including a Coulomb branch or a Higgs branch accompanied by a mixed branch. The latter case will be the main topic in this paper. There are no singular loci in the parameter space of this model, i.e. there are no non-compact Coulomb or mixed branches at phase boundaries. However, the grade restriction rule still exists since the variation of FI parameters (from positive to negative) is effectively given by their negative RG flow. The IR B-brane category is equivalent to the {\it big window category} that consists of the GLSM branes that satisfy the grade restriction rule. However, due to the rich vacuum structures in anomalous GLSMs, the IR limit of a B-brane could have components on either Higgs or Coulomb/mixed branches. The B-branes which only have Higgs branch components in the IR constitute a subcategory of the big window category, called the {\it small window category}. In summary, the B-brane category in an anomalous GLSM has the following structure:
\[
\begin{gathered}
\text{\textbf{Matrix factorization}}:\\
\text{B-branes in anomalous GLSM}
\\
\cup
\\
\textbf{Big  window}\ \mathbb{W}_{\text{big}}: 
\\
\text{B-branes in IR (both Higgs and mixed branches)}
\\
\cup
\\
\textbf{Small window}\ \mathbb{W}_{\text{small}}:
\\
\text{B-branes that only flow to Higgs branch}
\end{gathered}
\]

The general grade restriction rule for an abelian GLSM was first derived for anomaly-free cases in \cite{Herbst:2008jq} and later for anomalous cases in \cite{Clingempeel:2018iub}. Also see \cite{Acosta:2014cma,Galkin2016Gamma}. However, the GRR for non-abelian GLSMs is not well understood except for some  Pfaffian/Grassmannian type\cite{Eager2017,Donovan:2020jfh,Knapp:2023izn,Guo:2025yed} and quasi-symmetric models (GLSM with chiral matters in balanced representations and their duals) \cite{Donovan:2013gia,halpern2020combinatorial,Lin:2024fpz}. At this stage, there are no general results; one must instead resort to a case-by-case study. In particular, the analysis for non-abelian anomalous models has not been done so far.

\subsection{Summary of results}

In this paper, we initiate the investigation on the grade restriction rule and derived equivalence in the IR phases of anomalous non-abelian GLSMs. 

In the first part, we investigate the small window of the following type of non-abelian GLSMs: A $U(2)$ GLSM with chiral fields $\Phi_i\in\CC^{2n}$ in $n$-copies of fundamental representations and chiral fields $P_I\in\CC^K$ in $K$-copies of anti-determinantal representations, each with $\det U(2)$ weight $m_I$, 
\begin{equation}
    \begin{array}{c|cccccc|c}
     & \Phi_1 & \cdots & \Phi_n & P_1 & \cdots & P_K &
    \\\hline
    U(2) & \yng(1) &\cdots & \yng(1) & \det^{-m_1} & \cdots & \det^{-m_K} & \xi
    \end{array}
\end{equation}
with a renormalized FI parameter $\xi$ for the $U(2)$ gauge group. This model, which provides the grade restriction rule and the big/small windows, will serve as the local model on the chamber boundary when studying models with multiple FI parameters. 
The condition for this GLSM to be anomalous and asymptotically free (Fano) is
\begin{equation}
    m_1+\cdots +m_K <n,
\end{equation}
and the corresponding UV Fano target space at $\xi>0$ is the total space of determinantal line bundles over the Grassmannian:
\begin{equation}
   X_{\xi>0}= \Tot\left( \oplus_{I=1}^K {\cal O}(-m_I)\rightarrow Gr(2,n) \right),
\end{equation}
where $\cal{S}$ denotes the tautological bundle and ${\cal O}(-l)={\det}^l{\cal S}$. In the IR phase $\xi<0$, there will be a Higgs branch and several mixed branches. The geometry of the Higgs branch is given by a (possibly singular) GIT quotient\footnote{For generic $m_I$, the GIT quotient is a gerby $Gr(2,n)$ affine cone fibered over a weighted projective space.}. 
Meanwhile, there is also a mixed branch $\mathcal{C}_{\xi<0}$ at $\xi<0$. For the GLSM above, the mixed branch corresponds to the vacua where the maximal torus of the gauge group is partially broken. 

In the geometric phase, let $p: X_{\xi>0}\rightarrow Gr(2,n)$ be the projection map from the total space to the base space. The category of IR B-branes can be generated by pull-backs of the coherent sheaves on $Gr(2,n)$ under $p$, thus they are homogeneous bundles determined by $U(2)$ representations labeled by Young diagrams.

In Sec.~\ref{sec:small}, the small windows of the above model with $n=3,4$ will be analyzed since they are directly related to our derived equivalences. The $n=3$ models are dual to abelian models and our result is found to agree with \cite{Clingempeel:2018iub}. Thus this paper will focus on the anomalous restriction rule of the $Gr(2,4)$ models. The big window of $Gr(2,4)$ model was analyzed in \cite{Eager2017,Guo:2025yed}, which gives the restriction rule Eq.~\eqref{Aq} and a big window as the Lefschetz exceptional collection on $Gr(2,4)$:
\begin{equation}
\begin{aligned}
    \mathbb{W}_{\text{big}}=&\left\langle \cdot,\quad \overline{\yng(1)},\quad \overline{\yng(1,1)},\quad \overline{\yng(2,1)},\quad \overline{\yng(2,2)},\quad \overline{\yng(3,3)} \right\rangle
    \\
    =&\langle {\cal O}, \det{\cal S},{\det}^2{\cal S}, {\det}^3{\cal S},\ {\cal S},\ {\cal S}\otimes \det{\cal S} \rangle.
\end{aligned}
\end{equation}
Then by analyzing the saddle point contribution from the contour Eq.~\eqref{squarecontour}, the small window for each case is given by Eq.~\eqref{smallwindowN3}-~\eqref{smallwindowN4} as
\begin{itemize}
\item $n=3$
\begin{equation}\label{small3}
{\setstretch{2}
\begin{array}{ll}
m=1,2,3, & \mathbb{W}_{\text{small}}=\langle {\cal{O}},\ \det{\cal S}, \cdots, {\det}^m{\cal S} \rangle
\end{array}}
\end{equation}
\item $n=4$
\begin{equation}\label{small4}
{\setstretch{2}
\begin{array}{ll}
m=1, & \mathbb W_{\mathrm{small}}\:=\:\left\langle {\cal O},\ \det{\cal S},\ {\cal S} \right\rangle
\\
m=2,3, & \mathbb W_{\mathrm{small}}\:=\:\left\langle \det{\cal S},\ \cdots,\ {\det}^m{\cal S},\ {\cal S},\ {\cal S}\otimes \det{\cal S} \right\rangle\\
m=4, & \mathbb{W}_{\text{small}} = \langle {\cal O}, \det{\cal S},{\det}^2{\cal S}, {\det}^3{\cal S},\ {\cal S},\ {\cal S}\otimes \det{\cal S} \rangle
\end{array}
}
\end{equation}
\end{itemize}
where $m=m_1+\cdots+m_K$. 
The cases with $m=n$ are Calabi-Yau, and we find $\mathbb{W}_{\text{big}}=\mathbb{W}_{\text{small}}$ as expected.
In the non-Calabi-Yau cases, i.e. $m<n$, the small window is also tested by counting the Witten indices of the Higgs branch and the mixed branches. 

\medskip

Sec.~\ref{sec:example} is devoted to the application of small windows to the derived equivalence between two manifolds realized as the IR Higgs branches in different phases of a two-parameter GLSM. To apply the results of Sec.~\ref{sec:small}, $U(1)\times U(k)$ models with $k=1,2$ will be considered. We will derive a necessary condition on the representations of the gauge groups for the derived equivalence at the beginning of Sec.~\ref{sec:example}. We denote the renormalized FI parameters of the GLSM by $\xi_{1,2}$. The IR NLSM at $\xi_{1,2}>0$ has a Fano target space $\hat{X}$, and two other phases with either $\xi_1<0$ or $\xi_2<0$ consisting of Calabi-Yau Higgs branches $X_\pm$ and mixed branches $\mathcal{C}_\pm$. The derived equivalence between $X_+$ and $X_-$ is implemented by the variation of FI parameters and B-brane transport. B-branes as coherent sheaves that are supported on $X_+$ will be lifted to GLSM branes (matrix factorizations) through a small window category $\mathbb{W}_+$. They flow to complexes of sheaves on $\hat X$ in the $\xi_{1,2}>0$ phase. Then sheaves on $\hat X$ can also be lifted and projected to branes as sheaves supported on $X_-$ through another small window $\mathbb{W}_-$. The functor, sketched in Fig.~\ref{fig:idea}, induces the derived equivalence between $D^b(X_+)\cong\mathbb{W}_+$ and $D^b(X_-)\cong\mathbb{W}_-$.
\begin{figure}[!h]
    \centering
    \[
    \begindc{\commdiag}[300]
    \obj(1,3)[G]{GLSM}
    \obj(1,1)[xh]{$\hat X$}
    \obj(-1,0)[xp]{$X_+,\mathcal C_+$}
    \obj(3,0)[xm]{$X_-,\mathcal C_-$}

    \obj(-3,3)[uv]{UV}
    \obj(-3,0)[ir]{IR}
    \mor{uv}{ir}{}[\atright,\solidarrow]

    \mor{G}{xh}{$\hat\pi$}
    \mor{G}{xp}{$\pi_+$}
    \mor{G}{xm}{$\pi_-$}

    \mor{xp}{xh}{$\mathbb W_+$}[\atright,\dashArrow]
    \mor{xh}{xm}{$\mathbb W_-$}[\atright,\dashArrow]

    \enddc
    \]
    \caption{Schematic diagram for the derived equivalence. $\mathbb{W}_\pm$ denote the small windows, $\hat{\pi}$ and $\pi_{\pm}$ are the projection functors from the UV B-branes to the IR B-branes in the respective phase.}
    \label{fig:idea}
\end{figure}

One necessary condition for $X_+$ and $X_-$ to be derived equivalent under RG effects is that their Witten indices, that is, their Euler characteristics, should be equal to each other:
\begin{equation}
    \chi(X_+)=\chi(X_-)
\end{equation}
and the total Witten indices in different phases are also the same:
\begin{equation}
     \chi(\hat{X}) \:=\: \chi({X}_+) + \chi(\mathcal{C}_+) \:=\: \chi({X}_-) + \chi(\mathcal{C}_-)\,,
\end{equation}
where $\chi(\mathcal{C}_\pm)$ is the total contribution of the Coulomb/mixed branches. As will become clearer in Sec.~\ref{sec:example}, this imposes nontrivial conditions on $X_\pm$ and $\hat{X}$. In particular, the Witten indices of the mixed branches $\mathcal{C}_{\pm}$ can be computed by counting GLSM Coulomb vacua. The matching of the Witten indices of $\mathcal{C}_{\pm}$ induces the same constraint due to the fact that the total Witten index is always the same.

In conclusion, we construct the following families of GLSMs with gauge group $U(1)\times U(k)$, $k=1,2$ that realize a UV sigma model $\hat X$ and mixed IR phases with Higgs branches $X_\pm$. The corresponding target spaces $\hat X$ and $X_\pm$ are listed below: 
\begin{equation}
            \begin{array}{c|c|c}
            X_+ & \hat X & X_-
            \\\hline
            &  & \\
              \frac{\mathcal{O}_{\mathbb{P}^{M-1}}(-m/n)^{\oplus N}}{(\ZZ_{n/g_{nm}}\times \ZZ_{g_{nm}})}   
              & 
              \mathcal{O}_{\mathbb{P}^{M-1}\times\mathbb{P}^{N-1}}(-m,-n) 
              &
              \frac{\mathcal{O}_{\mathbb{P}^{N-1}}(-n/m)^{\oplus M}}{(\ZZ_{m/g_{nm}}\times \ZZ_{g_{nm}})}
            \\
            &  & \\
            \frac{\mathcal{X}_{\mathbb{P}^{M-1}}(2,4)}{(\ZZ_{n/g_{nm}}\times \ZZ_{g_{nm}})}   & 
            (\mathcal{L}^{\otimes m}\otimes(\wedge^2\mathcal{S})^4)_{\PP^{M-1}\times Gr_{24}}& 
            \frac{\mathcal{O}_{Gr_{24}}(-4/m)^{\oplus M}}{(\ZZ_{m/g_{m4}}\times \ZZ_{g_{m4}})}
            \\
            &  & \\
            \left[\mathcal{L}^2\otimes\mathcal{L}^{\perp}\right]_{\mathbb{P}^{2}}  
            & 
            (\mathcal{L}\otimes(\wedge^2\mathcal{S}))_{Fl_{123}}\mathcal{O}_{Fl_{123}}(-1,-\det) 
            & 
            \left[ \mathcal{S}\otimes \wedge^{2}\mathcal{S}\right]_{Gr_{23}}
            \\
            &  & \\
            \left[(\mathcal{L}^\perp/\mathcal{L})\otimes\mathcal{L}^2\right]_{\mathbb{P}^3} & (\mathcal{L}\otimes\wedge^2\mathcal{S})_{SFl_{124}} & \left[\mathcal{S}\otimes\wedge^2\mathcal{S}\right]_{SG_{24}}\\
            &  & \\\hline
            \end{array}
\end{equation}
where $g_{nm}=\gcd(n,m)$, and ${\cal{L}}$ and $\mathcal{S}$ denote respectively the (pull back of) tautological (line) bundle over projective space and the Grassmannian. Note that $X_\pm$ are local Calabi-Yau varieties. The derived equivalence of Calabi-Yau 5-folds via symplectic flag variety $SFl(1,2,4)$, known as the Abuaf flop, in the last example was studied in \cite{Segal_2016}. We also obtain an explicit description of the functor realizing the derived equivalence in each case through the method of brane transport.
\medskip

The organization of this paper is the following: In section~\ref{sec:review}, we will give a review of GLSM, hemisphere partition functions and brane transport. Then, we generalize the idea to $U(2)$ gauge theories and derive the window categories in section~\ref{sec:small}. In section~\ref{sec:example}, we first derive a necessary condition on the GLSMs for derived equivalence, and then explicitly study several concrete examples. We start with an abelian gauge theory with gauge group $U(1)\times U(1)$ to illustrate the idea, then we study the non-abelian theories with gauge group $U(1)\times U(2)$ including the theory corresponding to the derived equivalence between the Calabi-Yau fivefolds mentioned above. We give some concluding remarks and potential future directions in section~\ref{sec:conclusion}. The appendix includes supplementary materials including counting of Coulomb vacua and the analysis of the GLSMs for flag manifolds.

\section{Review on hemisphere partition functions and window categories}
\label{sec:review}
In this section, we give a short review of the hemisphere partition functions of GLSMs, following \cite{Herbst:2008jq,Hori:2013ika,Clingempeel:2018iub} and references therein. In particular, we review how we identify big/small window categories from the hemisphere partition function of B-branes. 

\subsection{GLSM and geometric phases}
The $2$d $\mathcal{N}=(2,2)$ gauged linear sigma model (GLSM) consists of the following data \cite{Witten:1993yc} (also see, for example, \cite{Hori:2019vkm}):
\begin{itemize}
    \item \textbf{(Gauge group and R-symmetry)} A compact complex Lie group $G$ and a $U(1)_R$ R-symmetry.

    \item \textbf{(Twisted chiral fields)} The vector multiplet valued in $\mathfrak{g}_{\mathbb{C}}$, the complex Lie algebra of $G$. In particular, we will denote the field strength by $\sigma \in \mathfrak{h}_G$ the Cartan subalgebra of $\mathfrak{g}$.

    \item \textbf{(Chiral fields)} A complex vector space $V$, constituted by chiral fields $\Phi_I$ (with scalar component denoted as $\phi_I$), that carries a $G$ representation $\rho_G$ and a $U(1)_R$ representation $R$, and a moment map: $\mu_G(\xi):V\rightarrow \mathfrak{g}^\vee$. In particular, the \textbf{gauge charge} $Q_I$ and \textbf{R-charge} $R_I$ are the Cartan weight of $G\times U(1)_R$ on each chiral field:\footnote{We denote the pairing between Cartan weights $Q^a_I$ and Cartan sub-algebra elements $\sigma_a$ as $Q_I(\sigma)=\sum_aQ_I^a\sigma_a$}
    \begin{equation}
    \begin{aligned}
        \rho_{G}(\sigma)\cdot\phi_I &=(\exp Q_I(\sigma))\phi_I\,,
        \\
        R(\lambda)\cdot \phi_I &=\lambda^{R_I}\phi_I,\quad \lambda\in U(1)_R \,.
    \end{aligned}
    \end{equation}

    \item \textbf{(Superpotential)} A generic quasi-homogeneous $G$-invariant polynomial $W(\phi)\in\mathrm{Sym}_G(V^\vee)$ with R-charge $2$: $W(R(\lambda) \phi)=\lambda^2 W(\phi)$. The polynomial coefficients are chiral parameters, but they will not be needed in this paper.

    \item \textbf{(Twisted superpotential)} A linear form of $\sigma$: $\widetilde{W}_{\mathrm{cl}}(\sigma,t)=t_{\text{cl}}(\sigma)$, with (classical) twisted-chiral parameters (FI-$\theta$ parameter) in the dual center of $G$: $t_{\text{cl}}=(\xi-i\theta)\in\mathfrak{z}^\vee_G$. 
\end{itemize}

The IR behavior of the GLSM depends on the vacuum expectation value of the scalar fields, which in turn are determined by the FI-parameters, thus the FI-parameters space (K\"ahler moduli space) is divided into several chambers called phases, each with a specific IR description in the asymptotic region, which has three possibilities \cite{Hori:2006dk,Hori:2011pd,Clingempeel:2018iub}, where $\sigma$ and $\phi$ below denote the vev of the scalar components of the vector multiplets and chiral multiplets respectively:
\begin{itemize}
\item {\bf Higgs branch}: The scalar fields in the vector multiplet have zero vev while the scalar fields in the chiral multiplets acquire nonzero vev whose IR behavior can be described by a nonlinear sigma model (NLSM), a Landau-Ginzburg (LG) model, or more generally a hybrid model. In this case the vev breaks the gauge group completely or to a finite subgroup.  
\item {\bf Coulomb branch}: In this case the scalar components of the vector multiplet with values in the Cartan subalgebra have nonzero vev and the gauge group is broken to its maximal torus. The low energy degrees of freedom is given by the scalar fields in the Cartan subalgebra constrained by an effective twisted superpotential. The vacuum states correspond to the critical points of the twisted superpotential.
\item {\bf Mixed branch}: On this branch, some of the Cartan scalar fields have nonzero vev, while some of the matter fields also have nonzero vev. The latter can be described by an NLSM, an LG model, or a hybrid model. The gauge group is broken to a nontrivial subgroup with positive rank of its maximal torus. See Sec. 2.4 of \cite{Clingempeel:2018iub} for more details on mixed branches.
\end{itemize}

On a Higgs branch or mixed branch where the IR theory is described by an NLSM, the target space is the classical bosonic vacuum of chiral fields w.r.t. the FI-parameter $\xi$. This bosonic vacuum is given by the critical locus of $W(\phi)$ in the GIT quotient \cite{thomas2005notes}
\begin{equation}
    X_\xi=\operatorname{Crit}(W)\cap(V//_\xi G)\,,
\end{equation}
where the critical locus of $W(\phi)$ (F-term equations) is
\begin{equation}
    \operatorname{Crit}(W):\ \{ \partial W(\phi)=0 \}\,,
\end{equation}
and the GIT quotient (or ``double quotient'') is the quotient stack of the semi-stable loci from $\mu_G$ (D-term equations):
\begin{equation}
    V//_\xi G:\quad [(V-\Delta_{\text{un}}(\xi))/G]:=[\mu_G^{-1}(\xi)/G]\,,\label{eq:GIT}
\end{equation}
where $\Delta_{\text{un}}(\xi)$ is the unstable locus with respect to the stability condition $\xi$, which is deleted from $\mu^{-1}_G(\xi)\subset V$ under the quotient. 

The Calabi-Yau (anomaly free) condition is that $\rho_G:\ G\rightarrow GL(V)$ factors through $SL(V)$, which means the sum of gauge charges is zero. As a consequence, variation of GLSM FI parameter has no renormalization effect and it forms a bona fide  ``stringy K\"{a}hler moduli'' of SCFTs. See also \cite{Morrison:1994fr,Aspinwall:2004jr}.

In this paper, we will mainly investigate anomalous GLSMs with geometric phases \cite{Herbst:2008jq,Clingempeel:2018iub}. The Calabi-Yau condition is replaced by the Fano condition that the sum of gauge charges (that is, the total weights of $\det G\cong U(1)$) is positive. This positive sum triggers an RG flow of the FI parameter (thus the K\"{a}hler class) such that $t=t_R(\Lambda)$ is renormalized w.r.t. a UV cut-off $\Lambda$ as in Eq.~\eqref{tR}. This has two consequences. First, a GLSM still has a Fano NLSM phase as the classical GIT quotient $\hat X:=X_{\xi>0}$ in the positive K\"{a}hler cone, since the corresponding gauge theory is asymptotically free and UV-complete \cite{Clingempeel:2018iub}. Second, the IR phases may consist not only of Higgs branches but also of mixed branches or Coulomb branches. Consider a specific RG flow direction on the moduli space given by an FI parameter $\zeta$, which parametrizes a one-dimensional subspace of the moduli space corresponding to the gauge subgroup $G_\zeta\subset G$ and gauge charge $Q_I^\zeta$. Then after the RG flow to $\zeta<0$, the classical IR phase from the GIT quotient will be one of the following two cases:
\begin{enumerate}
    \item Empty if $Q^\zeta_I>0$ for all $I$;
    \item A Higgs branch $X_{\zeta<0}$, if there exists $Q^\zeta_I<0$ for some $I$.
\end{enumerate}
In the first case, there is no classical solution to D-term equations when $\zeta<0$, while, for the second case, there are classical solutions but only to D-term equations associated with a subgroup of the gauge group. However, by considering quantum correction to the $\widetilde{W}_{\text{cl}}$, there will in fact be additional Coulomb vacua for both cases determined by the Weyl invariant loci of
\[
    \exp(\partial \widetilde{W}_{\text{eff}}(\sigma)) = 1\,,
\]
where $\widetilde{W}_{\text{eff}}(\sigma)$ is the \textbf{effective twisted superpotential} and it is obtained by one-loop corrections to the twisted superpotential after integrating out the chiral fields \cite{Witten:1993yc,Hori:2006dk}:\footnote{Note that a W-boson contribution from non-abelian $G$ was omitted in the references, but is recovered in \cite{Hori:2013ika,Honda:2013uca}.}
\begin{equation}
    \widetilde{W}_{\text{eff}}(\sigma,t)=-t_{\mathrm{cl}}(\sigma)+i\pi\sum_{\alpha>0}|\alpha(\sigma)|+\sum_{I=1}^{\dim V}Q_I(\sigma)\left[ \log \left(Q_I(\sigma)/\Lambda\right) -1\right],
\end{equation}
where $\alpha>0$ denotes the positive roots of $G$ and $\Lambda$ is the UV cut-off. The renormalized FI-parameters are defined by \textcolor{black}{absorbing the cut-off $\Lambda$ into FI-paramters as}
\begin{eqnarray}
\label{tR}
    t(\sigma)\equiv t_R(\Lambda,\sigma)=t_{\mathrm{cl}}(\sigma)-\log(\Lambda)\sum_{I=1}^{\dim V} Q_I(\sigma). 
\end{eqnarray}
After compensating for the quantum Coulomb vacua and the solution of mixed branches, the full IR phase for $\zeta <0$ can be described by
\begin{enumerate}
    \item Pure Coulomb branch if all $Q^\zeta_I>0$,
    \item A Higgs branch $X_{\zeta<0}$ and mixed branches $\mathcal{C}_{\zeta<0}$ from quantum Coulomb vacua, if there exists $Q^\zeta_I<0$.
\end{enumerate}
The crucial fact for our analysis in the mixed branches is that the Witten index is an RG flow invariant of the GLSM: In the UV, it is counted by the Euler characteristic of the Fano variety, while in the IR, it will be the sum of the Euler characteristic of the Higgs branch and the Witten indices of the mixed branches.

\subsection{B-branes in GLSM}

The GLSM boundary conditions that preserve the B-type supersymmetry are given by the triplet $\mathcal{B}=(\gamma,\mathbf{T},M)$, which is known as \textbf{GLSM equivariant matrix factorization} \cite{Herbst:2008jq,Hori:2013ika}:
\begin{itemize}
    \item \textbf{(Dirichlet b.c. of $\sigma$)} A middle-dimensional contour $\gamma\subset \mathfrak{g}_\CC$ that is a deformation of the real loci in $\mathfrak{g}_\CC$.

    \item \textbf{(Chan-Paton space)} A $\mathbb{Z}_2$-graded, finite dimensional $\mathrm{Sym}(V^\vee)$-module $M=M_0\oplus M_1$ which carries a representation of $G$ as $\rho_M:\ G\rightarrow GL(M)$ and a representation of $U(1)_R$ as $R_M:\ U(1)_R\rightarrow GL(M)$. Such an equivariant module $M$ splits into a direct sum of \textbf{Wilson line branes} \cite{Herbst:2008jq}:
    \begin{eqnarray}
        M=\bigoplus_{i=1}^{\dim M} \mathcal{W}(q_i)_{r_i}\,
    \end{eqnarray}
    where each Wilson line brane component of $M$ carries \textbf{boundary gauge charge} $q_i$ and \textbf{boundary R-charge} $r_i$ as the Cartan weights of $G\times U(1)_R$:\footnote{Again, $q(\sigma)$ denotes the pairing between weights $q$s and Cartan subalgebra elements $\sigma$s.}
    \begin{equation}
        \begin{aligned}
            \rho_M(\sigma)\cdot \mathcal{W}(q)_{r}&=(\exp q(\sigma))\mathcal{W}(q)_{r}\,,
            \\
            R_M(\lambda)\cdot \mathcal{W}(q)_{r} &= \lambda^{r} \mathcal{W}(q)_{r}\,.
        \end{aligned}
    \end{equation}

    \item \textbf{(Tachyon profile)} An odd equivariant matrix factorization of the superpotential that is an element $ \mathbf{T}(\phi)\in\mathrm{End}^1_{\mathrm{Sym}(V^\vee)}(M)$ such that
    \begin{equation}
        \begin{aligned}
            \rho_M(g)^{-1}\cdot \mathbf{T}(\rho_G(g)\phi)\cdot\rho_M(g)=\mathbf{T}(\phi)\,,
            \\
            R_M(\lambda)\cdot \mathbf{T}(R(\lambda) \phi)\cdot R_M(\lambda)^{-1}=\lambda \mathbf{T}(\phi)\,,
        \end{aligned}\label{eqn:QRep}
    \end{equation}  
    and satisfying
    \begin{eqnarray}
        \mathbf{T}(\phi)^2=W(\phi)\cdot\mathrm{Id}_M\,.
    \end{eqnarray}

\end{itemize}

If the superpotential $W(\phi)$ can be written as $W(\phi)=\sum_if_i(\phi)g_i(\phi)$ with a regular sequence $\{ f_i\}$ or $\{ g_i\}$ ({\it e.g.} they form a complete intersection), a presentation of $\mathbf{T}(\phi)$ can be constructed by
\begin{equation}
    \mathbf{T}(\phi)=\sum_i g_i(\phi)\eta_i+f_j(\phi)\overline{\eta}_j\,,
\end{equation}
where $\eta_i$ and $\overline\eta_j$ form a complex Clifford algebra 
\begin{equation}
    \{ \eta_i,\ \overline\eta_j \}=\delta_{ij}\, , \quad \{ \eta_i,\ \eta_j  \}= \{ \overline\eta_i,\ \overline\eta_j \}=0\, .
\end{equation}
The corresponding B-brane with a tachyon profile $\mathbf{T}(\phi)$ on the Clifford module $M=\bigoplus_k(\wedge^k\overline{\eta}_i)|q\rangle_r$ generated over Clifford vacuum $|q\rangle_r$ (which carries boundary gauge charge $q$ and R-charge $r$) can be written in the form of the following double-arrowed B-brane complex:
\begin{equation}
\begin{aligned}
(\mathbf{T},M)=&\xymatrix{
   \bigg(\cdots \ar@<2pt>[r]^-{\mathbf{T}} & \ar@<2pt>[l] \bigoplus_{i<j} \mathcal{W}(q_{ij})_{r+2}
 \ar@<2pt>[r]^-{\mathbf{T}} & \ar@<2pt>[l] \bigoplus_i\mathcal{W}(q_i)_{r+1} \ar@<2pt>[r]^-{\mathbf{T}} & \ar@<2pt>[l] \mathcal{W}(q)_r \bigg)}
\\
=&\xymatrix{
   \bigg(\overline{\eta}_1\cdots\overline{\eta}_m|q\rangle_r \ar@<2pt>[r]^-{g_i} & \ar@<2pt>[l]^-{f_i} \cdots \ar@<2pt>[r]^-{g_i} & \ar@<2pt>[l]^-{f_i} \bigoplus_{i<j}\overline{\eta}_i\overline{\eta}_j |q\rangle_r
 \ar@<2pt>[r]^-{g_i} & \ar@<2pt>[l]^-{f_i} \bigoplus_i\overline{\eta}_i|q\rangle_r \ar@<2pt>[r]^-{g_i} & \ar@<2pt>[l]^-{f_i} {|q\rangle_r} \bigg),
}
\end{aligned}
\end{equation}
such that each term is a sum of Wilson line branes of some gauge charges and R-charges determined by the representations of tachyon profile and Clifford vacuum state. Specifically, B-brane complex $(\mathbf{T},M)$ is a chain complex when $W=0$. In particular, one can take a factorization as a one-term complex with $\mathbf{T}=0$ and $M=\bigoplus_i\mathcal{W}(q_i)_0$ in any representation of $G$ with Cartan weights $q_i$, which represents a homogeneous vector bundle in the category $D^b(V//_\xi G)$. 

The GIT quotient structure of $X_\xi$ gives a similar structure of $MF_G(W)$ by the structure of $\Delta_{\text{un}}(\xi)$, see\cite{Herbst:2008jq,halpern2015derived,Ballard:2016ncw}. Suppose first $W=0$ and $X_\xi=V//_\xi G$. Fixing a K\"{a}hler cone of $\xi$, the Koszul complex of $\mathcal O_{\Delta_{\text{un}}(\xi)}$ is called \textbf{empty brane of $X_{\xi}$} since it is acyclic in $D^b(X_{\xi})$:
\begin{equation}
H^*(X_{\xi}, \mathcal O_{\Delta_{\text{un}}(\xi)})=0.
\end{equation}
Every phase has an associated set of empty branes and thus different structures of the B-brane category under GIT quotient. This correspondence between matrix factorization category (GLSM B-branes) and derived category of coherent sheaves on $X_\xi$ w.r.t. $\xi$ (NLSM B-branes) will factor through window categories:
\begin{equation}
   \pi_{\xi}^{(l)}:\quad  MF_G(W)\xrightarrow{}\mathbb{W}_l\cong D^b(X_{\xi}).\label{eqn:window}
\end{equation}
When $W\neq0$, the same argument is applicable since every object in $D^b(X_{\xi})$ has a lifted representative along $\pi_{\xi}^{(l)}$ in $MF_G(W)$ up to the subcategory generated by $\mathcal{O}_{\Delta_{\text{un}}(\xi)}$ and index $l$ \cite{Ballard:2016ncw}. 

It is worth remarking that such construction does not directly apply to a non-abelian GIT quotient, which is the main interest of this paper, due to the fact that $\Delta_{\text{un}}(\xi)$ is in general a non-complete intersection (hence, $\cO_{\Delta_{\text{un}}}$ is not a Koszul complex). For example, in the GIT quotient $Gr(k,n)=\{V=\mathrm{Mat}_{k\times n}(\CC)\}//U(k)$, the unstable locus $\Delta_{\text{un}}\subset V$ is the determinantal loci of rank $(k-1)$ and lower. The B-brane complexes on such loci are known as the Eagon-Northcott complexes \cite{eagon1962ideals}. In particular, on $Gr(2,4)$, the following exact sequences (and their variations) in $U(2)$ representations will be crucial for later analysis:\cite{weyman2003cohomology, Donovan:2013gia}
\begin{equation}\label{eqn:ENcpx}
\begin{aligned}
    &0\rightarrow\overline{\yng(3,1)} \rightarrow  \overline{\yng(2,1)}^{\oplus 4} \rightarrow  \overline{\yng(1,1)}^{\oplus 6} \rightarrow {\cO}\rightarrow0,
\\
    &0\rightarrow\overline{\yng(3,2)} \rightarrow  \overline{\yng(2,2)}^{\oplus 4} \rightarrow  \overline{\yng(1,1)}^{\oplus 4} \rightarrow \overline{\yng(1)}\rightarrow 0,
\\
    &0\rightarrow\overline{\yng(3,3)} \rightarrow  \overline{\yng(2,2)}^{\oplus 6} \rightarrow  \overline{\yng(2,1)}^{\oplus 4} \rightarrow \overline{\yng(2)}\rightarrow 0.
\end{aligned}
\end{equation}
The Young diagrams above stand for the images of the Schur functors of the corresponding Young diagrams acting on the tautological bundle. Each representation is a homogeneous bundle over $Gr(2,4)$, with chain maps between them as sections in $\phi_{I=1,\cdots,4}^{\alpha=1,2}\in \mathrm{Mat}_{2\times4}(\CC)$. Their lifted matrix factorizations indeed exist \cite{Lin:2024fpz}, however, it is neither obvious nor helpful to find the counterpart  $\mathbf{T}(\phi)$ by chain maps. We will not use the full matrix factorization for these loci. \textcolor{black}{As a final remark, in the rest of the paper, we may simply write a B-brane complex $(\mathbf{T},M)$ in terms of a chain complex with one-direction arrows (with R-charge assignment by homological degrees), due to the complexity of, and lack of need for, writing explicit tachyon profiles for most of B-branes. However, the relevant chain complexes can always be lifted to compatible matrix factorizations.}

In conclusion, the GLSM brane category, together with the appropriate window subcategories, organizes the categorical relations between the different phases. The variation of the FI parameter will not affect the B-type supersymmetry on the boundary because it is in the twisted chiral sector, however, when there is a topologically non-trivial path of deformation, the loop gives a monodromy action on B-brane category as ``derived symmetry''. See \cite{Aspinwall:2001zq,Herbst:2008jq,Aspinwall:2006ib,halpern2016autoequivalences, Guo:2025yed,Hori:202x,Lin:2024fpz}. Moreover, the connections between the derived categories of the phases are provided by the information about windows. 

To understand the IR behavior of window categories, we will introduce the effective twisted superpotential for B-branes by the hemisphere partition function, which will be the key to later analysis.

\subsection{Hemisphere partition function and window category}

The central charge of B-branes (A-branes) for $\mathcal{N}=(2,2)$ theories is defined by the partition function of the A-twisted (B-twisted) theory on a disk attached with an infinitely flat cylinder with its boundary condition as B-branes (A-branes) \cite{Cecotti:1991me,hori2000d}. Using supersymmetric localization techniques \cite{Hori:2013ika,Honda:2013uca,Sugishita:2013jca} the central charge for a B-brane $\mathcal{B}$ was computed in the context of A-twistable GLSMs as
\begin{equation}
Z_{\mathcal{B}}(t)\:=\:\int_\gamma  d^{h_G}\sigma \prod_{\alpha>0}\alpha(\sigma)\sinh\pi\alpha(\sigma)\ \prod_{I=1}^{\dim V}\Gamma\left(iQ_I(\sigma)+\frac{R_I}{2}\right)e^{it(\sigma)}f_\mathcal{B}(\sigma)\, ,
\label{ZB}
\end{equation}
where the \textbf{brane factor} of $\mathcal{B}$ is
\begin{equation}
f_{\mathcal{B}}(\sigma):=\mathrm{tr}_{M}\left(R_{M}(e^{i
\pi})\rho_{M}(e^{2\pi\sigma
} )\right)=\sum_{i=1}^{\dim M}e^{i\pi r_i}e^{2\pi q_i(\sigma)}.\label{brane factor}
\end{equation}
The FI-parameters $t(\sigma)\equiv t_R(\Lambda;\sigma)$ in an anomalous model are renormalized as discussed in Eq~\eqref{tR}. The symbol $\prod_{\alpha>0}$ denotes the product over the positive roots of $G$ and $h_{G}:=\mathrm{dim}(\mathfrak{t}_G)$.  The function Eq.~\eqref{ZB} is conjectured to coincide with the IR central charge as defined by \cite{Cecotti:1991me} for SCFTs, i.e. non-anomalous GLSMs, and to be related in a certain limit in the anomalous case (see \cite{Hori:2013ika,Hori:2019vkm}). It is important to remark that in Eq.~\eqref{ZB} the integration variable is dimensionless since it corresponds to $r\sigma$ (where $r$ is the radius of the hemisphere, or equivalently, the inverse of dynamical scale $\mu^{-1}$) but we denoted it $\sigma$ for simplicity. In addition, Eq.~\eqref{ZB} has a normalization of the form $C(r\Lambda)^{\frac{\hat{c}}{2}}$. We will ignore this factor in the following since it does not affect the window, but in general it should be considered for applications involving anomalous models \cite{Hori:2019vkm}.

We propose that the \textbf{B-brane effective twisted superpotential} $\widetilde W_{\text{eff}}(\sigma,q,t)$ is given by the Stirling's expansion of the integrand of Eq.~\eqref{ZB}:
\begin{equation}\label{eqn:expW}
\begin{aligned}
Z_\mathcal{B}(t) \sim&\int_\gamma d^{h_G}\sigma \exp {\widetilde{W}_{\text{eff}}(\sigma,q,t)}\\
=&\int_\gamma d^{h_G}\sigma\exp\bigg( \sum_{I=1}^{\dim V}iQ_I(\sigma)\big(\log Q_I(\sigma)-1\big)+\sum_{\alpha>0}\pi|\alpha(\sigma)|+it(\sigma)-\sum_{i=1}^{\dim M} 2\pi q_i(\sigma) \bigg)\,.
\end{aligned}
\end{equation}
This expression agrees with the result in \cite{Herbst:2008jq} for abelian GLSMs in the asymptotic region of $Z_{\mathcal{B}}$, but it has not been verified for general cases.\footnote{We thank the referee for pointing this out to us.} The real part of $\widetilde{W}_{\text{eff}}$ for a Wilson line brane factor $f_{\mathcal{W}(q)}(\sigma)=\exp2\pi q(\sigma)$ is given by\footnote{In principle, only a representation of $G$ can be considered as a B-brane, which is a gauge multiplet $M=\bigoplus_i\mathcal{W}(q_i)$ for higher rank representations. We do not need this requirement and a single component $\mathcal{W}(q)$ in $M$ can appear in the analysis of $\widetilde{W}_{\text{eff}}$. However, the resulting window should be written in representations at the end.}
\begin{equation}
\begin{aligned}
\mathrm{Re} \widetilde{W}_{\text{eff}}(\sigma,q,t)\:=\:&\sum_{\alpha>0}\pi|\alpha(\tau)|+\big( -\xi_R(\nu)+(\theta+2\pi q)(\tau) \big)\\
&+\sum_I Q_I(\nu)\big(\log|Q_I(\sigma)|-1\big)-|Q_I(\tau)|\left[\frac{\pi}{2}+\arctan\frac{Q_I(\nu)}{|Q_I(\tau)|}\right]\, ,
\end{aligned}
\end{equation}
with convention 
\[
\tau=\mathrm{Re}\,\sigma,\quad \nu=\mathrm{Im}\,\sigma,\quad \Arg(iz)=\operatorname{sgn}(\mathrm{Re}(z))\left(\frac{\pi}{2}+ \arctan\frac{\mathrm{Im} (z)}{\mathrm{Re} (z)} \right).
\]
where $\xi_R=\mathrm{Re}\, t$. Then the contour $\gamma_{\xi}$ w.r.t. a choice of K\"ahler cone containing $\xi$ is admissible if the integral is absolutely convergent along the asymptotics of $\gamma_{\xi}$:
\begin{equation}
    \mathrm{Re}\widetilde{W}_{\text{eff}}(\sigma,q,t)<0\quad\text{for all }\sigma\in\gamma_{\xi},\ |\sigma|\gg0.
\end{equation}
We also call $\gamma_{\xi}$ conditionally admissible if the integral is conditionally convergent. The integral in Eq.~\eqref{ZB} is evaluated as the sum of residues of gamma poles enveloped by $\gamma_{\xi}$. In a weak-coupling geometric phase $X_{\xi>0}$, semi-classical B-brane is a chain complex of coherent sheaves in $D^b(X_{\xi>0})$ and the leading term of $Z_{\mathcal B}$ is the mirror period \cite{Green:1996dd,Minasian:1997mm,Galkin2016Gamma,Acosta:2014cma}.
 
From now on, we will introduce step by step the concepts in brane transport in the rest of this section. We first assume that $\dim\mathfrak{z}_G=1$ is a one-parameter anomaly-free (CY) model, thus cut-off $\Lambda$ does not appear and the UV GLSM parameter $t$ forms a bona fide moduli space of IR SCFT. Boundary gauge charges are given by Cartan weights of $G$ as $q(\sigma)=\sum_\alpha q^{\alpha}\sigma_\alpha$ for $\alpha=1,\cdots,h_G$. Let $t(\iota)$ be a path parametrized by $\iota\in[-1,1]$ that passes through the common boundary of two K\"ahler cones labeled by $\xi_\pm$, if there exists a continuous family of admissible contours 
\begin{equation}
\begin{aligned}
    \gamma_\iota\subset\mathfrak{t}_G\times[-1,1],\quad \gamma_{\iota}|_{\pm1}=\gamma_\pm
\end{aligned}
\end{equation}
such that
\begin{equation}
    \mathrm{Re}\widetilde{W}_{\text{eff}}(\sigma,q,t(\iota))<0\ \text{for all }\iota \text{ and for all }\sigma\in\gamma_{\iota},\ |\sigma|\gg0,
\end{equation}
then this path gives an analytic continuation of $Z_{\mathcal W(q)}(t)$\footnote{Again, a single component $\mathcal W(q_i)$ in a B-brane gauge multiplet $M=\bigoplus_i\mathcal W(q_i)$.} via contour variation along $\gamma_\iota$, which is called \textbf{B-brane transport} via hemisphere partition function. 

The essence of B-brane transport is that such a family of admissible contours only exists for a specific range of $q^\alpha$ called the \textbf{grade restriction rule} \cite{Herbst:2008jq}. It is conjectured that each grade restriction rule corresponds to a set of generators of IR B-brane category, the same as the \textbf{window category} defined by GIT quotient in Eq.~\eqref{eqn:window}. The integer index $l$ corresponds to shifting $\theta$-angle in the boundary coupling factor\cite{Hori:2000ic} (it also appears in hemisphere partition function as $\sum_i\exp( it(\sigma)+2\pi q_i(\sigma))$):
\begin{equation}
    [\theta](\sum_\alpha\sigma_\alpha)+2\pi q(\sigma)=\sum_\alpha([\theta]-2\pi l)\sigma_\alpha+2\pi(q^\alpha+l)\sigma_\alpha,\quad [\theta]\in[0,2\pi),\quad l\in\ZZ.
\end{equation}
Thus all $\mathbb{W}_l$ are isomorphic but constitute different generators with all gauge charges $q^\alpha$ shifted by an integer $l$.

In a higher-parameter model, the trick of contour can be similarly applied for each boundary of K\"ahler cone by only varying the contour in corresponding gauge group sector. This is known as \textbf{band restriction rule}. In conclusion, for the variation of GIT quotient along any $U(1)_\zeta\subset G$ in a non-anomalous abelian GLSM, the abelian window is derived as the following rule:\cite{Herbst:2008jq}
\begin{equation}
    -\frac{Q^\zeta_{\text{tot}}}{2}<\frac{\theta_\zeta}{2\pi}+q^\zeta<\frac{Q^\zeta_{\text{tot}}}{2},\quad Q^\zeta_{\text{tot}}=\frac{1}{2}\sum_I|Q_I^\zeta|.
\end{equation}
The $\theta_\zeta$ is the $\theta$-angle for the FI parameter $\zeta$, and $q^\zeta$ is $U(1)_\zeta$ boundary gauge charge. With a suitable choice of $\theta_\zeta$ modulo integer $l$, the window category $\mathbb{W}_l$ is given by the matrix factorization module $M$ with $U(1)_\zeta$ representations within the above grade restriction rule. 

It is worth mentioning that finding such a rule for non-abelian variation of GIT quotient, even in one-parameter models, is more complicated. In particular, there may be more than one transportation path in the K\"ahler moduli space\cite{Hori:2006dk}. Nevertheless, there are sporadic investigations of some (compact and non-compact) non-abelian Calabi-Yau GLSMs in \cite{Eager2017,Knapp:2023izn,Lin:2024fpz,Guo:2025yed,Hori:202x}. All these windows define a derived equivalence between the B-brane category of different phases.

\medskip

For the anomalous case, the situation is slightly different due to the existence of mixed branch. In particular, the window will be distinguished as ``big window'' and ``small window'' along a boundary of K\"ahler cones.

Assume one FI parameter $\zeta$ varies (or flows under renormalization) between a UV Fano Higgs branch $X_{\zeta>0}$ and an IR phase containing Higgs branch $X_{\zeta<0}$. When $X_{\zeta>0}$ is pure Higgs branch, Eq.~\eqref{ZB} is computed by summing the residues of poles in Gamma functions with $Q_I^\zeta>0$ that are enveloped by an ``almost-universally admissible contour'' $\gamma_+$ \cite{Clingempeel:2018iub}. This contour converges for a K\"ahler cone containing $\zeta>0$ under the asymptotic region of $\sigma$'s, and any boundary gauge charge $q$ in $\mathcal{W}(q)$ is admissible. However, some coherent sheaves corresponding to B-branes on $X_{\zeta>0}$ satisfy geometric exact sequences (e.g. Eq.~\eqref{eqn:ENcpx}), which is exactly in the kernel of the functor Eq.~\eqref{eqn:window}. Thus, we assign a \textbf{big window category} $\mathbb{W}_{\text{big}}$ from the classical data of the GIT quotient for a UV Fano pure Higgs branch $X_{\zeta>0}$. It is shown in \cite{Clingempeel:2018iub} that when considering the $\zeta<0$ phase, it is necessary for $\mathcal B$ to be in the big window in Fano models.

When a brane is transported to $X_{\zeta<0}$, it will not always be a geometric brane in $D^b(X_{\zeta<0})$ since there are also Coulomb vacua. The criterion for it to flow to $D^b(X_{\zeta<0})$ is that there exists an admissible contour $\gamma_-$ in K\"ahler cone $\zeta<0$ such that Eq.~\eqref{ZB} has only contribution from residues. Note that this is not always the case in the mixed branch. By writing $Z_\mathcal B$ as in Eq.~\eqref{eqn:expW}, the saddle point of $\widetilde{W}_{\text{eff}}$, known as Lefschetz thimble, will also contribute to the integral. This is exactly the Coulomb vacua contribution \cite{Herbst:2008jq}. Thus by this criterion, there is another set of branes in the mixed branch called \textbf{small window category} $\mathbb{W}_{\text{small}}$, which is compatible with the image of $\pi_{\zeta<0}^{(l)}$.

Therefore, for the image to be in $\mathbb{W}_{\text{small}}$, it should be possible to choose the contour such that there is no contribution to the hemisphere partition function from Coulomb vacua. An analysis in \cite{Clingempeel:2018iub} gives the abelian grade restriction rule for $\mathbb{W}_{\text{big}}$ as
\begin{equation}
     -\frac{Q^\zeta_+}{2}<\frac{\theta_\zeta}{2\pi}+q^\zeta<\frac{Q^\zeta_{+}}{2},\quad Q^\zeta_{+}=\sum_{Q^\zeta_I>0}Q_I^\zeta,
\end{equation}
and for $\mathbb{W}_{\text{small}}$ as
\begin{equation}
     -\frac{Q^\zeta_-}{2}<\frac{\theta_\zeta}{2\pi}+q^\zeta<\frac{Q^\zeta_{-}}{2},\quad Q^\zeta_{-}=\sum_{Q^\zeta_I<0}|Q_I^\zeta|.
\end{equation}

In this paper, we will mainly investigate the small window on a class of $U(2)$ models, due to the complexity in higher rank. Note that $q=(q_1,q_2)$ is the Cartan weight of an $U(2)$ representation and thus $\mathcal{W}(q)$ may have rank higher than one. First, we will choose the conditional admissible contour as in Eq.~\eqref{squarecontour} \cite{Guo:2025yed,Hori:202x}. 
{\it The advantage of this contour is that $\mathrm{Re}\,\widetilde{W}_{\text{eff}}(\sigma,q,t)|_\gamma$ does not depend on $\xi_R$, thus the analysis is renormalization invariant.}
Next, it can be numerically tested that if saddle point equation $\partial_\sigma (\mathrm{Re}\,\widetilde{W}_{\text{eff}}(\sigma,q,t)|_\gamma)$ has a simultaneous solution for both elements $\sigma_1,\sigma_2$ in the $U(2)$ Cartan subalgebra. This is illustrated in Fig.~\ref{smallexample}. Then the small window category is given by choosing the set of admissible gauge charges $q$ from the solution, and thus the corresponding representations of $U(2)$. Finally, we will count the Witten index from Higgs branch and mixed branches to check the result on small window that it indeed counts the number of independent branes correctly.

\section{Small window of anomalous $U(2)$ theories}
\label{sec:small}

In this section, we investigate an anomalous GLSM with $U(2)$ gauge group and obtain their small window categories. This type of GLSM will arise as the effective theories (local model at the phase boundary between two chambers \cite{Hori:202x}) in section~\ref{sec:example}. In principle, this method could be generalized to cases with higher rank gauge groups which will be studied in future work. 

Consider a $U(2)$ GLSM with chiral fields $\Phi_i\in\CC^{2n}$ in $n$ copies of fundamental representation and chiral fields $P_I\in\CC^K$ in $K$-copies of anti-determinantal representations with $\det U(2)$ weight $-m_I$: \footnote{This field content is a non-abelian generalization of the hybrid model studied in, {\it e.g.} \cite{Bertolini:2013xga,Erkinger:2022sqs}, except in general we do not require $m_1 = m_2$, however, here we only have vanishing superpotentials.}
\begin{equation}
    \begin{array}{c|cccccc|c}
     & \Phi_1 & \cdots & \Phi_n & P_1 & \cdots & P_K &
    \\\hline
    U(2) & \yng(1) &\cdots & \yng(1) & \det^{-m_1} & \cdots & \det^{-m_K} & \xi
    \end{array}
\end{equation}
where $\xi$ is the renormalized FI parameter for the $U(2)$ gauge group. The Fano condition is
\begin{eqnarray}
    m_1+\cdots+m_K<n.
\end{eqnarray}
The superpotential is zero, thus we omit $U(1)_R$ charges here. However, to evaluate the normalized hemisphere partition function, a proper R-charge should be assigned such that the result is compactified \cite{Clingempeel:2018iub}. The D-term equation is given by the matrix equation
\begin{equation}
    \sum_{i=1}^n \Phi_i \Phi^\dagger_i-\mathbb{I}_{2\times2}\sum_{I=1}^K m_I |P_I|^2=\xi\mathbb{I}_{2\times2}.
\end{equation}
The unstable loci are easily read off as
\begin{equation}
    \Delta_{\text{un}}(\xi)=\left\{\begin{array}{lc}
         \{\Phi_i^a~ | ~\mathrm{rank }\Phi\leq1\}, & \xi>0, 
         \\
         \{ P_I~|~P_1=\cdots=P_K=0\}, & \xi<0. 
    \end{array}\right.
\end{equation}
and the GIT quotient for $\xi>0$ gives the total space of line bundles over the Grassmannian:
\begin{equation}
    X_{\xi>0}=\Tot\left( \oplus_{I=1}^K {\det}^{m_I}\mathcal{S}\rightarrow Gr(2,n) \right).
\end{equation}

In the IR phase $\xi<0$, there will be a Higgs branch and several mixed branches. The geometry of the Higgs branch is given by the (possibly singular) GIT quotient of $\xi<0$. For general $m_I$, the GIT quotient is a gerby $Gr(2,n)$ affine cone fibered over a weighted projective space. It is defined by the projection
\begin{equation}
    \pi:\ X_{\xi<0}\rightarrow \mathbb{WP}(m_1,\cdots,m_K)\label{eqn:X-}
\end{equation}
with each fiber $\pi^{-1}(p)$ being a $\ZZ_{g(p)}$ quotient stack of the affine cone of $Gr(2,n)$ with $\ZZ_{g(p)}$, the unbroken subgroup of $U(1)$, on each $p\in\mathbb{WP}(m_1,\cdots,m_K)$. Meanwhile, there is also a mixed branch $\mathcal{C}_{\xi<0}$ at $\xi<0$. For the GLSM above, the mixed branches correspond to massive Coulomb vacua from the solutions of the effective twisted superpotential. Note that in the case $n=3$, which is dual to an abelian theory, $CGr(2,3)\cong\mathcal{O}(-1)^{\oplus3}$.

Denote $\sigma_a$, $a=1,2$ the Cartan subalgebra of $U(2)$. The effective twisted superpotential $\widetilde{W}_{\rm eff}$ on the Coulomb branch is obtained by integrating out the massive chiral fields,
\begin{equation}
    \widetilde{W}_{\rm eff}\:=\: - t (\sigma_1 + \sigma_2) + i \pi + n \sum_{a=1}^2 \sigma_a \left[\log(\sigma_a) - 1 \right] - \sum_{j=1}^K (m_j \sigma_1 + m_j \sigma_2 ) \left[ \log(m_j \sigma_1 + m_j \sigma_2) -1 \right]\,.
\end{equation}
The Coulomb vacua are given by the equations $\exp(\partial \widetilde{W}_{\rm eff}) = 1$ ($q=\exp(-t)$), which give
\begin{equation}
    \sigma_a^n=-q\prod_{I=1}^K(-m_I\sigma_1-m_I\sigma_2)^{m_I},\quad a=1,2.
\end{equation}
Note that by defining $m=m_1+\cdots+m_K$, the equation above is essentially the same for any number of $p$ fields with total charge $-m$, up to a rescaling. Thus consider
\begin{equation}
    \sigma_1^n = - q (-m \sigma_1 - m \sigma_2)^m \, , \quad \sigma_2^n = - q (-m \sigma_1 - m \sigma_2)^m,
\end{equation}
where one has $\sigma_{1,2}\neq 0$ and $\sigma_1 \neq \pm \sigma_2$ to break $U(2)$ Weyl symmetry. The above two equations immediately give $\sigma_1^n = \sigma_2^n$, which means one can take
\begin{equation}
    \sigma_1 = \eta \sigma_2,
\end{equation}
with $\eta$ being a $n$-th root of unity, $\eta^n=1$. Recall that the effective twisted superpotential is obtained by integrating out the massive fields \cite{Hori:2006dk}, which, in this case, encode the following conditions
\[
    \sigma_i \neq 0, \quad \sigma_1 + \sigma_2 \neq 0, \ \text{and}\ \sigma_1 - \sigma_2 \neq 0,
\]
where the first condition comes from the massiveness of the fundamentals, the second one is due to massiveness of the $P$ fields, and the third one is because of the $W$-bosons. Therefore, we should have $\eta \neq \pm 1$. When $n$ is even $\eta$ has $n-2$ choices, while when $n$ is odd $\eta$ has $n-1$ choices. Then, given a choice of the above $\eta$, one needs to solve the following equation
\begin{equation}
      \sigma_2^n = -q(-m)^m (1+\eta)^m\sigma_2^m,
\end{equation}
and it has $(n-m)$ nonzero solutions. However, due to the Weyl symmetry between $\sigma_1$ and $\sigma_2$, the total number of Coulomb vacua, $N(n,m)$, should be
\begin{equation}
    N(n,m) = \left\lfloor\frac{n-1}{2}\right\rfloor (n-m) = 
    \begin{dcases}
        \frac{(n-2)(n-m)}{2} & n~ \text{even}\, ,\\
        \frac{(n-1)(n-m)}{2} & n~ \text{odd}\, .
    \end{dcases}\label{numCV1}
\end{equation}
Note that, for the case of $m=0$ (i.e. pure Coulomb branch), we do not need to exclude the solution such that $\sigma_1+\sigma_2 = 0$ since there are no P-fields to integrate out. Then, the above derivation needs modification and, therefore, $N(n,0) = \binom{n}{2}$ which gives $\chi(Gr(2,n))$.

The mixed branch will be analyzed in detail case by case since it highly depends on the effective twisted superpotential of the GLSM. In general, the mixed branch is a disjoint union of components labeled by the discrete Coulomb vacua $p\in \mathcal{V}_{C}$. Each component $\mathcal{H}_p$ is the Higgs vacuum manifold consisting of fields that remain massless at the Coulomb vacua $p$. Therefore, the contribution of this mixed branch to the Witten index can be computed as
\[
    I_{\rm mixed} \:=\: \sum_{p\in\mathcal{V}_C} \chi(\mathcal{H}_{p})\,.
\]
Whenever all $\mathcal{H}_p$'s are isomorphic to the same Higgs component $\mathcal{H}$, the Witten index of this mixed branch can be computed by the multiplication of the number of Coulomb vacua $|\mathcal{V}_C|$ and the Euler number of $\mathcal{H}$, namely,
\[
    I_{\rm mixed}\:=\: |\mathcal{V}_C|\ \chi(\mathcal{H})\,.
\]

The final result we will conclude for the GLSMs under consideration is the B-brane effective potential. The hemisphere partition function of the GLSM above is given by
\begin{equation}
\begin{aligned}\label{eqn:ZBmodel}
    Z_{\mathcal{B}}(t)=&\int_\gamma d^2\sigma\ (\sigma_1-\sigma_2)\sinh\pi(\sigma_1-\sigma_2)
    \\
    \times&\prod_{a=1,2}\Gamma(i\sigma_a+R_\phi/2)\prod_{I=1}^K\Gamma(-im_I(\sigma_1+\sigma_2)+R_{p_I}/2)e^{it(\sigma_1+\sigma_2)}f_{\mathcal{B}}(\sigma),
\end{aligned}
\end{equation}
where a generic R-charge for chiral fields is presented in the formula. Denote $\sigma_a=\tau_a+i\nu_a$ with real variables $\tau,\nu$. As discussed in \cite{Guo:2025yed}, the contour $\gamma$ can be taken as the following one:
\begin{equation}\label{squarecontour}
\gamma \:=\: \left\{\nu_1 \:=\: -\nu_2 \:=\: \tau_1^2 - \tau_2^2 \right\}\, ,
\end{equation}
then the following wedge conditions \cite{Hori:2013ika, Guo:2025yed, Hori:202x} \footnote{We would like to thank Mauricio Romo for sharing their results of \cite{Hori:202x} prior to publication through private communications, and Kentaro Hori for sharing some results in his talk. See the recording of the talk for more information: \url{https://scgp.stonybrook.edu/video_portal/video.php?id=5941}.}
\begin{align}
\tau_1 \:=\:0 \quad \Rightarrow& \quad \nu_1 \:\leq\: 0\,; \\
	\tau_2 \:=\:0 \quad \Rightarrow& \quad \nu_2 \:\leq\: 0\,; \\
	\tau_1 + \tau_2 \:=\:0 \quad \Rightarrow& \quad \nu_1 + \nu_2 \:\geq\: 0
\end{align}
are satisfied so the contour will not hit the singularities of poles. The evaluation of $Z_{\mathcal{B}}$ requires a proper R-charge assignment such that the result can be finite by taking the zero R-charge limit \cite{Clingempeel:2018iub}. Nevertheless, $Z_{\mathcal B}$ will not be evaluated in this paper, the R-charges will be turned off.

Denote $A_q(t):=\mathrm{Re}\, \widetilde{W}_{\text{eff}}(\sigma,q,t)$ the B-brane effective twisted superpotential Eq.~\eqref{eqn:expW} of a single component $\mathcal W(q)$, which for Eq.~\eqref{eqn:ZBmodel} is the following:
\begin{equation}
    \begin{aligned}
A_q(t) \:=\: & \pi|\tau_1-\tau_2|-\xi(\nu_1+\nu_2)+\theta(\tau_1+\tau_2)+2\pi(q_1\tau_1+q_2\tau_2) 
\\
& + \sum_I m_I(-\nu_1-\nu_2)\left(\log|m_I(\sigma_1+\sigma_2)|-1\right)-m_I|\tau_1+\tau_2|\left( \frac{\pi}{2}-\arctan\frac{\nu_1+\nu_2}{|\tau_1+\tau_2|} \right)
\\
&+ n\left[ \nu_1\left(\log|\sigma_1|-1\right)-|\tau_1|\left( \frac{\pi}{2}+\arctan\frac{\nu_1}{|\tau_1|} \right)  \right]
\\
&+ n\left[ \nu_2\left(\log|\sigma_2|-1\right)-|\tau_2|\left( \frac{\pi}{2}+\arctan\frac{\nu_2}{|\tau_2|} \right)  \right] \, .
\end{aligned}
\end{equation}
Restricting the above to the contour Eq.~\eqref{squarecontour}, we have
\begin{equation}
\begin{aligned}
A_q(t)|_\gamma\:=\: & \pi|\tau_1-\tau_2|+\left(\theta+2\pi q_1\right)\tau_1+\left(\theta+2\pi q_2\right)\tau_2- \pi\frac{m}{2}|\tau_1+\tau_2|-\pi\frac{n}{2}(|\tau_1|+|\tau_2|)
\\
&- n|\tau_1|\arctan\left(\frac{(\tau_1)^2-(\tau_2)^2}{|\tau_1|}\right)-n|\tau_2|\arctan\left(\frac{(\tau_1)^2-(\tau_2)^2}{|\tau_2|}\right)  
\\
&+((\tau_1)^2-(\tau_2)^2)\frac{n}{2}\log\frac{(\tau_1)^2+((\tau_1)^2-(\tau_2)^2)^2}{(\tau_2)^2+((\tau_1)^2-(\tau_2)^2)^2}\, .
\end{aligned}\label{Aq}
\end{equation}
The convergence condition under $|\tau_a|\rightarrow\infty$ is given by \cite{Guo:2025yed,Hori:202x}, which gives the grade restriction rule
\begin{equation}\label{eqn:GRR}
    \begin{gathered}
\left|\frac{\theta}{2\pi}+q_1+\frac{\theta}{2\pi}+q_2\right| \:\leq\: \frac{n+m}{2} \, ,
\\
\left|\frac{\theta}{2\pi}+q_\alpha\right| \:\leq\: \frac{2n-2+m}{4} \, ,
\\
\left|q_1-q_2\right| \:\leq\: \frac{n-2}{2}\, .
\end{gathered}
\end{equation}
Due to the discussion in \cite{Guo:2025yed}, it is necessary to replace an admissible condition ($<$) with the conditional admissible one ($\leq$) due to the lack of enough objects at the diagonal. For $n=4$, this criterion gives a Lefschetz exceptional collection for $Gr(2,4)$ in a wedge shape (See also Fig.\ref{fig:big}). The small window category consists of those branes $\mathcal{W}(q)$ with gauge charge $q$ in the grade restriction, such that the saddle point equation $\partial_\sigma (A_q(t)|_\gamma)$ has no solution. 

Thus with contour Eq.~\eqref{squarecontour}, the big and small windows in $U(2)$ models seem to be only determined by the total charge $m$ of all the $P$-fields. At present, the related analysis can currently be found only found in \cite{Guo:2025yed}. In spite of that, there are also difficulties in tackling the general saddle point equation by analytic methods. Nevertheless, all these perspectives reduce the problem to the discussion of the one determinantal field model in next section, where the $n=3,4$ cases will be elaborated numerically. We leave the general analysis for future work.

Since the number of $P$ fields in determinantal representation does not affect the small window, we only consider the $U(2)$ gauge theory with $n$ fundamental fields $\Phi_i$ and one determinantal field $P$ in the following:
\begin{equation}
\centering
\begin{tabular}{c|cc|c}
        & $\Phi_{i}$ & $P$   & \\ \hline
 $U(2)$ &${{\yng(1)}}$    & $\det^{-m}$  & $\xi$
\end{tabular} \label{eq:small1}
\end{equation}
where $\xi$ is the FI parameter. We will only focus on the cases with $n=3$ and $n=4$.

\subsection{Category of B-branes}\label{negativehiggs}

When $m=n=3$, the negative phase can be understood via the IR duality of \cite{Hori:2006dk, Hori:2011pd}. The $U(2)$ theory \eqref{eq:small1} with $m=n=3$ is dual to the abelian GLSM, which has gauge group $U(1)$, three chirals with charge $1$ and one chiral with charge $-3$. The dual abelian GLSM has a phase corresponding to the limit $\xi \gg 0$ with target $\mathrm{Tot}(\mathcal{O}(-3)\rightarrow \mathbb{P}^2)$, a noncompact Calabi-Yau manifold. The target space of the phase with $\xi \ll 0$ is the orbifold $\mathbb{C}^3/\mathbb{Z}_3$. The category of B-branes in this phase is the derived category $D^b(\mathbb{C}^3/\mathbb{Z}_3)$, which is equivalent to the derived category of graded modules of $\mathcal{A}=\mathbb{C}[x_1,x_2,x_3]$, i.e. $D(\mathrm{gr}\text{-}\mathcal{A})$. This category is generated by $\mathcal{A}_0, \mathcal{A}_1, \mathcal{A}_2$, where $\mathcal{A}_l$ denotes the graded $\mathcal{A}$-module $\mathcal{A}$ with grading $l$ (the $\mathbb{Z}_3$-charge). $\mathcal{A}_0, \mathcal{A}_1, \mathcal{A}_2$ are mapped to $\pi^*\mathcal{O}_{\mathbb{P}^3}, \pi^*\mathcal{O}_{\mathbb{P}^3}(1), \pi^*\mathcal{O}_{\mathbb{P}^3}(2)$ respectively via brane transport to the phase with $\xi \gg 0$, where $\pi$ is the projection onto the base $\mathbb{P}^3$. From the duality, we see the category of B-branes in the negative phase of the original $U(2)$ theory with $m=n=3$ is equivalent to $D(\mathrm{gr}\text{-}\mathcal{A}) = \langle \mathcal{A}_0, \mathcal{A}_1, \mathcal{A}_2 \rangle$. Because the orbifold group $\mathbb{Z}_3$ in this theory is a subgroup of $\det U(2)$, we see that $\mathcal{A}_i$ can be lifted to the GLSM brane $\mathcal{W}_{\det^{i+3 l}}$. The window category is generated by $\mathcal{W}_{\det^{3 l}}, \mathcal{W}_{\det^{3 l+1}}, \mathcal{W}_{\det^{3 l+2}}$ (The integer $l$ depends on the choice of the theta-angle), which are mapped to $\det(\mathcal{S}^\vee)^{3 l}, \det(\mathcal{S}^\vee)^{3 l+1}, \det(\mathcal{S}^\vee)^{3 l + 2}$ in the positive phase.

When $0<m<n=3$, there are $3-m$ Coulomb vacua in the negative phase and the category of B-branes of the Higgs branch, due to the $\mathbb{Z}_m$ orbifold, is generated by $m$ objects, namely $\mathcal{A}_0,\cdots,\mathcal{A}_{m-1}$. Accordingly, the small window category is generated by $\mathcal{W}_{\det^{m l}}, \mathcal{W}_{\det^{m l+1}},\cdots, \allowbreak \mathcal{W}_{\det^{m l+(m-1)}}$ for some integer $l$. The big window category remains the same as the window category of the $m=n=3$ theory.

The window category of the theory with $m=n=4$ is one of the models studied in \cite{Guo:2025yed}. The target space of the positive phase is the canonical line bundle $K_{Gr(2,4)}$ of the Grassmannian $Gr(2,4)$. By computing the twisted superpotential on the Coulomb branch, it can be shown there is one singular point on the K\"ahler moduli space, located at
\[
\xi = 10 \log2, \quad\theta = 0 \ {\rm mod} \ 2\pi.
\]
In the negative phase ($\xi \ll 0$), the $P$ field acquires a nonzero vev, which breaks $\det U(2)$ to a $\mathbb{Z}_4$ subgroup. Upon integrating out the fluctuation of $P$, we are left with a $SU(2) \times \mathbb{Z}_4$ gauge group, the $SU(2)$ fundamentals $X^a_i$ all have charge one under $\mathbb{Z}_4$. The Higgs branch of the negative phase can be viewed as a noncommutative resolution of the affine cone of $Gr(2,4)$ \cite{Guo:2025yed}. This phase is irregular in the sense of \cite{Hori:2006dk}, meaning that there is a noncompact quantum Coulomb branch in this phase. This irregularity should be distinguished from the noncompact Coulomb branch at the isolated singular locus of the K\"{a}hler moduli space discussed above.

According to the analysis of \cite{Guo:2025yed}, for a fixed path between the two phases and a fixed theta-angle, more than one choice of the window category are possible. This is due to the fact that there are empty branes that satisfy the grade restriction rule. These empty branes impose non-trivial relations among the branes satisfying the GRR. Therefore, the window category is the quotient of the branes satisfying the GRR by the relations imposed by these empty branes, and we have multiple choices for the representatives of the window category. For $m=n=4$ and $\theta \in (2 \pi l, 2 \pi (l+1)), l \in \mathbb{Z}$, we have two choices, which in the positive phase are given by
\begin{eqnarray}
\omega^{(1)}_{2l}&:=&\{\mathrm{det}^{-2}S,\mathrm{det}^{-1}S,\mathcal{O},\mathrm{det}S,S\otimes\mathrm{det}^{-1}S,S\otimes \mathrm{det}^{-2}S\}\otimes \mathrm{det}^{-l}S, \nonumber\\
\omega^{(2)}_{2l}&:=&\{ \mathrm{det}^{-2}S,\mathrm{det}^{-1}S,\mathcal{O},\mathrm{det}S,S,S\otimes\mathrm{det}^{-1}S\}\otimes\mathrm{det}^{-l}S,
\end{eqnarray}
where $S$ is the tautological bundle of $Gr(2,4)$ pulled back to the canonical line bundle by the projection map. In other words, the window category is generated by some $\det^{l_1}$ shift of $\langle \mathcal{W}_{\cdot}, \mathcal{W}_{{\yng(1,1)}}, \mathcal{W}_{{\yng(2,2)}}, \mathcal{W}_{{\yng(3,3)}} \rangle$ and some $\det^{l_2}$ shift of $\langle \mathcal{W}_{{\yng(1)}}, \mathcal{W}_{{\yng(2,1)}} \rangle$, the latter is equivalent to the $\det^{l_2}$ shift of the subcategory generated by branes mapped to the spinor bundles of $Gr(2,4)$\footnote{Notice that the spinor bundles of $Gr(2,4)$ viewed as a quadric in $\mathbb{P}^5$ are $S$ and $Q^\vee$, where $Q$ is the universal quotient bundle.}.

When $0<m<n=4$, the big window category remains the same as the window category of the $m=n=4$ case. There are $4-m$ Coulomb vacua in the negative phase, therefore we expect the small window category to be generated by some $\det^{l_1}$ shift of $\langle \mathcal{W}_{\cdot}, \mathcal{W}_{{\yng(1,1)}}, \cdots, \mathcal{W}_{(m-1,m-1)} \rangle$ and some $\det^{l_2}$ shift of $\langle \mathcal{W}_{{\yng(1)}}, \mathcal{W}_{{\yng(2,1)}} \rangle$, where $l_1 - l_2$ can be determined by analyzing the hemisphere partition function, and the overall shift is determined by the theta-angle. 

\subsection{Hemisphere partition function and the window categories}

\begin{figure}[h!]
\centering
\includegraphics[scale=0.35]{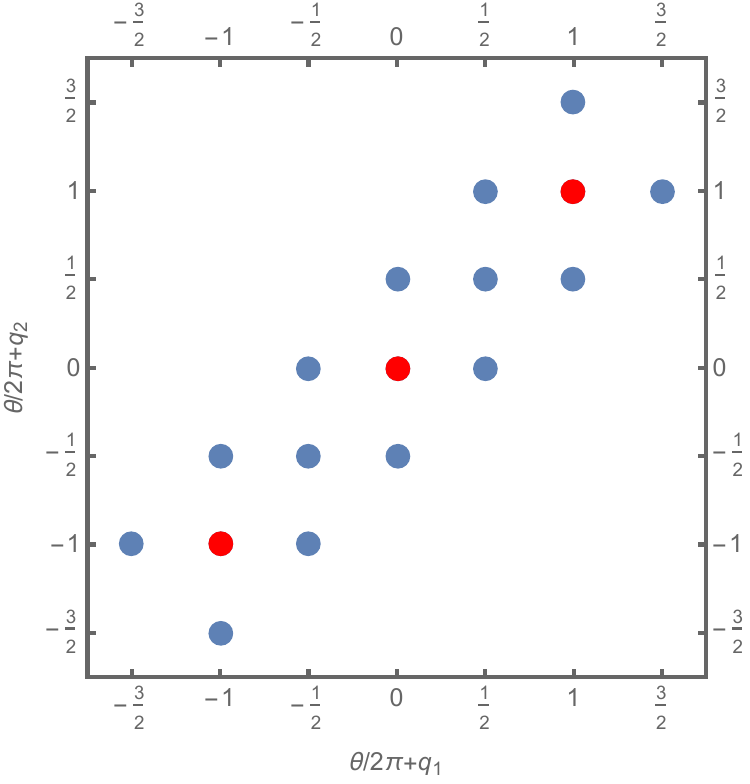}\hspace{2cm}
\includegraphics[scale=0.35]{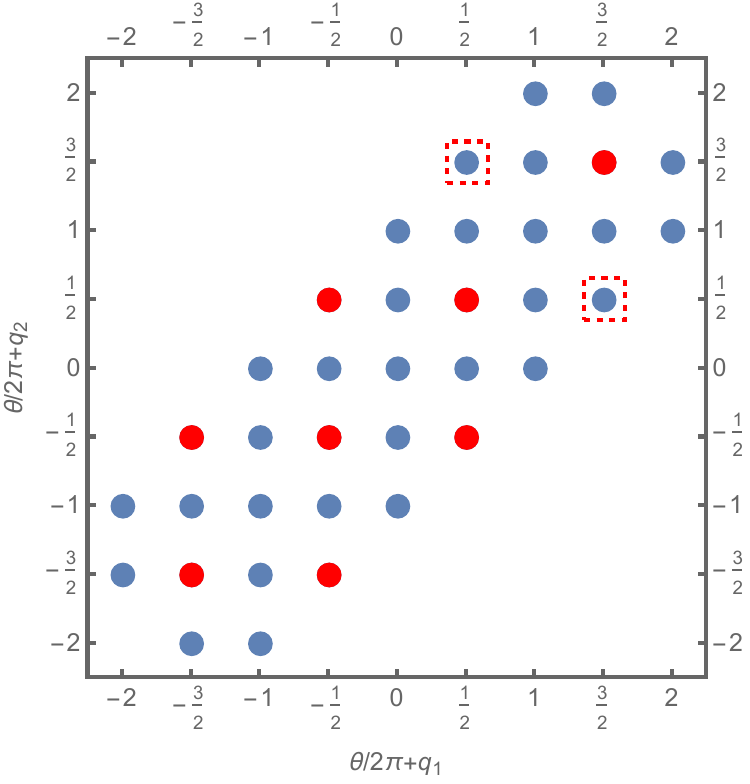}
\caption{Admissible charges in $\frac{1}{2}\mathbb Z$ and an integral choice of $\mathbb W_{\mathrm{big}}$ in red for $n=3$ (left) and $n=4$ (right). }
\label{fig:big}
\end{figure}

Now we continue the discussion of window categories from hemisphere partition function. It is argued in \cite{Guo:2025yed} that the condition Eq.~\eqref{eqn:GRR} is not restrictive enough and it gives more charges in the window than expected. Fortunately, the relations among the branes in the relaxed window are given by the Eagon-Northcott complex Eq.~\eqref{eqn:ENcpx} \cite{eagon1962ideals}, so the number of independent generators in the window will be exactly the Euler characteristic of $Gr(2,n)$. For $m=n$, the convergent region is shown in Fig.~\ref{fig:big}, in which the axes are parameterized by $q_a + \theta/{2\pi}$, where $q_1$ and $q_2$ are the boundary gauge charges of $U(2)$.
\\
The admissible charges are marked as blue dots in Fig.~\ref{fig:big}, among which the big window category consists of charges marked red, which correspond to representations
\begin{equation}
    \mathbb W_{\mathrm{big}}\:=\:\left\lbrace \begin{array}{cl}
    \left\langle \cdot,\ {\yng(1)},\ {\yng(1,1)} \right\rangle\, , &  n=3 \, ,
    \\\\
     \left\langle \cdot,\ {\yng(1)},\ {\yng(1,1)},\ {\yng(2)},\ {\yng(2,2)},\ {\yng(3,3)} \right\rangle\, ,    & n=4\, .
    \end{array}\right.
\end{equation}
They correspond to the semi-orthogonal decomposition of the derived category of $\PP^2(\cong Gr(2,3))$ and $Gr(2,4)$ respectively. Note that for $n=4$, there are multiple equivalent choices of representative off-diagonal charges, which are related to one another by binding with the empty branes \cite{Guo:2025yed}. In Fig.~\ref{fig:big}, we highlight one such choice with red dots, while the corresponding redundant charges are enclosed by red dashed boxes.

\begin{figure}[h!]
\centering
\includegraphics[scale=0.4]{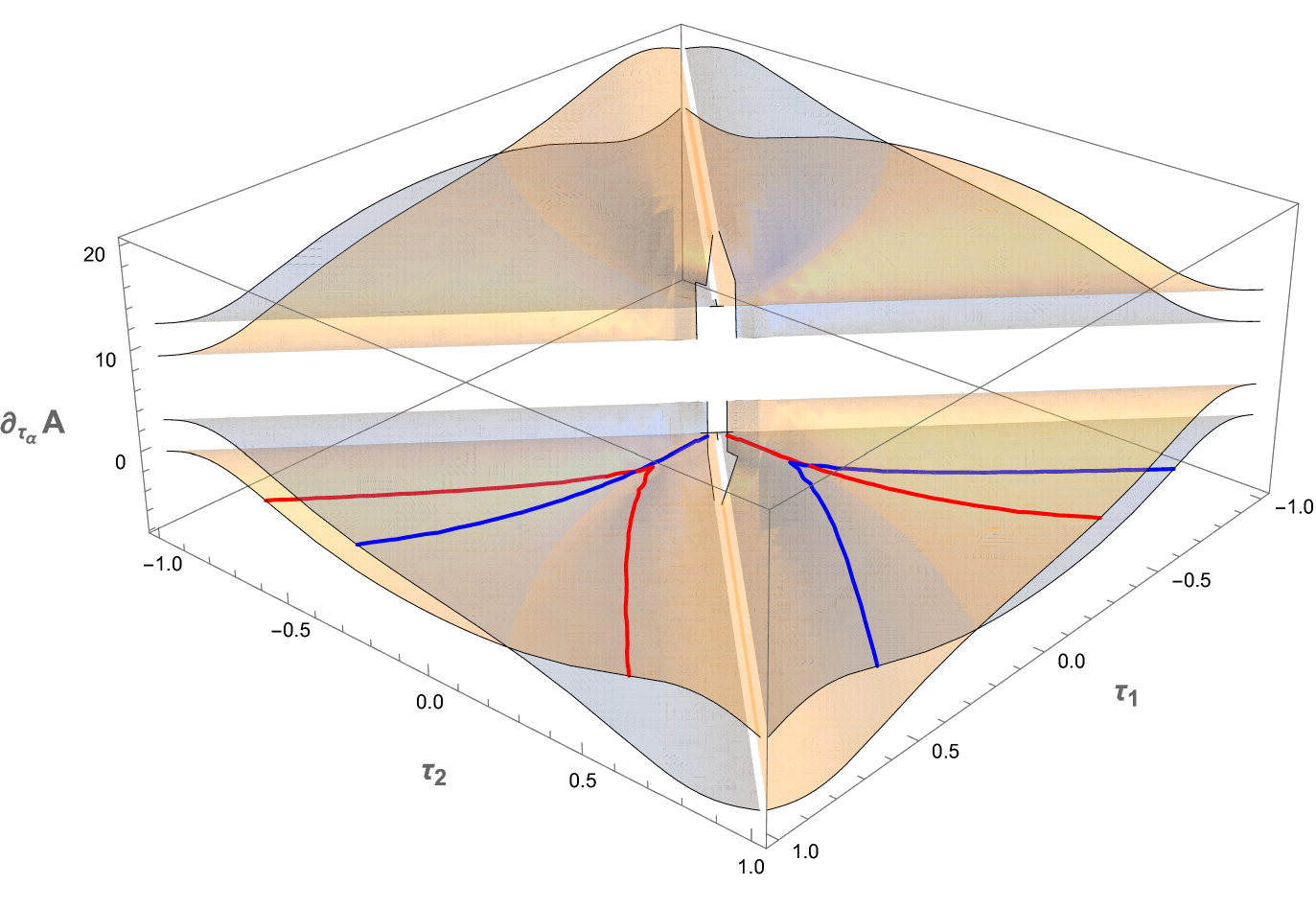}
\caption{Numerical illustration of $\partial_{\tau_a} A_q(\tau_1,\tau_2)$ and zero contours. The light yellow sheet is the plot for $\partial_{\tau_1}A_q(\tau_1,\tau_2)$ while the red contour indicates the solutions to $\partial_{\tau_1}A_q(\tau_1,\tau_2) = 0$. The light blue sheet and the blue contour are for $\partial_{\tau_2}A_q(\tau_1,\tau_2)$ and its zero loci. The results for $n=4$ are provided in the supplementary material.}
\label{smallexample}
\end{figure}

Next, in order to obtain the small window for $m<n$, one should pick out the solutions to the saddle point equations
\begin{equation}\label{dA0}
\partial_{\tau_a}A_q(\tau_1,\tau_2)\:=\:0,\quad a=1,2 \, ,
\end{equation}
among the admissible charges on the contour $\gamma$, \eqref{squarecontour}. More explicitly, they are
\begin{equation}\label{dA1}
    \begin{aligned}
    \partial_{\tau_1}A_q\vert_\gamma \:=\:&\left(2\pi q_1+\theta\right)+n\tau_1\log\left(\frac{(\tau_1)^2+((\tau_1)^2-(\tau_2)^2)^2}{(\tau_2)^2+((\tau_1)^2-(\tau_2)^2)^2}\right)-n\tan^{-1}\left(\frac{(\tau_1)^2-(\tau_2)^2}{\tau_1}\right)
        \\
        &-2n\frac{\tau_1(\tau_1-\tau_2)(\tau_1+\tau_2)(((\tau_1)^2-(\tau_2)^2)^2-(\tau_2)^2)}{((\tau_1)^4+(\tau_2)^4+(\tau_1)^2(1-2(\tau_2)^2))((\tau_2)^2+((\tau_1)^2-(\tau_2)^2)^2)}
        \\
        &-\frac{\pi}{2}\left( n\operatorname{sign}(\tau_1)+m\operatorname{sign}(\tau_1+\tau_2)-2\operatorname{sign}(\tau_1-\tau_2) \right) \, ,
    \end{aligned}
\end{equation}
and
\begin{equation}\label{dA2}
    \begin{aligned}
    \partial_{\tau_2}A_q\vert_\gamma\:=\:&\left(2\pi q_2+\theta\right)+n\tau_2\log\left(\frac{(\tau_2)^2+((\tau_1)^2-(\tau_2)^2)^2}{(\tau_1)^2+((\tau_1)^2-(\tau_2)^2)^2}\right)-n\tan^{-1}\left(\frac{(\tau_1)^2-(\tau_2)^2}{\tau_2}\right)
        \\
        &-2n\frac{\tau_2(\tau_2-\tau_1)(\tau_1+\tau_2)(((\tau_2)^2-(\tau_1)^2)^2-(\tau_1)^2)}{((\tau_1)^4+(\tau_2)^4+(\tau_1)^2(1-2(\tau_2)^2))((\tau_2)^2+((\tau_1)^2-(\tau_2)^2)^2)}
        \\
        &-\frac{\pi}{2}\left( n\operatorname{sign}(\tau_2)+m\operatorname{sign}(\tau_1+\tau_2)+2\operatorname{sign}(\tau_1-\tau_2) \right)\, .
    \end{aligned}
\end{equation}
To see when Eq.~\eqref{dA0} has no solution, we apply a numerical method. For example, for $n=3,m=1$ with $q_a+\theta/2\pi=(-1/2,-1/2)$, $\partial_{\tau_1} A_q(\tau_1,\tau_2)=0$ and $\partial_{\tau_2} A_q(\tau_1,\tau_2)=0$ are plotted as the red and blue contours in Fig.~\ref{smallexample}. Then, the transversal intersection between the red and blue contours implies the solutions to the Coulomb vacuum equations, and therefore the corresponding representation should be ruled out from the small window.

\begin{figure}[!h]
\centering
\subfigure[n=3,m=1]{\includegraphics[scale=0.35]{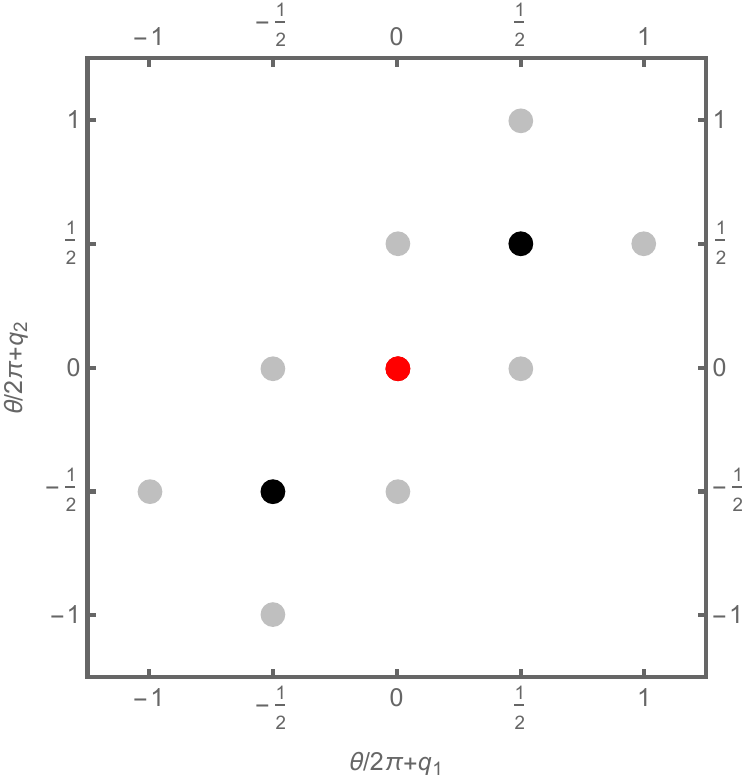}}
\hspace{1.5cm}
\subfigure[n=3,m=2]{\includegraphics[scale=0.35]{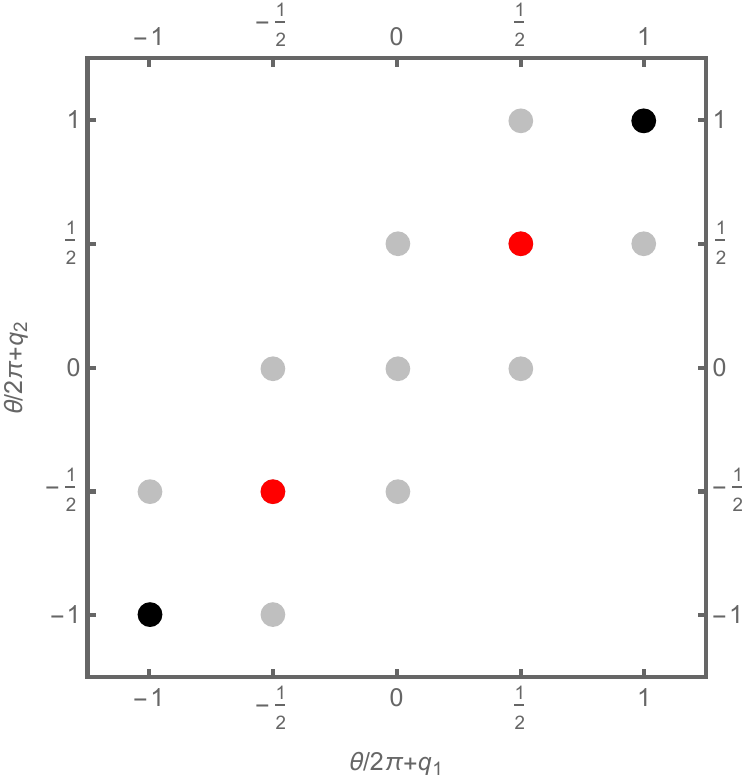}} \\
\subfigure[n=4,m=1]{\includegraphics[scale=0.35]{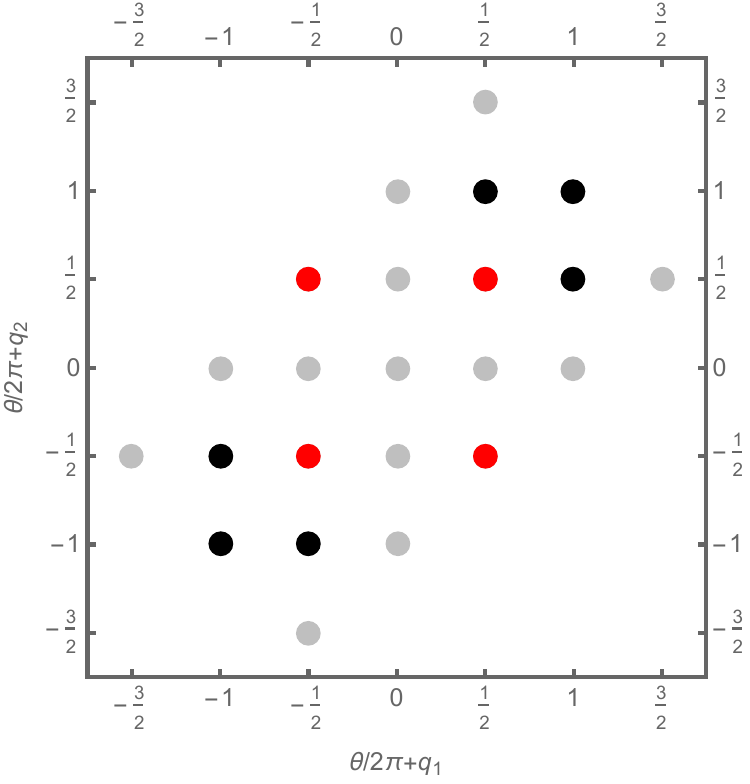}}
\hspace{0.5cm}
\subfigure[n=4,m=2]{\includegraphics[scale=0.35]{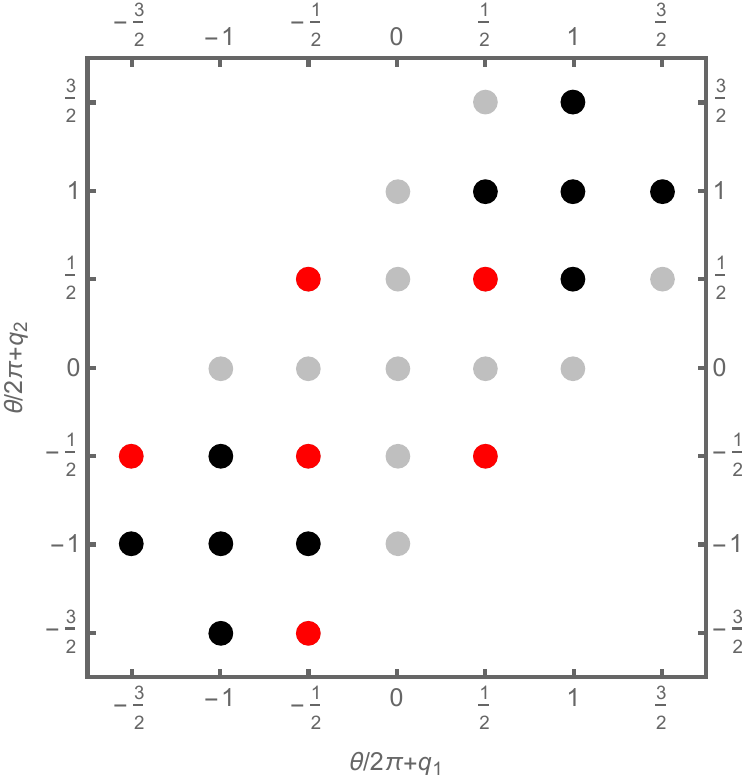}}
\hspace{0.5cm}
\subfigure[n=4,m=3]{\includegraphics[scale=0.35]{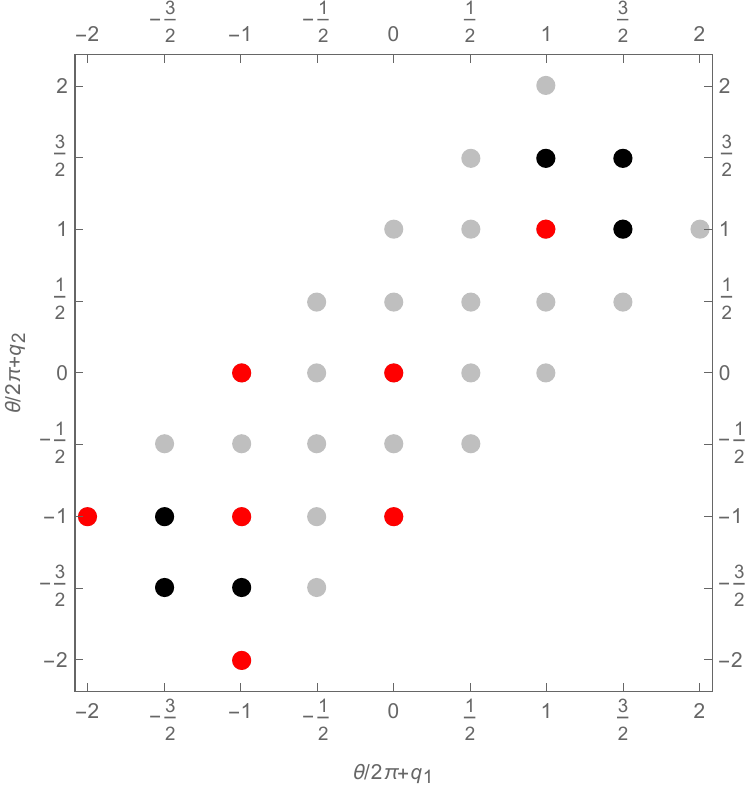}}
\caption{Saddle point solutions (black) and the admissible integer charges in small window (red).}
\label{smallwindowresult}
\end{figure}

We plot and examine the results by hand. The results for $n=4$ are provided in the supplementary material for readers' reference \footnote{One can find the supplementary material (small\_results.pdf) from the arXiv Tex source on \url{https://arxiv.org/abs/2312.05740}.}. They are summarized in Fig.~\ref{smallwindowresult}, where the gray background indicates the convergent region from Eq.~\eqref{eqn:GRR}, black points are the charges such that Eq.~\eqref{dA0} has saddle point solutions, which should be ruled out, and the red points correspond to the charges in the small window. 

According to the numerical results, the small windows of $U(2)$ GLSM with one determinantal field are given as follows:
\begin{itemize}
\item $n=3$
\begin{equation}\label{smallwindowN3}
{\setstretch{2}
\begin{array}{ll}
m=1, & \mathbb W_{\mathrm{small}}\:=\:\left\langle \cdot \right\rangle
\\
m=2, & \mathbb W_{\mathrm{small}}\:=\:\left\langle \cdot,\quad {\yng(1,1)} \right\rangle
\\
m=3, & \mathbb W_{\mathrm{small}}\:=\:\left\langle \cdot,\quad {\yng(1,1)},\quad {\yng(2,2)} \right\rangle
\end{array}
}
\end{equation}
\item $n=4$
\begin{equation}
{\setstretch{2}
\begin{array}{ll}\label{smallwindowN4}
m=1, & \mathbb W_{\mathrm{small}}\:=\:\left\langle \cdot,\quad \yng(1),\quad \yng(1,1) \right\rangle
\\
m=2, & \mathbb W_{\mathrm{small}}\:=\:\left\langle \yng(1),\quad \yng(1,1),\quad \yng(2,1),\quad \yng(2,2) \right\rangle
\\
m=3, & \mathbb W_{\mathrm{small}}\:=\:\left\langle \yng(1),\quad \yng(1,1),\quad \yng(2,1),\quad \yng(2,2),\quad \yng(3,3) \right\rangle
\\
m=4, & \mathbb W_{\mathrm{small}}= \left\langle \cdot,\quad \yng(1),\quad \yng(1,1),\quad \yng(2,1),\quad \yng(2,2),\quad \yng(3,3) \right\rangle
\end{array}
}
\end{equation}
\end{itemize}
If one counts the number of generators in the small windows, one can conclude that, in the cases of $n=3$ and $n=4$,
\[
\chi_- \:=\: \text{the number of generators of~} \mathbb W_{\mathrm{small}} \:=\:\left\lbrace\begin{array}{ll}
m\, , & n=3\, , \\ m+2\, , & n=4\, ,
\end{array} \right.
\]
and together with the number of Coulomb vacua \eqref{numCV1}, one can see that they match the total Witten index and the above analysis is consistent:
\begin{equation}
\label{eq:consist}
\chi_+ \:=\: \text{the number of generators of~}\mathbb W_{\mathrm{big}} \:=\: \chi_- +\left\lfloor \frac{n-1}{2} \right\rfloor(n-m) \:=\: \binom{n}{2}\, .
\end{equation}

\section{Derived equivalences in two-parameter models}
\label{sec:example}
In this section, we will implement the method outlined in the introduction and use results from the previous section to study the derived equivalence in concrete examples. In section \ref{sec:necessarycond}, we first derive a necessary condition for derived equivalence between two phases of a GLSM with $U(1)\times U(1)$ or $U(1) \times U(2)$ gauge groups. Then as a warmup, we first start with an abelian example in section~\ref{example1}, which highlights many of the features of this analysis. In sections~\ref{example2}, \ref{example3} and \ref{example4}, we then study three non-abelian examples and analyze the corresponding derived equivalences. In each example, we will use our results on brane transport to give a description of the functor realizing the derived equivalence.

For all the examples considered in this section, we will always use $\hat{X}$ to denote the target space of the geometric phase of the GLSM, while $X_\pm$ as the target spaces of the Higgs branches, accompanied by mixed branches $\mathcal{C}_{\pm}$, in two different phases of the same GLSM. The phase diagram can be schematically presented as in Fig. \ref{phasediagram}.

\begin{figure}[!h]
    \centering
    \tikzset{every picture/.style={line width=0.75pt}} 
    \begin{tikzpicture}[x=0.75pt,y=0.75pt,yscale=-0.6,xscale=0.6]
    \draw [line width=1.5]    (125,240) -- (290,265) -- (450,265) ;
    \draw [shift={(454,265)}, rotate = 180] [fill={rgb, 255:red, 0; green, 0; blue, 0 }  ][line width=0.08]  [draw opacity=0] (11.61,-5.58) -- (0,0) -- (11.61,5.58) -- cycle    ;
    \draw [line width=1.5]    (290,265) -- (290,120) ;
    \draw [line width=1.5]    (290,265) -- (200,380) ;
    \draw [shift={(290,114.83)}, rotate = 90] [fill={rgb, 255:red, 0; green, 0; blue, 0 }  ][line width=0.08]  [draw opacity=0] (11.61,-5.58) -- (0,0) -- (11.61,5.58) -- cycle    ;
    \draw [line width=1.5]    (290,265) -- (440,370) ;
    \draw (373,185) node [anchor=north west][inner sep=0.75pt]  [font=\normalsize]  {$\hat{X}$};
    \draw (466,250) node [anchor=north west][inner sep=0.75pt]    {$\xi_{1}$};
    \draw (300,115) node [anchor=north west][inner sep=0.75pt]    {$\xi_{2}$};
    
    \draw (373,285) node [anchor=north west][inner sep=0.75pt]    {$X_-$};
    \draw  [draw opacity=0][fill={rgb, 255:red, 255; green, 255; blue, 255 }  ,fill opacity=1 ]  (277,315) -- (304,315) -- (304,350) -- (277,350) -- cycle  ;
    \draw (200,185) node [anchor=north west][inner sep=0.75pt]  [font=\normalsize]  {$X_+$};
    \end{tikzpicture}
    \caption{Schematic phase diagram of a two-parameter GLSM. $\hat{X}$ and $X_\pm$ denote the target spaces of the Higgs branches in the corresponding phases. The goal is to study the conditions on the derived equivalence between $X_+$ and $X_-$.}
    \label{phasediagram}
\end{figure}

The goal is to study the derived equivalence between $X_+$ and $X_-$ by embedding them in different phases of a GLSM for $\hat{X}$. One necessary condition for such an equivalence relation is that their Euler characteristics should match, namely,
\[
    \chi(X_+) = \chi(X_-)\,.
\]
The Witten index of a 2d GLSM is invariant across phases and, in our setup, it should equal the Euler characteristic of $\hat{X}$, therefore one can conclude that
\[
    \chi(\hat{X}) = \chi(X_+) + \chi(\mathcal{C}_{+}) = \chi(X_-) + \chi(\mathcal{C}_{-})\,,
\]
where $\chi(\mathcal{C}_{\pm})$ counts the contributions of the mixed branches, and hence the above necessary condition is equivalent to 
\begin{equation}\label{matchindex}
    \chi(\mathcal{C}_{+}) = \chi(\mathcal{C}_{-})\,.
\end{equation}
One can directly check this condition by counting the number of nonzero Coulomb vacua from the effective twisted superpotential in different phases. 

\subsection{Necessary conditions for the derived equivalence}\label{sec:necessarycond}
The goal of this subsection is to derive a necessary condition for the derived equivalence in the $U(1)\times U(1)$ and $U(1)\times U(2)$ gauge theories under consideration. As discussed above, derived equivalence implies the condition \eqref{matchindex}, which will yield constraints on the matter content of the GLSM\footnote{This method has been used in \cite[appendix~B]{Chen:2020iyo} to study the phases of two-parameter anomalous models and we will also use this strategy in this section.} (Eq.~\eqref{constraintu1u1} and \eqref{constraintu1u2} below).

The method we will use below to count the number of nonzero Coulomb vacua\footnote{By nonzero Coulomb vacua, we mean the vacua in which at least one component of the adjoint scalar field $\sigma$ is nonzero.} is that we first turn on twisted masses for the matter fields and then analyze the behavior of the solutions to the vacuum equations in the massless limit.

\subsubsection{$U(1)\times U(1)$ gauge theory}
Let us consider a general $U(1)\times U(1)$ gauge theory with chiral matter specified below
\begin{equation}\label{tab:generalU1U1}
\centering
\begin{tabular}{c|cc|c}
       & $\Phi_{i}$  & $P$ &  \\ \hline
 $U(1)_1$ &$Q^{(1)}_i$    & $-m$ & $\xi_1$\\
 $U(1)_2$ &$Q_i^{(2)}$   & $-n$ & $\xi_2$
\end{tabular}
\end{equation}
where $i=1,\cdots,M$. We further assume that the charge assignment satisfies the Fano condition,
\[
    \sum_{i=1}^M Q^{(1)}_i > m\,, \quad \sum_{i=1}^M Q^{(2)}_i > n\,,
\]
and assume that the $2\times 2$ charge matrices, for every $i\in\{1,\dots,M\}$,
\[
    \begin{bmatrix}
        Q^{(1)}_{i} & m \\
        Q^{(2)}_{i} & n \\
    \end{bmatrix}
\]
are all non-degenerate. Let $m_i$ and $\tau$ be the twisted masses of $\Phi_i$ and $P$ respectively. Denote by $\sigma_a$ the scalar component of the field strength $\Sigma_a$ corresponding to $U(1)_a$ for $a=1,2$, then the Coulomb vacua equations can be read off from the effective twisted superpotential as
\begin{align}
        & \prod_{i=1}^M (Q_i^{(1)} \sigma_1 + Q_i^{(2)} \sigma_2 - m_i)^{Q_i^{(1)}}  = q_1 ( - m  \sigma_1 - n \sigma_2 - \tau )^m \,,\label{eq:u1u1-1}\\
        & \prod_{i=1}^M (Q_i^{(1)} \sigma_1 + Q_i^{(2)} \sigma_2- m_i)^{Q_i^{(2)}}   = q_2 (-m  \sigma_1 - n \sigma_2 - \tau)^n \,.\label{eq:u1u1-2}
\end{align}
Here $q_i = e^{-t_i}$ where $t_i$ are the FI-theta parameters associated with $U(1)_i$. Generally, the total number of solutions is
\[
    \left(\sum_{i=1}^M Q_i^{(1)} \right) \times \left(\sum_{i=1}^M Q_i^{(2)} \right)\,,
\]
which gives the total number of Coulomb vacua if we are lying in the phase of a pure Coulomb branch. Let us further analyze the structure of solutions to equations \eqref{eq:u1u1-1} and \eqref{eq:u1u1-2} in other phases.

\begin{itemize}
    \item Let us first consider the phase where $\xi_1<0$ while $\xi_2 \rightarrow + \infty$, {\it i.e.} Eq.~\eqref{eq:u1u1-2} becomes
    \begin{equation}\label{2ndEOMu1u1}
        \prod_{i=1}^M (Q_i^{(1)} \sigma_1 + Q_i^{(2)} \sigma_2 - m_i)^{Q_i^{(2)}}  = 0\,,
    \end{equation}
    implying that solutions $\sigma_1$ and $\sigma_2$ should satisfy the following relation
    \begin{equation}
        Q_a^{(1)} \sigma_1 + Q_a^{(2)} \sigma_2 - m_a = 0\,,
    \end{equation}
    for some $a\in\{1,\dots,M\}$, and there are a total of
    \begin{equation}
        \sum_{i=1}^M Q^{(2)}_i 
    \end{equation}
    choices for such a relation due to Eq.~\eqref{2ndEOMu1u1}. Choosing $Q_a^{(1)} \sigma_1 + Q_a^{(2)} \sigma_2 - m_a = 0$ (which also implies that $Q_a^{(2)}\neq 0$, otherwise we do not have this choice from \eqref{eq:u1u1-2}, we want to plug it into Eq.~\eqref{eq:u1u1-1} to find all possible nonzero solutions. There are two cases depending on whether $Q_a^{(1)}$ is zero or not. 
    \begin{itemize}
        \item When $Q_a^{(1)}\neq 0$, then the factor $(Q_a^{(1)} \sigma_1 + Q_a^{(2)} \sigma_2 - m_a)$ must also appear in the LHS of \eqref{eq:u1u1-1}, which means the LHS of \eqref{eq:u1u1-1} must be zero. Then, one obtains another equation
        \begin{equation}
            m \sigma_1 + n\sigma_2 =-\tau\,.
        \end{equation}
        Because of the non-degeneracy of the charge matrices by assumption, it is straightforward to obtain
        \begin{equation}
            \begin{bmatrix}
                \sigma_1 \\
                \sigma_2
            \end{bmatrix} = 
            {\begin{bmatrix}
                Q^{(1)}_a & Q^{(2)}_a\\
                m         & n
            \end{bmatrix}}^{-1}
            \begin{bmatrix}
                m_a \\
                -\tau
            \end{bmatrix}
        \end{equation}
        which will become zero in the massless limit $m_a \rightarrow 0$ and $\tau \rightarrow 0$. In other words, there are no non-vanishing solutions in the massless limit in this case.
        \item When $Q_a^{(1)}=0$, then the equation $Q_a^{(1)} \sigma_1 + Q_a^{(2)} \sigma_2 - m_a = 0$ is simplified and we obtain 
        \begin{equation}
            \sigma_2 = \frac{m_a}{Q^{(2)}_a}\,.
        \end{equation}
        Then \eqref{eq:u1u1-1} becomes
        \begin{equation}
            \prod_{i=1}^M \left( Q^{(1)}_i \sigma_1 + \frac{Q^{(2)}_i}{Q^{(2)}_a} m_a - m_i \right)^{Q^{(1)}_i} \:=\: q_1 \left( -m \sigma_1 - \frac{n}{Q^{(2)}_a}m_a - \tau\right)^m\,,
        \end{equation}
        and, in the massless limit, the equation above is reduced to the following form
        \begin{equation}
            \sigma_1^{\sum_{i=1}^M Q^{(1)}_i - m} \:\propto\: q_1\,,
        \end{equation}
        with nonzero $\sigma_1$ and it has $\left(\sum_{i=1}^M Q^{(1)}_i - m\right)$ nonzero solutions.
    \end{itemize}
    The above analysis applies to every $a \in\{1,\dots, M\}$.

    To conclude, in this phase, the total number of nonzero solutions is given by
    \begin{equation}\label{eq:u1u1-1-n0}
        \left(\sum_{i=1}^M Q^{(2)}_i \right)\times \left( \sum_{i=1}^M Q^{(1)}_i - m \right)\,.
    \end{equation}
    \item Another phase of interest is when $\xi_1 \rightarrow +\infty$ while $\xi_2 <0$. Following the same analysis as above with only minor changes, one can find that the total number of nonzero solutions is given by
    \begin{equation}\label{eq:u1u1-2-n0}
        \left(\sum_{i=1}^M Q^{(1)}_i \right)\times \left( \sum_{i=1}^M Q^{(2)}_i - n \right)\,.
    \end{equation}
\end{itemize}
In summary, the necessary condition on the GLSM \eqref{tab:generalU1U1} for the derived equivalence between these two phases, obtained by matching the number of nonzero Coulomb vacua, is 
\begin{equation}\label{constraintu1u1}
    \left(\sum_{i=1}^M Q^{(2)}_i  \right)\times \left( \sum_{i=1}^M Q^{(1)}_i - m \right)
    \:=\: \left(\sum_{i=1}^M Q^{(1)}_i \right)\times \left( \sum_{i=1}^M Q^{(2)}_i  - n \right)\,.
\end{equation}

\subsubsection{$U(1)\times U(2)$ gauge theory}
Let us now consider the GLSM with $U(1) \times U(2)$ gauge group and the following matter content
\[
\centering
\begin{tabular}{c|cccc|c}
       & $\Phi_{i}$  & $Z^\alpha_{j}$ & 
       $Y^\alpha_k$ &
       $P_s$ &  \\ \hline
 $U(1)$ &$Q_i$  & $0$ & $\tilde{Q}_k$ & $-m_s$ & $\xi_1$\\
 $U(2)$ &$\cdot$  & ${\yng(1)}$ & $\overline{\yng(1)}$ & ${\rm det}^{-n_a}$ & $\xi_2$
\end{tabular} \label{eq:abelianexample}
\]
for $i=1,\dots,M_0$, $j=1,\dots,M_1$, $k=1,\cdots,M_2$ and $s=1,\dots,N$. We assume $Q_i>0, \tilde{Q}_k>0, m_s \geq 0, n_s >0$ and the Fano condition
\begin{equation}
\sum_{i=1}^{M_0} Q_i + 2 \sum_{k=1}^{M_2} \tilde{Q}_k \geq \sum_{s=1}^N m_s,\quad M_1 \geq M_2 + \sum_{s=1}^N n_s.
\end{equation}
If equality holds in any of the two conditions above, the theory is non-anomalous in the corresponding direction.

Let $\sigma$ be the scalar component field of the $U(1)$ vector multiplet and $\sigma_1, \sigma_2$ be the diagonal entries of the diagonalized scalar component of the $U(2)$ vector multiplet. For clarity of exposition, we turn on twisted masses $\tau_i$, $t_j$ and $\tilde{t}_k$ for $\Phi_i$, $Z^\alpha_j$ and $Y^\alpha_k$ respectively, but keep in mind that we are counting the solutions in the limit where all the twisted masses vanish. The equations of motion on the Coulomb branch read
\begin{align}
&\prod_{i=1}^{M_0} (Q_i \sigma - \tau_i)^{Q_i} \prod_{k=1}^{M_2} \prod_{\alpha=1}^2 (\tilde{Q}_k \sigma - \sigma_\alpha-\tilde{t}_k)^{\tilde{Q}_k} = q_1 \prod_{s=1}^N (-m_s \sigma - n_s (\sigma_1+\sigma_2))^{m_s},\label{EOMgeneral1}\\
&\prod_{j=1}^{M_1} (\sigma_\alpha-t_j) = -q_2 \prod_{k=1}^{M_2}(\tilde{Q}_k \sigma - \sigma_\alpha - \tilde{t}_k) \prod_{s=1}^N (-m_s \sigma - n_s(\sigma_1+\sigma_2))^{n_s}\label{EOMgeneral2},
\end{align}
where $q_1 = \exp{(-\xi_1+i\theta_1)}, q_2 = \exp{(-\xi_2+i\theta_2)}$. The phase with $\xi_1 > 0, \xi_2>0$ is the geometric phase where only Higgs branch exists. Now we compute the number of nonzero solutions\footnote{By nonzero solution, we mean a solution such that at least one of $\sigma$, $\sigma_1$, $\sigma_2$ is nonzero.} of \eqref{EOMgeneral1} and \eqref{EOMgeneral2} in the following two phases adjacent to the geometric phase: (i) $\xi_1 < 0, \xi_2 \gg 0$; (ii) $\xi_2 < 0, \xi_1 \gg 0$. 

\begin{itemize}
\item Phase (i): $\xi_1<0, \xi_2 \rightarrow +\infty$. In the $q_2 \rightarrow 0$ limit, Eq.~\eqref{EOMgeneral2} becomes $\prod_{j=1}^{M_1} (\sigma_\alpha -t_j) = 0$, so there are ${M_1 \choose 2}$ choices for $(\sigma_1, \sigma_2)$, modulo the $\mathbb{Z}_2$ Weyl symmetry that exchanges $\sigma_1$ and $\sigma_2$. With each of the choices, \eqref{EOMgeneral1} becomes
\begin{equation}\label{eq1phase1}
\prod_{i=1}^{M_0}(Q_i \sigma)^{Q_i} \prod_{k=1}^{M_2} (\tilde{Q}_k \sigma)^{2 \tilde{Q}_k} = q_1 \prod_{s=1}^N(-m_s \sigma)^{m_s}
\end{equation}
in the limit where all the twisted masses tend to zero.
Eq.~\eqref{eq1phase1} indicates that there are $\sum_{i=1}^{M_0} Q_i + 2 \sum_{k=1}^{M_2} \tilde{Q}_k - \sum_{s=1}^N m_s$ nonzero solutions for $\sigma$. We conclude that there are
\begin{equation}\label{Nsolphase1}
{M_1 \choose 2} \left( \sum_{i=1}^{M_0} Q_i + 2 \sum_{k=1}^{M_2} \tilde{Q}_k - \sum_{s=1}^N m_s \right)
\end{equation}
nonzero solutions in this phase.

\item Phase (ii): $\xi_2<0, \xi_1 \rightarrow +\infty$. In the $q_1 \rightarrow 0$ limit, Eq.~\eqref{EOMgeneral1} reduces to
\begin{equation}\label{eq1phase2}
\prod_{i=1}^{M_0} (Q_i \sigma - \tau_i)^{Q_i} \prod_{k=1}^{M_2} \prod_{\alpha=1}^2 (\tilde{Q}_k \sigma - \sigma_\alpha - \tilde{t}_k)^{\tilde{Q}_k} = 0.
\end{equation}
We have the following two cases to consider:
\begin{itemize}
\item If $\sigma = \tau_i/Q_i$ for some $i$, there are $\sum_{i=1}^{M_0} Q_i$ solutions for $\sigma$. In this case, Eq, \eqref{EOMgeneral2} becomes
\begin{equation}\label{eq2phase2}
\sigma_\alpha^{M_1} = -q_2 (-\sigma_\alpha)^{M_2} \prod_{s=1}^N (-n_s (\sigma_1+\sigma_2))^{n_s},\quad \alpha=1,2.
\end{equation}
Under the constraints $\sigma_1 \neq \sigma_2$ and $\sigma_1 \neq0$ or $\sigma_2 \neq 0$, Eq.~\eqref{EOMgeneral2} has $\lfloor \frac{M_1-1}{2} \rfloor (M_1-M_2-\sum_{s=1}^N n_s)$ solutions modulo the $\mathbb{Z}_2$ symmetry. This can be seen from the fact that $\sigma_2 = \omega \sigma_1$ due to $\sigma_1^{M_1} = \sigma_2^{M_1}$, where $\omega^{M_1} =1$. Notice that $\omega \neq \pm 1$ as a result of the constraints, so there are $\lfloor \frac{M_1-1}{2} \rfloor$ choices for $\omega$. For a fixed $\omega$, Eq.~\eqref{eq2phase2} has $M_1-M_2-\sum_{s=1}^N n_s$ nonzero solutions for $\sigma_1$. So in total, there are
\begin{equation}\label{Nsol1}
\left(\sum_i^{M_0} Q_i \right) \left\lfloor \frac{M_1-1}{2} \right\rfloor \left(M_1-M_2-\sum_{s=1}^N n_s\right)
\end{equation}
solutions in the case $\sigma = \tau_i/Q_i$ for some $i=1,\cdots, M_0$.

\item
From Eq.~\eqref{eq1phase2}, it is also possible to have $\sigma = (\sigma_\alpha+\tilde{t}_k)/\tilde{Q}_k$ for $\alpha =1,2$ and $k=1,\cdots,M_2$. Due to the $\mathbb{Z}_2$ symmetry, we may take $\sigma = (\sigma_1+\tilde{t}_k)/\tilde{Q}_k$. There are therefore $\sum_{k=1}^{M_2} \tilde{Q}_k$ choices for $\sigma$ in this case. For each choice of $\sigma$, Eq.~\eqref{EOMgeneral2} with $\alpha =1$ becomes $\prod_{j=1}^{M_1} (\sigma_1 -t_j) = 0$, so there are $M_1$ choices for $\sigma_1$. Eq.~\eqref{EOMgeneral2} with $\alpha =2$ then becomes
\[
\sigma^{M_1} = -q_2 (-\sigma_2)^{M_2} \prod_{s=1}^N(-n_s \sigma_2)^{n_s}
\]
in the limit where all the twisted masses vanish. Consequently there are $M_1-M_2-\sum_{s=1}^N n_s$ nonzero solutions for $\sigma_2$. Therefore, the number of nonzero solutions in this case is
\begin{equation}\label{Nsol2}
M_1 \left( \sum_{k=1}^{M_2} \tilde{Q}_k \right) \left( M_1 -M_2-\sum_{s=1}^N n_s \right).
\end{equation}
\end{itemize}
Combining Eq.~\eqref{Nsol1} and \eqref{Nsol2}, we see the total number of nonzero solutions in this phase is
\begin{equation}\label{Nsolphase2}
\left[\left\lfloor \frac{M_1-1}{2} \right\rfloor \left(\sum_i^{M_0} Q_i \right) +M_1 \left( \sum_{k=1}^{M_2} \tilde{Q}_k \right) \right] \left( M_1 -M_2-\sum_{s=1}^N n_s \right).
\end{equation}
\end{itemize}

In conclusion, by counting the number of nonzero solutions to the equations of motion on the Coulomb branch in the two phases, we see a necessary condition for the two phases to realize a derived equivalence is
\begin{multline}\label{constraintu1u2}
{M_1 \choose 2} \left( \sum_{i=1}^{M_0} Q_i + 2 \sum_{k=1}^{M_2} \tilde{Q}_k - \sum_{s=1}^N m_s \right) \\ = \left[\left\lfloor \frac{M_1-1}{2} \right\rfloor \left(\sum_i^{M_0} Q_i \right) +M_1 \left( \sum_{k=1}^{M_2} \tilde{Q}_k \right) \right] \left( M_1 -M_2-\sum_{s=1}^N n_s \right)    
\end{multline}
from Eq.~\eqref{Nsolphase1} and \eqref{Nsolphase2}.

\subsection{Orbibundle over projective space}\label{example1}
As the first example, we will consider the derived equivalence between two orbibundles, $X_+$ and $X_-$, which are defined as the following quotient stacks
\begin{equation}
\begin{aligned}
    &X_+ \:=\: [{\rm Tot}(\mathcal{O}(-n/m)^{\oplus M} \to \mathbb{P}^{N-1})/(\mathbb{Z}_{m'}\times\mathbb{Z}_{g})]\, , \\
    &X_- \:=\: [{\rm Tot}(\mathcal{O}(-m/n)^{\oplus N} \to \mathbb{P}^{M-1})/(\mathbb{Z}_{n'}\times\mathbb{Z}_{g})]\, ,
\end{aligned}
\end{equation}
where $g=\gcd(m,n)$, $n'=n/g$, $m'=m/g$. $X_+$ and $X_-$ can be realized in two different phases of a GLSM with $\hat{X}$ as its target space in the pure geometric phase, where
\begin{equation}
\hat{X}\:=\:{\rm Tot}(\mathcal{O}(-m,-n) \to \mathbb{P}^{M-1} \times \mathbb{P}^{N-1})\, .
\end{equation}
In order to ensure the derived equivalence between $X_+$ and $X_-$, the constraint $m N = n M$ must hold due to Eq.~\eqref{constraintu1u1} , which is also the Calabi-Yau condition for both $X_+$ and $X_-$.

\subsubsection{GLSM construction}

To realize $\hat{X}$ , we consider the GLSM with $U(1) \times U(1)$ gauge group and it is defined by the following matter content for the chiral fields 
\begin{equation}\label{eq:abelianexample}
\centering
\begin{tabular}{c|ccc|c}
       & $\Phi_a$  & $Y_{i}$ & $P$ &  \\ \hline
 $U(1)_1$ &$1$  & $0$  & $-m$ & $\xi_1$\\
 $U(1)_2$ &$0$  & $1$  & $-n$ & $\xi_2$
\end{tabular} 
\end{equation}
with $a = 1,\dots,M$ and $i = 1,\dots,N$. $\xi_1$ and $\xi_2$ are the FI parameters. We assume $m<M, n<N$ so the axial $U(1)$ R-symmetry has anomaly, i.e. the theory is anomalous. There is no superpotential in this model. This is a special case of the $U(1)\times U(1)$ gauge theory defined in \eqref{tab:generalU1U1}. Then, the Coulomb vacuum equations follow directly from \eqref{eq:u1u1-1} and \eqref{eq:u1u1-2} in the massless limit, namely
\begin{equation}\label{CoeqEX1}
    \begin{aligned}
        \sigma_1^M &\:=\: q_1 (-m \sigma_1 - n \sigma_2 )^m\, , \\
        \sigma_2^N &\:=\: q_2 (-m \sigma_1 - n \sigma_2 )^n\, ,
    \end{aligned}
\end{equation}
with $q_a = \exp(-t_a)$ and the $t_a$'s are the FI-$\theta$ parameters. The $D$-term equations are also recoded below, which will be useful for identifying the Higgs components of the different phases,
\begin{equation}\label{DeqEX1}
    \sum_{a=1}^M |\Phi_a|^2 - m |P|^2 \:=\: \xi_1\,,\quad \sum_{i=1}^N |Y_i|^2 - n |P|^2 \:=\: \xi_2\,.
\end{equation}

\subsubsection{Phases}
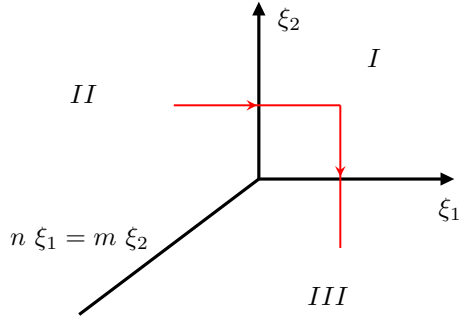
\begin{figure}[!h]
    \centering
    \tikzset{every picture/.style={line width=0.75pt}} 
    \begin{tikzpicture}[x=0.75pt,y=0.75pt,yscale=-0.6,xscale=0.6,decoration={markings, mark= at position 0.5 with {\arrow{stealth}}, mark= at position 4cm with {\arrow{stealth}}} ]
    \draw [line width=1.2]    (311.5,232.02) -- (471.83,232.02) ;
    \draw [shift={(475.83,232.02)}, rotate = 180] [fill={rgb, 255:red, 0; green, 0; blue, 0 }  ][line width=0.08]  [draw opacity=0] (11.61,-5.58) -- (0,0) -- (11.61,5.58) -- cycle    ;
    
    \draw [line width=1.2]    (311.5,232.02) -- (311.5,86.83) ;
    \draw [shift={(311.5,82.83)}, rotate = 90] [fill={rgb, 255:red, 0; green, 0; blue, 0 }  ][line width=0.08]  [draw opacity=0] (11.61,-5.58) -- (0,0) -- (11.61,5.58) -- cycle    ;
    \draw [line width=1.2]    (160.5,346) -- (311.5,232.02) ;
     
    \draw [draw = red, postaction={decorate}] (240,170) -- (380,170); 
    \draw [draw = red, postaction={decorate}] (380,170) -- (380,290);

    \draw (400,120) node [anchor=north west][inner sep=0.75pt]    {$I$};
    \draw (460,245) node [anchor=north west][inner sep=0.75pt]    {$\xi_{1}$};
    \draw (325,85) node [anchor=north west][inner sep=0.75pt]    {$\xi_{2}$};
    \draw (350,320) node [anchor=north west][inner sep=0.75pt]    {$III$};
    \draw (150,150) node [anchor=north west][inner sep=0.75pt]  [font=\normalsize]  {$II$};
    \draw (100,270) node [anchor=north west][inner sep=0.75pt]  [font=\normalsize]  {$n\ \xi_{1} =m\ \xi_{2}$};
    
    \end{tikzpicture}
    \caption{Phases for the abelian two-parameter model \eqref{eq:abelianexample}.}
    \label{fig:abelianphases}
\end{figure}

As depicted in Fig.~\ref{fig:abelianphases}, this GLSM has three different phases: $\{\xi_1 \gg 0,\xi_2 \gg 0\}$, $\{\xi_1 \ll 0, n\xi_1 \ll m\xi_2\}$ and $\{\xi_2 \ll 0, n\xi_1 \gg m\xi_2\}$. Let us go through these phases and analyze their IR behavior.

\paragraph{Phase I: $\xi_1 \gg 0$ and $\xi_2 \gg 0$.} This phase is a pure geometric phase, and the target space is
\[
   \hat{X} \:=\: {\rm Tot}(\mathcal{O}(-m,-n) \to \mathbb{P}^{M-1} \times \mathbb{P}^{N-1})\, ,
\]
which can be read off from the D-term equations \eqref{DeqEX1}. One can also confirm this by analyzing the asymptotic behavior of the Coulomb branch equations \eqref{CoeqEX1} in the deep phase limit, $\xi_1 \to + \infty$ and $\xi_2 \to + \infty$.
In this limit, the only solution to the Coulomb vacuum equations is $\sigma_1 = \sigma_2 = 0$, which indicates that this phase only has a pure Higgs branch.

Then the Witten index in this phase only has contributions from the Euler characteristic of the target space $\hat{X}$, which is $M N$. 

\paragraph{Phase II: $\xi_1 \ll 0$ and $n \xi_1 \ll m \xi_2$.} This phase consists of a Higgs branch and several mixed branches. One can see this by looking at the behavior of solutions to the Coulomb branch equations \eqref{CoeqEX1} in the limit $\xi_1 \to - \infty$ and $\xi_2 \to +\infty$ such that $n \xi_1 < m \xi_2$. In this limit, $\sigma_2 \to 0$ and then the first equation becomes
\[
    \sigma_1^M = q_1 (-m)^m \sigma_1^m \, ,
\]
which has total $M$ solutions and $m$ of them are zero solutions while the other $(M-m)$ solutions are non-zero. Namely, now we have $m$ zero solutions $(\sigma_1,\sigma_2) = (0,0)$ and $(M-m)$ solutions with $(\sigma_1,\sigma_2) = (\ast, 0)$ where $\ast$ is one of the solutions to equation $\sigma_1^{M-m} = q_1 (-m)^m$.

The solution $(\sigma_1,\sigma_2) = (0,0)$ suggests that there exists a pure Higgs branch, which can be read off from the D-term equations. To describe the Higgs branch, one can change the basis for the Lie algebra of the gauge group and obtain
\[
\centering
\begin{tabular}{c|ccc|c}
       & $\Phi_a$  & $Y_{i}$  &P & \\ \hline
 $n' U(1)_1 - m' U(1)_2$& $n'$  & $-m'$ & 0 & $n' \xi_1 - m' \xi_2$
 \\
 $\mu U(1)_1+\nu U(1)_2$ & $\mu$ & $\nu$ & $-g$ &  $\mu\xi_1+\nu\xi_2$
\end{tabular}
\]
where $g = \mathrm{gcd}(m,n)$, and $\mu$ and $\nu$ is a pair of integers such that\footnote{This is to ensure that the transformation given by the matrix $\left( \begin{array}{cc} n' & -m' \\ \mu & \nu \end{array}\right)$ is in $SL(2,\ZZ)$. The choice of $(\mu, \nu)$ is not unique, but a different choice, say $(\mu+ l n',\nu - l m')$ for some integer $l$, amounts to combining $\mu U(1)_1+\nu U(1)_2$ with $l$ multiples of $n' U(1)_1 - m' U(1)_2$, so we may select one choice and fix $(\mu,\nu)$.} $\mu m'+\nu n'=1$. Since $\xi_1 \ll 0$, the $P$ field should take the non-vanishing expectation value $\left<P\right> = \sqrt{-\xi_1 / m}$. As such, the gauge group breaks to a subgroup $U(1)_s \times \mathbb{Z}_g$, where $U(1)_s := n' U(1)_1 - m' U(1)_2$ and its associated FI parameter is $\xi_s:= n' \xi_1 - m' \xi_2$. Once the fluctuation of the $P$ field is integrated out, the remaining massless degrees of freedom are
\begin{equation}\label{masslessEX1}
\centering
\begin{tabular}{c|cc|c}
       & $\Phi_a$  & $Y_{i}$  & \\ \hline
 $U(1)_s$& $n'$  & $-m'$  &  $\xi_s$ \\
 $\mathbb{Z}_g$ & $\mu$ & $\nu$ & 
\end{tabular}
\end{equation}
and because $\xi_s \ll 0$, one can immediately find out that the pure Higgs branch has target space
\[
    X_+ \:=\: [{\rm Tot}(\mathcal{O}(-n'/m')^{\oplus M} \to \mathbb{P}^{N-1})/(\ZZ_{m'}\times \ZZ_g)]\, ,
\]
where the orbifold group acts trivally on the base space, and its action on the fiber can be read off from \eqref{masslessEX1}.
The Euler characteristic of $X_+$ can be computed as
\[
    \chi(X_+) \:=\: |\ZZ_{m'}\times \ZZ_g| \cdot \chi(\mathbb{P}^{N-1}) \:=\: m N\,.
\]

\begin{figure}[!h]
    \centering
    \tikzset{every picture/.style={line width=0.75pt}}
    \begin{tikzpicture}[x=0.75pt,y=0.75pt,yscale=-0.6,xscale=0.6]
    \draw  [line width=1.5]  (309.01,103.94) -- (622.64,103.94) -- (353.2,235.06) -- (39.56,235.06) -- cycle ;
    \draw [line width=1.5]  [dash pattern={on 5.63pt off 4.5pt}]  (325,70) -- (325,180) ;
    \draw [line width=1.5]  [dash pattern={on 5.63pt off 4.5pt}]  (480,40) -- (480,120) ;
    \draw [line width=1.5]  [dash pattern={on 5.63pt off 4.5pt}]  (180,120) -- (180,200) ;
    \draw [line width=1.5]  [dash pattern={on 5.63pt off 4.5pt}]  (360,140) -- (360,220) ;
    \fill (325,180) circle[radius=4pt];
    \fill (480,120) circle[radius=3pt];
    \fill (180,200)  circle[radius=3pt];
    \fill (360,220)  circle[radius=3pt];
    \draw (250,200) node [anchor=north west][inner sep=0.75pt]  [font=\Huge,rotate=-10] [align=left] {...};
    \draw (300,35) node [anchor=north west][inner sep=0.75pt]    {$X_+$};
    \draw (250,160) node [anchor=north west][inner sep=0.75pt]    {$\sigma_1=0$};
    \draw (460,15) node [anchor=north west][inner sep=0.75pt]    {$\mathbb{P}^{N-1}$};
    \draw (160,90) node [anchor=north west][inner sep=0.75pt]    {$\mathbb{P}^{N-1}$};
    \draw (340,110) node [anchor=north west][inner sep=0.75pt]    {$\mathbb{P}^{N-1}$};
    \draw (566,80) node [anchor=north west][inner sep=0.75pt]  [font=\large]  {$\sigma _{1}$};

    \end{tikzpicture}
    \caption{Mixed structure of Phase~II.}
    \label{fig:phase2abelianmodel}
\end{figure}
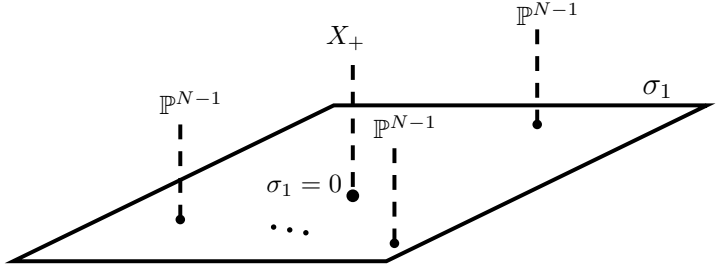

At the nonzero solutions $(\sigma_1,\sigma_2)=(\ast,0)$, the fields $\Phi_a$ and $P$ will be massive and should be integrated out. What's left is an effective GLSM for $\mathbb{P}^{N-1}$ defined by $U(1)_2$ with $N$ chiral fields $Y_i$ of charge $1$. In summary, the whole structure of this phase can be described as in Fig.~\ref{fig:phase2abelianmodel}.

Now, let us count the Witten index as a consistency check. There are two types of contributions: one is from the pure Higgs branch $X_+$ and the other one is from the $(M-m)$ mixed Higgs-Coulomb branches ($M-m$ copies of $\mathbb{P}^{N-1}$). Therefore, the total Witten index is the sum of these two contributions
\[
    m N + (M-m) N = M N\, ,
\]
matching the result in Phase~I.

\paragraph{Phase III: $\xi_2 \ll 0$ and $n \xi_1 \gg m \xi_2$.}
This phase has a similar mixed structure as Phase~II, including: one Higgs branch, which is $X_- = [{\rm Tot}(\mathcal{O}(-m'/n')^{\oplus N} \to \mathbb{P}^{M-1})/\ZZ_{n'}\times \ZZ_g]$, and $(N-n)$ mixed Higgs-Coulomb branches, namely $\mathbb{P}^{M-1}$ located at $(N-n)$ non-zero points on the $\sigma_2$-plane. We skipped the details here, but one can also check the matching of the total Witten index 
\[
    n M + (N-n) M = M N.
\]

\subsubsection{Effective theories on the phase boundaries}
Now we want to implement the brane transport along the path illustrated in Fig.~\ref{fig:abelianphases} to find a functor realizing the equivalence between $D^b(X_+)$ and $D^b(X_-)$. Apparently, this requires the understanding of the effective local models at the phase boundaries, $\{\xi_1=0, \xi_2 \gg 0\}$ and $\{\xi_1 \gg 0,\xi_2=0\}$, and the small window categories thereof.

The local model at the phase boundary $\{\xi_1=0, \xi_2 \gg 0\}$ has the following matter content
\begin{equation}\label{local+EX1}
\centering
\begin{tabular}{c|cc|c}
       & $\Phi_a$ & $P$ &  \\ \hline
 $U(1)_1$ &$1$  & $-m$ & $\xi_1$
\end{tabular} 
\end{equation}
Therefore, the small window category consists of the branes with $U(1)_1$ charges $q_1$ satisfying
\[
\left| q_1 + \frac{\theta_1}{2 \pi} \right|< \frac{m}{2},
\]
which is termed band restriction rule in \cite{Herbst:2008jq}. Thus, with a suitable choice of the $\theta$-angle, the small window category at the phase boundary $\{\xi_1=0, \xi_2 \gg 0 \}$ is generated by branes of the form $\mathcal{W}(i,q)$, where $0 \leq i \leq m-1, q \in \ZZ$. 

The local model at the phase boundary $\{\xi_1 \gg 0,\xi_2=0\}$ has the following matter content
\begin{equation}\label{local-EX1}
\centering
\begin{tabular}{c|cc|c}
       & $Y_i$ & $P$ &  \\ \hline
 $U(1)_2$ & $1$  & $-n$ & $\xi_2$
\end{tabular} 
\end{equation}
Therefore, the small window category consists of the branes with $U(1)_2$ charges $q_2$ satisfying the band restriction rule
\[
\left| q_2 + \frac{\theta_2}{2 \pi} \right|< \frac{n}{2}.
\]
Thus, with a suitable choice of the theta-angle, the small window category at the phase boundary $\{\xi_1 \gg 0, \xi_2 = 0 \}$ is generated by branes of the form $\mathcal{W}(q,j)$, where $-n+1 \leq j \leq 0, q \in \ZZ$.

\subsubsection{Brane transport and derived equivalence}\label{sec:transEX1}
We are interested in studying the derived equivalence between $X_+$ and $X_-$. One necessary condition is that their Euler characteristics coincide, namely $m N = n M$. As discussed before, we have realized them in different phases of the GLSM for $\hat{X}={\rm Tot}(\mathcal{O}(-m,-n) \to \mathbb{P}^{M-1} \times \mathbb{P}^{N-1})$. In the GLSM construction, $X_{\pm}$ are accompanied by the mixed Higgs-Coulomb branches $\mathcal{C}_{\pm}$ respectively. In this example,
\[
\begin{aligned}
    &\mathcal{C}_+ \:=\: \left\{(M-m) \text{ copies of } \mathbb{P}^{N-1}\ \text{located at $(M-m)$ nonzero points on $\sigma_1$-plane} \right\}\, , \\
    &\mathcal{C}_-\:=\: \left\{(N-n) \text{ copies of } \mathbb{P}^{M-1}\ \text{located at $(N-n)$ nonzero points on $\sigma_2$-plane} \right\} \, . 
\end{aligned}
\]
The Witten index contributions from $\mathcal{C}_{\pm}$ should also match, {\it i.e.} $(M-m)N = (N-n)M$, which is exactly the same as $m N = n M$ as one would expect\footnote{In later examples, direct computation of the Euler characteristics of some pure Higgs branches might be difficult and so we might instead compute the Witten indices of the mixed branches to derive the matching conditions for the Euler characteristics of the pure Higgs branches.}. One also immediately recognizes that the condition,  $m N = n M$, is also the Calabi-Yau condition for both $X_+$ and $X_-$. 

The deleted sets on the phases are given by
\begin{equation}
    \begin{aligned}
     X_+:&\ \lbrace Y=0 \rbrace\cup \lbrace P=0 \rbrace
     \\
     \hat{X}:&\ \lbrace Y=0 \rbrace\cup \lbrace \Phi=0 \rbrace
     \\
     X_-:&\ \lbrace \Phi=0 \rbrace\cup \lbrace P=0 \rbrace 
    \end{aligned}\label{eptbrane}
\end{equation}
and the corresponding Koszul complexes of empty branes on these deleted sets are 
\begin{equation}
\begin{gathered}
    \mathcal E_Y:\quad \mathcal{W}(q,-N)\xrightarrow{Y}\mathcal{W}(q,-N+1)^{\oplus N}\xrightarrow{Y}\cdots\xrightarrow{Y}\mathcal{W}(q,-1)^{\oplus N}\xrightarrow{Y}\mathcal{W}(q,0),
    \\
    \mathcal E_\Phi:\quad \mathcal{W}(-M,q)\xrightarrow{\Phi}\mathcal{W}(-M+1,q)^{\oplus M}\xrightarrow{\Phi}\cdots\xrightarrow{\Phi}\mathcal{W}(-1,q)^{\oplus M}\xrightarrow{\Phi}\mathcal{W}(0,q),
    \\
    \mathcal{E}_P:\quad \mathcal{W}(m,n)\xrightarrow{P}\mathcal{W}(0,0),
\end{gathered}
\end{equation}
from which $\mathcal E_P$ will be used to grade restrict UV brane complexes into the small windows. As we assume $m<M, n<N$, the FI parameters are subject to renormalization. If the FI parameters $(\xi_1,\xi_2)$ are above the line $n \xi_1 = m \xi_2$, then they will end up in the phase of $X_+$ in the IR limit under RG flow. The RG flow will take a UV GLSM brane $\mathcal W(p_1,p_2)$ to an IR brane on $X_+$ as\footnote{Here and in the following, we will use $\cO_X(q)$ to denote $\pi^*\cO_{\PP^{r}}(q)$ for $X = \mathrm{Tot}(\mathcal{F} \stackrel{\pi}{\rightarrow} \PP^{r} )$, where $\mathcal F$ is a vector bundle or an orbibundle.}
\begin{equation}\label{proj+EX1}
\mathcal W(p_1,p_2)\xrightarrow{\pi_+} \mathcal O_{X_+}\left(\frac{p_2 m'-p_1 n'}{m'}\right)_{s},
\end{equation}
where $s \equiv \mu p_1+\nu p_2 ~(\mathrm{mod}~g)$ labels the twisted sector of $\mathbb Z_g$. On the other hand, if the FI parameters $(\xi_1,\xi_2)$ are below the line $n \xi_1 = m \xi_2$, then they will end up in the phase of $X_-$ in the IR limit. 
A UV GLSM brane $\mathcal W(p_1,p_2)$ will descend to an IR brane on $X_-$ as
\begin{equation}\label{proj-EX1}
\mathcal W(p_1,p_2)\xrightarrow{\pi_-} \mathcal O_{X_-}\left(\frac{p_1n'-p_2m'}{n'}\right)_{s}.
\end{equation}
Notice that if we use $\mathcal{E}_P$ to replace $\mathcal{W}(p_1,p_2)$ with an IR equivalent brane $\mathcal{W}(p_1- l m,p_2- l n)$ for some integer $l$, then both $i=q_2m'-q_1n'$ and $s \equiv \mu (p_1 - l m) + \nu (p_2 - l n) \equiv \mu p_1 + \nu p_2~ (\mathrm{mod}~g)$ are invariant, therefore the IR images under $\pi_+$ and $\pi_-$ are indeed unchanged. 

Consequently, the brane transport between $D^b(X_+)$ and $D^b(X_-)$ is given by the following procedure:
\begin{itemize}
\item For any $i \in \{ 0,1,\cdots, m' N -1 \}, s \in \ZZ_g $, choose a pair of integers $(p_1, p_2)$ such that $i = p_2 m' - p_1 n'$ and $s \equiv \mu p_1 + \nu p_2~(\mathrm{mod}~g)$. For example,
\[
p_1=-\nu i+m's +lm,\ p_2=\mu i+n's+ln,\ \forall~ l\in\mathbb Z.
\]
\item Then, from Eq.~\eqref{proj+EX1}, The lift of a brane $\mathcal O_{X_+}\left(\frac{i}{m'}\right)_{s}$ in the phase of $X_+$ (above the line $n\xi_1 = m \xi_2$) is
\begin{equation}
    \mathcal O_{X_+}\left(\frac{i}{m'}\right)_{s}\xrightarrow{\pi_+^{-1}}\mathcal W(p_1,p_2),
\end{equation}

\item Use the IR empty brane $\mathcal{E}_P$ to grade restrict $\mathcal{W}(p_1,p_2)$ so the resulting brane is in the small window category at the phase boundary $\{\xi_1=0, \xi_2 \gg 0 \}$, i.e. $p_1=0,\cdots,m-1$, so it can pass through this phase boundary. 
\item Once in the phase $\{ \xi_1 \gg 0, \xi_2 \gg 0 \}$, we can move it below the line $n \xi_1 = m \xi_2$ and use the IR empty brane $\mathcal{E}_P$ to grade restrict it such that it is in the small window category at the phase boundary $\{ \xi_1 \gg 0 , \xi_2 = 0 \}$, i.e. $p_2 \in \{ -n+1, -n+2, \cdots, 1,0 \}$. Then the brane can be transported to the phase of $X_-$.
\item Due to Eq.~\eqref{proj-EX1}, the brane can be projected to give the image:
\[
\mathcal W(\overline{p}_1,\overline{p}_2) \xrightarrow{\pi_-}\mathcal O_{X_-}\left(-\frac{i}{n'}\right)_{s},
\]
where $\overline{p}_1$ and $\overline{p}_2$ are the resulting gauge charges after the grade restriction has been applied.

\end{itemize}

In conclusion, with our choice of the small windows, the brane transport yields the following map between the generators of $D^b(X_+)$ and $D^b(X_-)$:
\[
\mathcal O_{X_+}\left(\frac{i}{m'}\right)_{s} \Rightarrow \mathcal O_{X_-}\left(-\frac{i}{n'}\right)_{s}
\]
for $i = 0,1,\cdots, m'N-1, ~s \in \ZZ_g$.

\paragraph{Example:} $m=4,n=6$, $M=6$, $N=9$. \\
In this case, $g = \gcd(m,n) = 2, m'=2,n'=3$ and we can take $\mu = -1, \nu =1$. The geometries are
\[
\begin{gathered}
X_+= [\mathrm{Tot}(\mathcal O(-3/2)^{\oplus 6}\rightarrow\mathbb P^8)/(\mathbb Z_2\times \mathbb Z_2)]\, ,
\\
X_-= [\mathrm{Tot}(\mathcal O(-2/3)^{\oplus 9}\rightarrow\mathbb P^5)/(\mathbb Z_3\times\mathbb Z_2)]\, .
\end{gathered}
\]
The small window at the phase boundary $\{\xi_1=0, \xi_2 \gg 0\}$ is generated by $\mathcal{W}(i,p_2)$, where $i=0,1,2,3$ and $p_2$ is any $U(1)_2$ charge. The small window at the phase boundary $\{\xi_1 \gg 0, \xi_2 = 0\}$ is generated by $\mathcal{W}(p_1,j)$, where $j=-5,-4,\cdots,0$ and $p_1$ is any $U(1)_1$ charge.

The derived category $D^b(X_+)$ is generated by $\cO_{X_+}\left(\frac{i}{2}\right)_s$ with $i=0,1,\cdots,17,~s = 0,1$. The derived category $D^b(X_-)$ is generated by $\cO_{X_-}\left(-\frac{j}{3}\right)_t$ with $j=0,1,\cdots,17,~t = 0,1$. 
For any $i$ and $s$ in the range indicated above, take
\[
p_1=-i+2s,\quad p_2=-i+3s.
\]
The lift of $\cO_{X_+}\left(\frac{i}{2}\right)_s$ is $\mathcal{W}(p_1+4l,p_2+6l)$ for any integer $l$. We can first choose $l$ such that $p_1+4l \in \{0,1,2,3\}$. After the brane is transported to the phase $\{\xi_1 \gg 0, \xi_2 \gg 0\}$, we can choose another integer $l'$ such that $p_2+6l + 6l' \in \{ -5,-4,\cdots,0\}$ so the grade restricted brane can be transported to the phase of $X_-$, where its IR image is $\cO_{X_-}\left(-\frac{i}{3}\right)_s$.

\subsection{Product of a projective space and a Grassmannian}\label{example2}
In the previous section, we have presented our strategy for showing the derived equivalence in an abelian gauge theory. Now, in this section, we start to consider a non-abelian example and show the derived equivalence between $X_+$, a vector bundle over Grassmannian as defined by Eq.~(\ref{eq:ex2X+}), and $X_-$, a fibre bundle over projective space as defined by Eq.~(\ref{eq:ex2X-}). We follow a similar construction as discussed in the previous section and find that $X_+$ and $X_-$ can be realized in two different phases of the GLSM for $\hat{X}$, which is a fiber bundle over the product of a projective space and a Grassmannian, Eq.~(\ref{eq:ex2Xhat}).

\subsubsection{GLSM construction}
The GLSM has the gauge group $U(1) \times U(2)$ and the following matter content
\begin{equation}\label{glsmEX2}
\centering
\begin{tabular}{c|ccc|c}
       & $Y_a$  & $\Phi^{\alpha}_{j}$ & $P$    \\ \hline
 $U(1)$& 1  & 0  &$-m$ & $\xi_1$\\
 $U(2)$ & $\cdot$ & ${\yng(1)}$   & $\det^{-n}$ & $\xi_2$
\end{tabular}
\end{equation}
where $\alpha=1,2$ is the $U(2)$ color index, $a=1,\cdots,M$ and $j=1,\dots,N$ are flavor indices, $\yng(1)$ stands for the fundamental representation of $U(2)$, and we assume $m \leq M, n \leq N$. When $m=M$ and $n=N$, the theory is non-anomalous, i.e. the theory is Calabi-Yau, otherwise the theory is anomalous. In this model, the superpotential $W=0$. The Coulomb vacuum equations are 
\begin{equation}\label{cvEX2}
    \begin{aligned}
        \sigma^M &= q_1 (-m \sigma - n \sigma_1 - n \sigma_2)^m\,, \\
        \sigma_\alpha^{N} &= -q_2(-m \sigma - n \sigma_1 -n \sigma_2)^n\,, ~~\alpha =1,2\, ,
    \end{aligned}
\end{equation}
where $\sigma$ is the scalar component of the gauge field strength associated with $U(1)$, while $\sigma_\alpha, \alpha=1,2,$ are those associated with the Cartan subalgebra of $U(2)$. In the above, $q_s:= \exp(-t_s)$ and $t_s$ is the FI-$\theta$ parameter corresponding to $\xi_s$.

\subsubsection{Phases}
The phase diagram for this GLSM is the same as the previous example, {\it i.e.}~Fig.~\ref{fig:abelianphases}. Clearly, in the $\xi_1 \gg 0, \xi_2 \gg 0$ region, there is only a pure Higgs branch with target space
\begin{equation}\label{eq:ex2Xhat}
\hat{X} \:=\: \mathrm{Tot}\left( \mathcal{O}_{\mathbb{P}^{M-1}}(-m) \otimes \left( \det \mathcal{S} \right)^{\otimes n} \rightarrow \mathbb{P}^{M-1} \times Gr(2,N) \right)\,,
\end{equation}
The total Witten index in this phase equals the Euler characteristic of $\hat{X}$, which is 
\begin{equation}\label{indexEX2}
    \chi(\hat{X}) \:=\: M  \binom{N}{2}\,.
\end{equation}

If either $\xi_1 \ll 0$ or $\xi_2 \ll 0$, the $P$ field gets a nonzero vev due to the D-term equations and breaks the $U(1) \times \det U(2)$ gauge group to a subgroup $U(1)_s \times \ZZ_{\gcd(m,n)}$ with the FI parameter of $U(1)_s$ given by $\xi_s := n' \xi_1 - m' \xi_2$, where $m'=m/{\gcd(m,n)}, n'=n/{\gcd(m,n)}$. Under the $U(1)_s$ gauge group, the $SU(2)$ invariant fields $Y_a$ and the baryon fields $B_{ij}=\varepsilon_{\alpha\beta} \Phi^\alpha_i \Phi^\beta_j$ have charges 
\[
\centering
\begin{tabular}{c|cc}
  &  $Y_a$ & $B_{ij}$ \\ \hline
$U(1)_s$ & $n'$ & $-m'$  
\end{tabular}
\]
When $\xi_s \ll 0, \xi_1 \ll 0$, $Y_a$ and $B_{ij}$ become the fiber and base coordinates of
\begin{equation}\label{productX+}
\mathrm{Tot}\left(\mathcal{O}(-n/m)^{\oplus M} \rightarrow \mathbb{P}^{{N \choose 2}-1}\right)
\end{equation}
respectively at low energy scale. The relations satisfied by the $B_{ij}$'s then tell us that in this phase, there is a Higgs branch whose target space $X_+$ is an orbifold of the vector bundle \eqref{productX+} restricted to the image of the Pl\"ucker embedding of $Gr(2,N)$ in $\mathbb{P}^{{N \choose 2}-1}$. Therefore, $X_+$ is defined as \footnote{In fact, $\mathcal{O}(-n/m)$ should be interpreted as the pullback of the orbibundle $\mathcal{O}(-n/m)$ on $\mathbb{P}^{\binom{N}{2}-1}$ under the Pl\"ucker embedding and here we abuse the notation to avoid clumsy symbols.} 
\begin{equation} \label{eq:ex2X+}
    X_+ \:=\: [\mathrm{Tot}\left( \mathcal{O}(-n/m)^{\oplus M} \rightarrow Gr(2,N) \right)/(\ZZ_{m'} \times \ZZ_{\gcd(m,n)})]\,,
\end{equation}
The Euler characteristic of $X_+$ can be computed directly,
$$\chi(X_+) \:=\: m \binom{N}{2}\,.$$
In this region, there are also mixed branches, as can be read off from non-vanishing Coulomb vacua when $\xi_s \ll 0, \xi_1 \ll 0$. When we take $\xi_s \to - \infty$, one can see from \eqref{cvEX2} that
\[
    \sigma_\alpha^{Nm} \:=\: - q_s \sigma^{Mn} \to 0 \,,
\]
which means $\sigma_\alpha \to 0$. The remaining Coulomb vacua equation is 
\[
    \sigma^M \:=\: q_1 (-m)^m \sigma^m\, ,
\]
and this equation suggests that $\sigma$ has zero solutions, which is consistent with the existence of the pure Higgs branch, namely $X_+$. In addition, this equation also has $(M-m)$ nonzero solutions, which indicates the existence of the mixed branches. One can find that the effective theory corresponding to each mixed branch can be described by a NLSM with target space the Grassmannian $Gr(2,N)$. Therefore, the mixed branch is
\[
\mathcal{C}_+ \:=\: \left\{(M-m) \text{ copies of } Gr(2,N) \text{ located at } (M-m) \text{ nonzero points on $\sigma$-plane}  \right\}\,.
\]
One can check that the total Witten index equals
\[
    \chi(X_+) + \chi(\mathcal{C}_+) \:=\: m \binom{N}{2} + (M-m) \binom{N}{2} \:=\: M \binom{N}{2}\,,
\]
which is consistent with the total Witten index \eqref{indexEX2}.

When $\xi_s \gg 0, \xi_2 \ll 0$, $X_a$ and $B_{ij}$ become the base and fiber coordinates of
\begin{equation}
\mathrm{Tot}\left(\mathcal{O}(-m/n)^{\oplus{N \choose 2}} \rightarrow \mathbb{P}^{M-1} \right)
\end{equation}
respectively at low energy scale. The relations satisfied by the $B_{ij}$'s then tell us that in this phase, there is a Higgs branch whose target space $X_-$ is an orbifold of the fiber bundle over $\mathbb{P}^{M-1}$ with each fiber being the affine cone of $Gr(2,N)$ (we denote the affine cone by $CGr(2,N)$). Let us call the fiber bundle $\mathcal{X}(2,N)$. In short the target space in this phase is the quotient stack
\begin{equation} \label{eq:ex2X-}
    X_- \:=\: \left[\left(\mathrm{Tot}\left( \mathcal{X}(2,N) \rightarrow \mathbb{P}^{M-1}\right) \right)/(\mathbb{Z}_{n'} \times \ZZ_{\gcd(m,n)}) \right]\,.
\end{equation}
Note that $X_-$ is not an ordinary smooth fiber bundle and it is not straightforward to calculate its Euler characteristic, but we can easily obtain the Euler characteristic of $X_-$ by subtracting the mixed branch contributions from the total Witten index. One way to do this computation is to take the standpoint that each fiber arises as the Higgs branch of the negative phase of the $U(2)$ GLSM with $N$ fundamentals and one field in the $\det^{-n}$ representation. The positive phase of this $U(2)$ theory has a pure Higgs branch which realizes a vector bundle over the Grassmannian $Gr(2,N)$, so the total Witten index is ${N \choose 2}$. We see that there are $(N-n)\lfloor \frac{N-1}{2} \rfloor$ Coulomb vacua in the negative phase of the $U(2)$ theory. Therefore the Witten index of the Higgs branch of the negative phase of this $U(2)$ theory is ${N \choose 2} - (N-n)\left\lfloor \frac{N-1}{2} \right\rfloor$. Consequently, the Witten index of $X_-$ is
\[
   \chi(X_-)  \:=\:   M\binom{N}{2} - M(N-n)\left\lfloor \frac{N-1}{2} \right\rfloor \,.
\]

The mixed Higgs-Coulomb branch can be found following the same procedure as before. In the limit $\xi_s \to +\infty$, one will find the mixed branches located at $\sigma=0$ and $(N-n)\lfloor \frac{N-1}{2} \rfloor$ non-vanishing solutions for $(\sigma_1, \sigma_2)$. Further, on each such nonvanishing point in the $\{\sigma_1,\sigma_2\}$-plane, there is a projective space $\mathbb{P}^{M-1}$. Namely, we have
\[
\mathcal{C}_- \:=\: \left\{(N-n)\left\lfloor \frac{N-1}{2} \right\rfloor \text{ copies of } \mathbb{P}^{M-1} \text{ located at the nonzero points on $\{\sigma_1,\sigma_2\}$-plane}  \right\}\,.
\]
Thus, this leads to the matching of the Witten indices
\[
   \chi(X_-) \:=\: \chi(\hat{X}) - \chi(\mathcal{C}_-)\, .
\]
In particular,
\[
\chi(X_-) \:=\: \begin{dcases}
    \frac{ n M (N-1)}{2}, &N \text{ is odd}\,, \\
    \frac{nM(N-2)}{2} +\frac{MN}{2}, &N \text{ is even}\,.
\end{dcases}
\]
Comparing $\chi(X_+)$ and $\chi(X_-)$, the derived equivalence between $X_+$ and $X_-$ suggests 
\begin{equation}\label{eq:ex2eulermatch}
\begin{dcases}
    m N \:=\: nM, &N \text{ is odd}\,, \\
     m N (N-1) - n M ( N  - 2 ) \:=\: MN, &N \text{ is even}\,.
\end{dcases}
\end{equation}
Note that the condition with $N$ odd is also the Calabi-Yau condition for $X_{\pm}$, while $\hat{X}$ is not necessarily a Calabi-Yau. When $N$ is even, if we further apply the Calabi-Yau condition for $X_{\pm}$, {\it i.e. $m N = nM$}, then the second equation becomes
\[
    (N-n)M \:=\: 0\, ,
\]
namely, we should have $M=m$ and $N=n$, which is also the Calabi-Yau condition for $\hat{X}$.

In particular, when $N=3$, the first equation gives $n M=3 m$. When $N=4$, the second equation gives $(n+2)M=6m$ and in this case, if the Calabi-Yau condition is further applied, we can only take $M=m$ and $n=N=4$. In the following, we will consider the cases of $X_{\pm}$ being Calabi-Yau varieties.

\subsubsection{Effective theories on the phase boundaries}\label{eftEX2}
The local model at the phase boundary of $\{ \xi_1=0, \xi_2 \gg 0\}$ is a $U(1)$ gauge theory with the following matter content
\begin{equation}\label{local+EX2}
\centering
\begin{tabular}{c|cc}
  &  $Y_a$ & $P$ \\ \hline
$U(1)$ & 1 & $-m$  
\end{tabular}
\end{equation}
The small window of this abelian GLSM is
\begin{equation}\label{GRREX2}
    \left| q_1+\frac{\theta_1}{2\pi}\right |<\frac{m}{2}\, ,
\end{equation}
where $q_1$ and $\theta_1$ are the gauge charge and theta angle of $U(1)$. Eq.~\eqref{GRREX2} suggests that the window category at the phase boundary $\{\xi_1 = 0, \xi_2 \gg 0\}$ can be chosen to be generated by branes of the form $\mathcal{W}(q_1, \lambda)$, where $q_1 \in \{ 0, 1, \cdots m-1 \}$ and $\lambda$ is any $U(2)$ representation. 

The local model at the phase boundary of $\{ \xi_2=0, \xi_1 \gg 0\}$ is the $U(2)$ gauge theory with the following matter content
\begin{equation}\label{local-EX2}
\centering
\begin{tabular}{c|cc}
  &  $\Phi^\alpha_j$ & $P$ \\ \hline
$U(2)$ & ${\yng(1)}$   & $\det^{-n}$
\end{tabular}
\end{equation}
For the cases with $n \leq N$, $N=3$ and $4$, this model has been studied in section~\ref{sec:small} and the small window is given by Eq.~(\ref{smallwindowN3}) and Eq.~(\ref{smallwindowN4}). Therefore, the small window category at the phase boundary $\{ \xi_2=0, \xi_1 \gg 0\}$ can be chosen to be generated by branes of the form $\mathcal W(q_1, \lambda)$, where $\lambda$ is given by Eq.~(\ref{smallwindowN3}) and Eq.~(\ref{smallwindowN4}) in the cases of $N=3$ and $N=4$ respectively and $q_1$ is any $U(1)$ charge.

\subsubsection{Brane transport and derived equivalence}
The theory \eqref{glsmEX2} with $N=3$ is dual in the IR limit to an abelian theory with matter content
\[
\centering
\begin{tabular}{c|ccc|c}
       & $Y_a$  & $\Phi_{j}$ & $P$    \\ \hline
 $U(1)$& 1  & 0  &$-m$ & $\xi_1$\\
 $U(1)$ & 0 &  1  & $-n$ & $\xi_2$
\end{tabular}
\]
which reduces to one of the cases studied in section~\ref{example1}.
In this section, we focus on the behavior of brane transport in the case of $N=4$.

As discussed before, the Calabi-Yau condition for $X_{\pm}$ requires that $n=N=4$ and $m=M$. In this case, $\hat X$ is also Calabi-Yau, and we have the derived equivalence 
\[
D^b(X_+) \cong D^b(\hat{X}) \cong D^b(X_-),
\]
where the geometries in the three phases are
\[
\begin{aligned}
&X_+ \:=\: \left[\mathrm{Tot}\left(\mathcal O(-4/m)^{\oplus m} \stackrel{\pi_+}{\rightarrow} Gr(2,4) \right)/(\ZZ_{m/\gcd(m,4)} \times \ZZ_{\gcd(m,4)}) \right],
\\
&X_-\:=\: \left[ \left( \mathcal{X}(2,4) \stackrel{\pi_-}{\rightarrow} \mathbb P^{m-1} \right) / (\ZZ_{4/\gcd(m,4)} \times \ZZ_{\gcd(m,4)}) \right],
\\
&\hat{X} \:=\: \mathrm{Tot} \left( \mathcal O_{\PP_{m-1}}(-m)\otimes(\wedge^2 \cS)^{\otimes 4}) \stackrel{\pi}{\rightarrow}\mathbb P^{m-1}\times Gr(2,4) \right),
\end{aligned}
\]
where $\mathcal{X}(2,4)$ is the fiber bundle with fiber $CGr(2,4)$, which is the affine cone of $Gr(2,4)$. In this case, the axial $U(1)$ R-symmetry is non-anomalous, the small window category and the big window category coincide. Moreover, because the FI parameters are not renormalized, the brane transport amounts to the following steps\footnote{Here we take the transport to be from the phase of $X_+$ to the phase of $X_-$ for example, and the transport in the opposite direction follows the same procedure in the reverse order.}: 
\begin{itemize}
\item For any brane $B_+$ in $D^b(X_+)$, lift it to a GLSM brane $\mathcal W_+$ such that it is in the window category of the local model \eqref{local+EX2}, so it can be smoothly transported to the phase of $\hat{X}$.
\item Once in the phase of $\hat{X}$, we can read off its image in $D^b(\hat{X})$.
\item Use empty branes to grade restrict $\mathcal W_+$ and obtain an IR equivalent brane $\mathcal W_-$, which is in the window category of the local model \eqref{local-EX2}.
\item $\mathcal W_-$ can be smoothly transported to the phase of $X_-$, where its image in $D^b(X_-)$ can be determined.
\end{itemize}

Let us investigate some examples to see how the above steps are implemented. Assume that $M=m=3$ for simplicity. In this case, the empty branes for grade restriction at the phase boundary  $\{\xi_1 \gg 0, \xi_2 = 0\}$ are derived from exact sequences on $Gr(2,4)$ generated by Eagon-Northcott type complexes, such as\footnote{The Young diagram $\lambda$ is a shorthand notation for $\Sigma_{\lambda} \mathcal{S}^\vee$, where $\Sigma_{\lambda}$ is the Schur functor associated with $\lambda$.}
\begin{equation}\label{sequence1}
\overline{\yng(3,2)} \rightarrow  \overline{\yng(2,2)}^{\oplus 4} \rightarrow  \overline{ \yng(1,1)}^{\oplus 4} \rightarrow \overline{ \yng(1)},
\end{equation}
\begin{equation}\label{sequence2}
\overline{\yng(4,4)} \rightarrow \overline{ \yng(3,3)}^{\oplus 6} \rightarrow \overline{ \yng(3,2)}^{\oplus 4} \rightarrow \overline{ \yng(2,1)}^{\oplus 4} \rightarrow \overline{ \yng(1,1)}^{\oplus 6} \rightarrow \cO.
\end{equation}
On the other hand, the empty branes for grade restriction at the phase boundary  $\{\xi_1 = 0, \xi_2 \gg 0\}$ are of the form
\[
W(q_1,\lambda) \rightarrow W(q_1+1,\lambda)^{\oplus 3} \rightarrow W(q_1+2,\lambda)^{\oplus 3} \rightarrow W(q_1+3,\lambda),
\]
where the arrows are maps given by $Y_a, a=1,2,3$, and $\lambda$ is any $U(2)$ representation. 
Notice that a GLSM brane of the form $\mathcal{W}(z_1,\det^{z_2})$ is projected to $\cO_{X_+}((3 z_2 - 4 z_1)/3)$ in the phase of $X_+$, and to $\cO_{X_-}((4 z_1-3 z_2)/4)$ in the phase of $X_-$, as explained in section~\ref{sec:transEX1}.
\begin{itemize}
\item $\mathcal{O}_{X_+}(1)$ can be lifted to the GLSM brane $\mathcal W(0, \det)$, which is in the window category of \eqref{local+EX2} so it can be transported to the phase $\xi_1 \gg 0, \xi_2 \gg 0$. Then it projects to $\pi^*(\cO_{\PP^2} \otimes \wedge^2 \mathcal{S}^\vee)$ on $\hat{X}$. Because $\mathcal W(0, \det)$ is also in the window category of \eqref{local-EX2}, it can be transported to the phase of $X_-$, where it projects to $\cO_{X_-}(-3/4)$.
\item $\mathcal{S}^\vee \otimes \cO_{X_+}(1/3)$ can be lifted to the GLSM brane $\mathcal W(3, {\yng(1)}\otimes\det^2)$, which is in the window category of \eqref{local+EX2} so it can be transported to the phase $\xi_1 \gg 0, \xi_2 \gg 0$, where it projects to $\pi^*(\cO_{\PP^2}(3) \otimes (\wedge^2 \mathcal{S}^\vee)^{\otimes 2}\otimes \mathcal{S}^\vee)$ on $\hat{X}$. As $\mathcal W(3, {\yng(1)}\otimes\det^2)$ is not in the window category of \eqref{local-EX2}, we need to grade restrict it. This can be done by using the exact sequence \eqref{sequence1}. The grade restricted brane is then
\[
\mathcal{W}(3,{\det}^2)^{\oplus 4} \rightarrow \mathcal{W}(3,\det)^{\oplus 4} \rightarrow \mathcal{W}(3,{ \yng(1)}),
\]
which is now in the window category so it can be transported to give the image
\[
\cO_{X_-}\left(\frac{3}{2}\right)^{\oplus 4} \rightarrow \cO_{X_-}\left(\frac{9}{4}\right)^{\oplus 4} \rightarrow \cO_{X_-}\left(3\right) \otimes \tilde{\mathcal S}^\vee.
\]
\item $\cO_{X_+}(4/3)$ can be lifted to the GLSM brane $\mathcal W(2, \det^4)$, which is in the window category of \eqref{local+EX2} so it can be transported to the phase $\xi_1 \gg 0, \xi_2 \gg 0$, where it projects to $\pi^*(\cO_{\PP^2}(2) \otimes (\wedge^2 \mathcal{S}^\vee)^{\otimes 4})$ on $\hat{X}$. $\mathcal W(3, {\yng(1)}\otimes\det^2)$ is not in the window category of \eqref{local-EX2}, but it can be grade restricted with the help of the cone of the exact sequences \eqref{sequence2} and four copies of \eqref{sequence1}, in which the representation ${\yng(3,2)}$ is cancelled. The grade restricted brane reads
\[
\mathcal{W}(2,{\det}^3)^{\oplus 6} \rightarrow \mathcal{W}(2,{\det}^2)^{\oplus 16} \rightarrow \mathcal{W}(2,\det)^{\oplus 16} \rightarrow \mathcal{W}(2,{\yng(1)})^{\oplus 4},
\]
which is now in the window category so it can be transported to give the image
\[
\cO_{X_-}\left(-\frac{1}{4}\right)^{\oplus 6} \rightarrow \cO_{X_-}\left(\frac{1}{2}\right)^{\oplus 16} \rightarrow \cO_{X_-}\left(\frac{5}{4}\right)^{\oplus 16} \rightarrow \left(\cO_{X_-}\left(2\right) \otimes \tilde{\mathcal S}^\vee \right)^{\oplus 4}.
\]
\item Let us look at an example of brane transport in the reverse direction. Starting with the brane $\cO_{X_-}(3/4)$ in the phase of $X_-$, we get a GLSM brane $\mathcal{W}(3,\det^3)$, which is in the window category of \eqref{local-EX2}, so it can be transported to the phase of $\hat{X}$ and projected to $\pi^*(\cO_{\PP^2}(3) \otimes (\wedge^2 \mathcal{S}^\vee)^{\otimes 3})$. Brane $\mathcal{W}(3,\det^3)$ is not in the window category of \eqref{local+EX2}, but can be grade restricted using the exact sequence
\[
\mathcal{W}(0,{\det}^3) \rightarrow \mathcal{W}(1,{\det}^3)^{\oplus 3} \rightarrow \mathcal{W}(2,{\det}^3)^{\oplus 3} \rightarrow \mathcal{W}(3,{\det}^3).
\]
Therefore, the image of the brane transport in the phase of $X_+$ is
\[
\cO_{X_+}(3) \rightarrow \cO_{X_+}\left( \frac{5}{3} \right)^{\oplus 3} \rightarrow \cO_{X_+}\left( \frac{1}{3} \right)^{\oplus 3}.
\]
\end{itemize}
Other cases can be analyzed in the same way.

\subsection{Two-step flag manifold}\label{example3}
In this subsection, we discuss the derived equivalence between $X_+$ and $X_-$ defined as
\[
X_+ \:=\: {\rm Tot}[\mathcal{S}\otimes \det\mathcal{S} \to Gr(2,N)], \quad \quad 
    X_- \:=\: {\rm Tot}[\mathcal{L}^2\otimes \left(\mathcal{O}^{\oplus N}/\mathcal{L}\right) \to \mathbb{P}^{N-1} ]\, ,
\]
where $\mathcal{S}$ and $\mathcal{L}$ are the tautological bundles over $Gr(2,N)$ and $\mathbb{P}^{N-1}$ respectively. 
It turns out that $X_+$ and $X_-$ can be constructed by a $U(1) \times U(2)$ GLSM whose target space in the pure geometric phase is
\[
\hat{X} \:=\: {\rm Tot}[\mathcal{L}\otimes\det\mathcal{S}\to Fl(1,2,N)]\, ,
\]
where $\mathcal{L}$ and $\mathcal{S}$ are the first and second tautological bundles of the two-step flag manifold $Fl(1,2,N)$. 

Note that some analysis in this section can also be applied to the two-step flag manifolds without fiber directions. Therefore, following the GLSM constructions for flag manifolds \cite{Donagi:2007hi,Ohmori:2018qza,Guo:2018iyr}, we have included some discussions on $Fl(1,2,N)$
in appendix~\ref{app:flagphase} for comparison. 

\subsubsection{GLSM construction for $\hat{X}$}
The GLSM data for $\hat{X}$ are given as follows
\begin{equation}\label{glsmEX3}
\centering
\begin{tabular}{c|ccc|c}
       & $\Phi_{\alpha}$  & $Y^{\alpha}_{i}$ & $P$    \\ \hline
 $U(1)$& 1  & 0  &$-1$ & $\xi_1$\\
 $U(2)$ & $\overline{{\yng(1)}}$  & ${\yng(1)}$   & $\det^{-1}$ & $\xi_2$
\end{tabular}
\end{equation}
with the color index $\alpha=1,2$ and the flavor index $i=1,\dots,N$. To obtain nontrivial geometry, we need $N>2$, which also makes the theory anomalous. The D-terms associated with $U(1)$ and $U(2)$ are 
\begin{equation}\label{DEX3}
\begin{aligned}
    D &\sim \sum_{\alpha=1}^2 |\Phi_\alpha|^2 - |P|^2 - \xi_1\, , \\
    D^{\alpha}_\beta &\sim - \overline{\Phi}^\alpha \Phi_\beta + \sum_{i}\overline{Y}_{\beta,i} Y^{\alpha}_i - |P|^2 \delta^\alpha_\beta - \xi_2 \delta^\alpha_\beta\, . 
\end{aligned}
\end{equation}
The equations for Coulomb vacua derived from the effective twisted superpotential are
\begin{equation}\label{coulombEX3}
\begin{aligned}
   & \prod_{\alpha=1}^{2}\left( \sigma^{(1)} - \sigma^{(2)}_{\alpha}\right) = - q_1 \left( \sigma^{(1)} +  \sigma^{(2)}_1 +  \sigma^{(2)}_2  \right)\, , \\
   & {\sigma^{(2)}_\alpha}^N = q_2 \left(\sigma^{(1)} - \sigma^{(2)}_{\alpha}\right) \left( \sigma^{(1)} +  \sigma^{(2)}_1 +  \sigma^{(2)}_2  \right)\, ,
\end{aligned}
\end{equation}
for $\alpha = 1,2$ and $q_a = \exp(-t_a)$ with $t_1$ and $t_2$ the complex FI parameters associated with $U(1)$ and $U(2)$ respectively. In the above equations, $\sigma^{(1)}$ is the scalar component of the $U(1)$ gauge field strength, while $\sigma_{\alpha}^{(2)}$ are the scalar components of the Cartan part of $U(2)$ gauge field strength.

\subsubsection{Phase structure}\label{phaseEX3}

\begin{figure}[!h]
    \centering
    \tikzset{every picture/.style={line width=0.75pt}} 
    \begin{tikzpicture}[x=0.75pt,y=0.75pt,yscale=-0.6,xscale=0.6]
    \draw [line width=1.5]    (125,265) -- (260,265) -- (450,265) ;
    \draw [shift={(454,265)}, rotate = 180] [fill={rgb, 255:red, 0; green, 0; blue, 0 }  ][line width=0.08]  [draw opacity=0] (11.61,-5.58) -- (0,0) -- (11.61,5.58) -- cycle    ;
    \draw [line width=1.5]    (290,265) -- (290,120) ;
    \draw [shift={(290,114.83)}, rotate = 90] [fill={rgb, 255:red, 0; green, 0; blue, 0 }  ][line width=0.08]  [draw opacity=0] (11.61,-5.58) -- (0,0) -- (11.61,5.58) -- cycle    ;
    \draw [line width=1.5]    (290,265) -- (440,330) ;
    \draw (373,185) node [anchor=north west][inner sep=0.75pt]  [font=\normalsize]  {$I$};
    \draw (466,250) node [anchor=north west][inner sep=0.75pt]    {$\xi_{1}$};
    \draw (300,115) node [anchor=north west][inner sep=0.75pt]    {$\xi_{2}$};

    \draw (373,275) node [anchor=north west][inner sep=0.75pt]    {$III$};
    \draw  [draw opacity=0][fill={rgb, 255:red, 255; green, 255; blue, 255 }  ,fill opacity=1 ]  (277,315) -- (304,315) -- (304,350) -- (277,350) -- cycle  ;
    \draw (190,185) node [anchor=north west][inner sep=0.75pt]  [font=\normalsize]  {$II$};
    \draw (435,340) node [anchor=north west][inner sep=0.75pt]    {$\xi_1 + 2 \xi_2 = 0$};
    \end{tikzpicture}
    \caption{Phase diagram.}
    \label{fig:phases}
\end{figure}
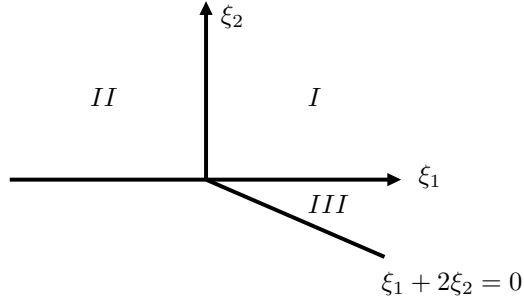

The phase structure is shown in Fig.~\ref{fig:phases}. For the purpose of this paper, we are only interested in Phase~I, ~II and ~III below.

\paragraph{Phase I: $\xi_1 \gg 0$ and $\xi_2 \gg 0$.} In this phase, the gauge groups are totally broken. Following the analysis of the asymptotic behavior of solutions to Coulomb branch equations in the limit of $\xi_1 \to +\infty$ and $\xi_2 \to + \infty$, one can see that the only solution for $\sigma^{(1)}$ and $\sigma^{(2)}_\alpha$ is the zero solution. Therefore, this phase only has a Higgs branch, whose target space is determined by the D-term equations $\{D=0,D^\alpha_\beta = 0\}$, which is the space $\hat{X}$ as expected,
\[
    \hat{X} \:=\: {\rm Tot}[\mathcal{L}\otimes\det\mathcal{S}\to Fl(1,2,N)]\, .
\]
The base $Fl(1,2,N)$ is determined by $\Phi_\alpha$ and $Y^{\alpha}_i$ (see e.g. \cite{Donagi:2007hi} for more details) and the fiber direction is described by the field $P$.

\paragraph{Phase II: $\xi_1 \ll 0$ and $\xi_2 \gg 0$.} Let us first analyze the asymptotic behavior of the Coulomb vacua in the large $\xi$ region, namely, the limit $\xi_1 \to -\infty$ and $\xi_2 \to +\infty$. In this limit, $\sigma^{(2)}_{\alpha} \to 0$ and the first equation of motion becomes 
\[
    \left(\sigma^{(1)}\right)^2 = - q_1 \sigma^{(1)}\, ,
\]
which has two solutions, $\sigma^{(1)} = 0$ and $\sigma^{(1)} = - q_1$. Therefore, there is a Higgs branch suggested by the solution $\sigma^{(1)} = \sigma^{(2)}_{\alpha} = 0$ and a mixed branch at $\sigma^{(1)} = - q_1$ and $\sigma^{(2)}_{\alpha} = 0$.

The pure Higgs branch can be described by an effective $U(2)$ GLSM defined by fundamentals $X^{\alpha}_i$ and composite fields $Z_{\alpha}:= P \Phi_\alpha$, namely, $Z_{\alpha}$ is in the representation $\det^{-1} \otimes \bold{\bar{2}}$ \footnote{Similar analysis for pure flag manifolds has been used in \cite[section~2.1]{Donagi:2007hi}, in which they have defined the (forgetful) projection map. Here, we slightly generalize this idea to discuss bundles on flag manifolds.}. One can read off the geometry from the corresponding D-term equation, which is 
\[
    X_+\:=\: {\rm Tot}[\mathcal{S}\otimes \det\mathcal{S} \to Gr(2,N)]\, .
\]
For the mixed branch, since $\Phi_\alpha$ and $P$ are charged under $U(1)$, they are massive on this branch and should be integrated out. What is left is an effective $U(2)$ GLSM for the Grassmannian $Gr(2,N)$, defined by $N$ chiral fields $Y^{\alpha}_i$ in the fundamental representation of $U(2)$ gauge group. In other words, the mixed branch in this phase is
\[
    \mathcal{C}_+ \:=\: \left\{ \text{one } Gr(2,N) \text{ located at one non-zero point on the } \sigma\text{-plane}\right\}\, .
\]
The contributions to the total Witten index from the mixed branch $\mathcal{C}_+$ is easy to count, which is just the Euler characteristic of $Gr(2,N)$, {\it i.e.}
$  \binom{N}{2}\, .$

\paragraph{Phase III: $\xi_1 \gg 0$ and $\xi_1 + 2\xi_2 \gg 0$.} Again, let us get some hints from the asymptotic behavior of the Coulomb vacua equations in this phase. In the limit, $\xi_1 \to +\infty$, $\prod_{\alpha=1}^{2}\left( \sigma^{(1)} - \sigma^{(2)}_{\alpha}\right) = 0$ and so there are two solutions for $\sigma^{(1)}$, namely $\sigma^{(1)} = \sigma^{(2)}_1$ or $\sigma^{(1)} = \sigma^{(2)}_2$.
Also, note that
\[
    \left( \sigma^{(2)}_1 \sigma^{(2)}_2\right)^N = q_2^2 \left( \prod_{\alpha=1}^{2}\left( \sigma^{(1)} - \sigma^{(2)}_{\alpha}\right) \right)  \left( \sigma^{(1)} +  \sigma^{(2)}_1 +  \sigma^{(2)}_2  \right) = -q_1 q_2^2  \left( \sigma^{(1)} +  \sigma^{(2)}_1 +  \sigma^{(2)}_2  \right) ^2 \, ,
\]
and, due to $\xi_1+2\xi_2 >0$, $q_1 q_2^2 \to 0$ in the large $\xi$ region of this phase, we shall have the solution $ \sigma^{(2)}_1  = 0$ or $ \sigma^{(2)}_2 = 0 $. We choose $\sigma^{(2)}_1  = 0$ for convenience in the following discussion. Due to the Weyl symmetry, the case of $\sigma^{(2)}_2 = 0$ follows the same analysis but will not contribute more to the Witten index. If take $\sigma^{(1)} = \sigma^{(2)}_1 = 0$, Eq.~\eqref{coulombEX3} reduces to an equation for $\sigma^{(2)}_2$ taking the following form,
\[
    \left(\sigma^{(2)}_2\right)^N = -q_2\,  \left(\sigma^{(2)}_2\right)^2\, .
\]
This equation has $N-2$ non-vanishing solutions and the rest is $\sigma^{(2)}_2 = 0$. While if take $\sigma^{(1)} = \sigma^{(2)}_2$, then the remaining equation for $\sigma^{(2)}_2$ only has zero solutions. In summary, the asymptotic behavior of the Coulomb vacua suggests that this phase consists of a Higgs branch together with $(N-2)$ mixed branches.

Now let us give more details on the Higgs branch and the mixed branches in phase III. To describe the effective theory, we follow the strategy described in \cite[section~2.2]{Donagi:2007hi} by Pl\"{u}cker embedding into a product of projective space. See also some similar discussions in appendix~\ref{app:flagphase}. We need to introduce two sets of $SU(2)$ invariant coordinates: $W_i := \Phi_{\alpha}Y^{\alpha}_i$ and $B_{ij} := \epsilon_{\alpha\beta}Y^{\alpha}_{i}Y^{\beta}_{j}$. The effective theory is a gauged nonlinear sigma model defined by
\[
\centering
\begin{tabular}{c|ccc}
       & $W_i$  & $B_{ij}$ & $P$    \\ \hline
 $U(1)$& 1  & 0  &$-1$ \\ 
 $\det {U(2})$ & $0$  & $1$   & $-1$ \\ 
\end{tabular}
\]
with nontrivial nonlinear relations for $N\geq 4$ among coordinates $B_{ij}$'s determined by their definitions. Moreover, the definition of $W_i$ suggests that $W_i$ is a vector in the vector space spanned by $Y^1$ and $Y^2$. However, different from the example discussed in appendix~\ref{app:flagphase}, in this example both $D$-term equations will have solutions.

Because $\xi_2 \ll 0$, $P$ acquires a nonzero vev and breaks the gauge group to $U(1) - \det U(2)$ whose FI-parameter is large and positive. Moreover, due to the D-term equations from \eqref{DEX3},
\[
\begin{aligned}
    \sum_{i=1}^N |W_i|^2 &\:=\: \sum_{i=1}^N \sum_{\alpha,\beta} \overline{\Phi}^\beta \Phi_\alpha \overline{Y}_{\beta,i} Y^\alpha_i = \sum_{\alpha,\beta} \overline{\Phi}^\beta \Phi_\alpha (\overline{\Phi}^\alpha \Phi_\beta + |P|^2 \delta^\alpha_\beta + \xi_2 \delta^\alpha_\beta),\\
    &\:=\: \sum_{\alpha=1}^2 |\Phi_\alpha|^2 ( \sum_{\beta=1}^2 |\Phi_\beta|^2 + |P|^2 + \xi_2) = (2 |P|^2+\xi_1 + \xi_2) \sum_{\alpha=1}^2 |\Phi_\alpha|^2 >0\, ,
\end{aligned}   
\]
thus the $W_i$'s cannot vanish simultaneously in this phase.
Therefore the $W_i$'s become the homogeneous coordinates of $\mathbb{P}^{N-1}$, and the $B_{ij}$'s represent the vector subspace in $\mathbb{C}^N$ spanned by $Y^1$ and $Y^2$ which contains $W_i$. Since such a subspace can be determined by a vector in $\mathbb{C}^N$ which is either a nonzero vector in a direction different from that of $Z_i$ or a zero vector (In the latter case, it simply means $Y^1$ and $Y^2$ are linearly dependent so their span is one dimensional), we see the $B_{ij}$'s furnish the vector bundle $\mathcal{L}^2\otimes \left(\mathcal{O}^{\oplus N}/\mathcal{L}\right)$ over $\mathbb{P}^{N-1}$. The extra $\mathcal{L}$ factor is due to the fact that $B_{ij}$ has $U(1) - \det U(2)$ charge $-1$. Therefore, we have
\[
    X_-\:=\: \mathrm{Tot}\left(\mathcal{L}^2\otimes \left(\mathcal{O}^{\oplus N}/\mathcal{L}\right) \rightarrow \mathbb{P}^{N-1} \right)\, ,
\]
and
\[
    \mathcal{C}_- \:=\: \left\{ (N-2) \text{ copies of } \mathbb{P}^{N-1} \text{ located at } (N-2) \text{~non-zero points in the } \sigma \text{-plane} \right\}\, .
\]

Since we will discuss the derived equivalence between $X_+$ and $X_-$, the first requirement is the matching of their Euler characteristics. Instead of direct computation, we compare the Witten indices of the mixed branches. Since the total Witten index should be invariant across different phases, matching of contributions from mixed branches will indicate the Euler characteristics of the Higgs branches are the same. In the current example, we need
\[
    \chi(Gr(2,N)) \:=\: (N-2) \chi(\mathbb{P}^{N-1})\, ,
\]
in agreement with Eq.~\eqref{constraintu1u2}. Therefore $N=3$ is the only choice in this case.

\subsubsection{Effective theories on the phase boundaries}
The local model at the phase boundary $\{ \xi_1=0, \xi_2 \gg 0\}$ is the $U(1)$ gauge theory with the following matter content
\begin{equation}\label{loc+EX3}
\centering
\begin{tabular}{c|cc}
  &  $\Phi_\alpha$ & $P$ \\ \hline
$U(1)$ & 1 & $-1$  
\end{tabular}
\end{equation}
The small window of this model consists of only one $U(1)$ charge. By a suitable choice of theta angle, this charge can be taken to be zero.

The local model at the phase boundary $\{ \xi_2=0, \xi_1 \gg 0\}$ is the $U(2)$ gauge theory with the following matter content
\begin{equation}\label{loc-EX3}
\centering
\begin{tabular}{c|cc}
  &  $Y_i^\alpha$ & $P$ \\ \hline
$U(2)$ & ${\yng(1)}$   & $\det^{-1}$
\end{tabular}
\end{equation}
The small window is given by Eq.~\eqref{smallwindowN3}, from the $m=1$ case, we see the small window consists of a single $U(2)$ representation with dimensional one. We can choose the theta angle such that the representation in the small window is the trivial representation.

\subsubsection{Brane transport and derived equivalence}

Now we can compute the functor for the equivalence of $D^b(X_+)$ and $D^b(X_-)$. Let $\pi_\pm$ be the projection of the vector bundle $X_\pm$ onto the corresponding base space. We choose the generators of $D^b(X_+)$ to be $\pi_+^*\cO_{Gr(2,3)}$, $\pi_+^* \cS_{Gr(2,3)}^\vee$ and $\pi_+^* \wedge^2 \cS_{Gr(2,3)}^\vee$.

$\pi_+^*\cO_{Gr(2,3)}$ can be lifted to the GLSM brane $\mathcal{W}(0,\cdot)$, which fits in the small windows of both \eqref{loc+EX3} and \eqref{loc-EX3}, so it can be directly transported and projected to give the image $\pi_-^*\cO_{\mathbb{P}^2}$.

$\pi_+^* \wedge^2 \cS_{Gr(2,3)}^\vee$ can be lifted to the GLSM brane $\mathcal{W}(0,\wedge^2 {\yng(1)})$, which fits in the small window of the local model \eqref{loc+EX3} at the first boundary. Due to the empty brane
\[
{\mathcal{W}(0,\wedge^2 {\yng(1)})} \stackrel{P}{\longrightarrow} \mathcal{W}(-1,\cdot),
\]
it is IR equivalent to $\mathcal{W}(-1,\cdot)$, which is then in the small window of the local model \eqref{loc-EX3} at the second boundary. Consequently, it can be projected to give the image $\pi_-^*\cO_{\mathbb{P}^2}(-1)$.

For $\pi_+^* \cS_{Gr(2,3)}^\vee$, we need another empty brane because the $P$ field cannot turn a fundamental representation into a trivial representation. Instead, we use the IR empty brane given by the short exact sequence\footnote{Note that $\Phi_1$ and $\Phi_2$ cannot both be zero, otherwise the $W_i$'s would vanish simultaneously.}
\begin{equation}\label{emptyEX3}
\xymatrix{
\mathcal{W}(-1,\det)
\ar@<0.5ex>[rr]^-{\left( {\footnotesize \begin{array}{c} \Phi_1 \\ \Phi_2 \end{array} }\right)}
&& { \mathcal{W}(0, { \yng(1)}) }
\ar@<0.5ex>[rr]^-{(\Phi_2,-\Phi_1)}
&& \mathcal{W}(1, \cdot)
}\, .
\end{equation}
Together with the empty brane
\[
\mathcal{W}(-1,\det) \stackrel{P}{\longrightarrow} \mathcal{W}(-2,\cdot)\,,
\]
we get the IR equivalence
\[
\mathcal{W}(0,{ \yng(1)}) \cong \text{Cone} \left( \mathcal{W}(1,\cdot)[-1] \rightarrow \mathcal{W}(-2,\cdot) \right)\, .
\]
Therefore the IR image in the phase $X_-$ is the complex $\text{Cone} (\pi_-^*\cO_{\mathbb{P}^2}(1)[-1] \rightarrow \pi_-^*\cO_{\mathbb{P}^2}(-2))$.

In summary, the brane transport with the choice of small windows mentioned above gives rise to the following map between the generators of $D^b(X_+)$ and $D^b(X_-)$:
\[
\begin{aligned}
\pi_+^*\cO_{Gr(2,3)} &\Longrightarrow \pi_-^*\cO_{\mathbb{P}^2}\, , \\
\pi_+^* \cS_{Gr(2,3)}^\vee &\Longrightarrow \text{Cone} \left(\pi_-^*\cO_{\mathbb{P}^2}(1)[-1] \rightarrow \pi_-^*\cO_{\mathbb{P}^2}(-2) \right)\, , \\
\pi_+^* \wedge^2 \cS_{Gr(2,3)}^\vee &\Longrightarrow \pi_-^*\cO_{\mathbb{P}^2}(-1)\, . \\
\end{aligned}
\]

\subsubsection{An equivalent abelian GLSM description}

In the case $N=3$, the manifolds $X_+$ and $X_-$ are actually isomorphic. This can be seen from the fact that under the isomorphism $Gr(k,N) \rightarrow Gr(N-k,N)$, the tautological bundle $\mathcal{S}$ is mapped to $\mathcal{Q}^\vee$, where $\mathcal{Q}$ is the universal quotient bundle, and $\mathcal{Q}$ is mapped to $\mathcal{S}^\vee$. Therefore, we have
\[
\begin{split}
&X_- \:=\: {\rm Tot}[\mathcal{L}^{\otimes 2}\otimes \left(\mathcal{O}^{\oplus 3}/\mathcal{L}\right) \to \mathbb{P}^{2} ] \cong {\rm Tot}[(\mathcal{Q^\vee})^{\otimes 2} \otimes \mathcal{S}^\vee \rightarrow Gr(2,3)]\\ &\cong {\rm Tot}[(\mathrm{det} \mathcal{S})^{\otimes 2} \otimes (\mathrm{det}\mathcal{S}^\vee \otimes \mathcal{S}) \rightarrow Gr(2,3)]
\cong {\rm Tot}[(\mathrm{det}\mathcal{S}) \otimes \mathcal{S} \rightarrow Gr(2,3)] = X_+,
\end{split}
\]
where we have used $\mathrm{det}{\mathcal{Q}^\vee} \cong \det \mathcal{S}$ and $\mathcal{S}^\vee \cong \mathrm{det}\mathcal{S}^\vee \otimes \mathcal{S}$ on $Gr(2,3)$. Actually, both $X_+$ and $X_-$ are isomorphic to the cotangent bundle of $\mathbb{P}^2$ because
\[
{\rm Tot}[\mathcal{L}^{\otimes 2}\otimes \left(\mathcal{O}^{\oplus 3}/\mathcal{L}\right) \to \mathbb{P}^{2} ] \cong {\rm Tot}[\mathcal{L}^{\otimes2}\otimes \mathcal{Q} \rightarrow \mathbb{P}^2] \cong {\rm Tot}[\mathcal{L}\otimes \mathcal{Q}^\vee\rightarrow \mathbb{P}^2] \cong T^*\mathbb{P}^2.
\]

For $N=3$, there is an equivalent abelian GLSM that is dual to the $U(1)\times U(2)$ theory defined in \eqref{glsmEX3}. This dual theory is a $U(1)\times U(1)$ gauge theory with the following chiral fields
\[
\centering
\begin{tabular}{c|cccc|c}
       & $Z_i$  & $B_{i}$ & $P$ & $Q$    \\ \hline
 $U(1)_1$& 1  & 0  &$-1$ &$-1$ & $\xi_1$\\
 $U(1)_2$ & 0 & 1  &$-1$ & $-1$ & $\xi_2$
\end{tabular}
\]
for $i=1,2,3$, and this theory is also equipped with the following superpotential
\begin{equation}
    W \:=\: Q \sum_{i=1}^3 Z_i B_i\,.
\end{equation}

The classical vacuum equations are given by D-term equations
\begin{equation}
    \sum_{i=1}^3 |Z_i|^2 - |P|^2 - |Q|^2 \:=\: \xi_1 \,,\qquad \sum_{i=1}^3 |B_i|^2 - |P|^2 - |Q|^2 \:=\: \xi_2 \,,
\end{equation}
and the F-term equations
\begin{equation}
    \sum_{i=1}^3 Z_i B_i \:=\:0\,, \qquad  Q Z_j \:=\: 0\,, \qquad Q B_j\:=\: 0\,, \qquad \text{for } j =1,2,3.
\end{equation}
Combined with the D-term equations, one can immediately see that, in the generic interior of each phase, the solution to the F-term equations should have $Q=0$, otherwise it will force us to consider the special loci with $\xi_1 = \xi_2$. 

The Coulomb vacua equations can be derived from the effective twisted
superpotential
\begin{equation}
    {\sigma_a}^3 \:=\: q_a (\sigma_1 + \sigma_2)^2 \,,\quad  \text{for } a =1,2\,.
\end{equation}
Following the same procedures, these three phases can be analyzed with the abovementioned D-term and F-term equations, and the Coulomb vacua equations. For comparison, we only summarize the target spaces of the three phases of our interest below:
\begin{equation*}
\begin{array}{rl}
    \text{Phase~I } (\xi_1 \gg 0,\xi_2 \gg 0): \qquad &  {\rm Tot}[ \mathcal{O}(-1,-1) \longrightarrow \mathfrak{B}]\,, \\
    \text{Phase~II } (\xi_1 \ll 0,\xi_2 \gg \xi_1): \qquad & T^*\mathbb{P}^2 \,,\\
    \text{Phase~III } (\xi_2 \ll 0,\xi_1 \gg \xi_2): \qquad & T^*\mathbb{P}^2 \,,
\end{array}
\end{equation*}
where the base space $\mathfrak{B}$ is defined as
\begin{equation}
    \mathfrak{B} = \Big\{([Z_i], [B_i]) \in \mathbb{P}^2\times \mathbb{P}^2 \mid \sum_i Z_i B_i = 0\Big\}\,,
\end{equation}
and one can see that $\mathfrak{B}\cong Fl(1,2;3)$. We can naturally view $Z_i$ and $B_i$ live in the dual vector spaces, and the relation $\sum_i Z_i B_i = 0$ comes from their pairing. 

\subsection{Derived equivalence between Calabi-Yau 5-folds}\label{example4}

In this section, we generalize the example considered in the previous section by introducing a symplectic structure, which is realized by a superpotential. We will consider $\hat{X}$ as the total space of a fiber bundle over the two-step symplectic flag manifold $SF(1,2,N)$,
\[
SF(1,2,N) \:=\: \{ \mathbb{C} \subset \mathbb{C}^2 \subset V ~|~ V \text{ is an $N$-dimensional symplectic vector space }, \mathbb{C}^2 \text{ is isotropic } \}.
\]
When $N=4$, this becomes Segal's example in \cite{Segal_2016}, which has
\[
   \hat{X} \:=\: {\rm Tot}[\cL \otimes \wedge^2 \mathcal{S} \to SF(1,2,4)]\, ,
\]
where $\mathcal{L}$ and $\mathcal{S}$ are the first and second tautological bundles over $SF(1,2,4)$. The Calabi-Yau fivefolds, $X_+$ and $X_-$, are defined as
\[
    X_+ \:=\: {\rm Tot}[\cS\otimes \wedge^2S\to SGr(2,4) ]\ , 
\]
and
\[
    X_- \:=\: {\rm Tot}[(\cL^{\perp}/\cL)\otimes\cL^2 \to \PP^3 ]\ ,
\]
where $\cL^{\perp}$ is the symplectic orthogonal of $\cL$ with respect to the symplectic structure on $\cO^{\oplus 4}$.
This example was studied in detail in \cite{Segal_2016} from the mathematical perspective, in which it was shown that there are birational maps from $\hat{X}$ to $X_+$ and $X_-$ which make $X_+$ and $X_-$ birational, and further arguments showed that there is a derived equivalence between $X_+$ and $X_-$. Here we use a GLSM construction to realize this equivalence and derive the functor between $D^b(X_+)$ and $D^b(X_-)$ induced by brane transport.

\subsubsection{GLSM for $\hat{X}$}
The GLSM for $\hat{X}$ is a $U(1)\times U(2)$ gauge theory with chiral fields $\Phi_\alpha$, $Y^\alpha_i$, $Q_{\alpha \beta}$ and $P_{\alpha \beta}$ with $\alpha = 1,2$ and $i=1,2,3,4$. These fields transform under the gauge group action as follows
\[
\centering
\begin{tabular}{c|cccc|c}
       & $\Phi_{\alpha}$  & $Y^{\alpha}_{i}$ & $Q_{\alpha\beta}$ & $P_{\alpha\beta}$   \\ \hline
 $U(1)$& 1  & 0  &0 &$-1$ & $\xi_1$\\
 $U(2)$ & $\overline{{\yng(1)}}$  & ${\yng(1)}$ & $\wedge^2\overline{{\yng(1)}}$  & $\wedge^2\overline{{\yng(1)}}$ & $\xi_2$
\end{tabular}
\]
and there is a superpotential
\begin{equation}\label{superpotential}
    W \:=\: \sum_{i,j}\sum_{\alpha,\beta} Q_{\alpha\beta} Y_i^{\alpha}Y_{j}^{\beta} \Omega^{ij} \:\equiv\: \sum_{i,j}\sum_{\alpha,\beta}Q_{\alpha\beta} Y_i^{\alpha} \wedge Y_j^{\beta}\, ,
\end{equation}
where $\Omega^{ij}$ is a symplectic form in $\CC^4$. From the matter content, we see that the theory is anomalous. The $D$-term equations read off
\be\label{DeqEX4}
    \begin{aligned}
        \sum_\alpha |\Phi_\alpha|^2 - \sum_{\alpha<\beta}|P_{\alpha\beta}|^2 \:=\: \sum_\alpha |\Phi_\alpha|^2 - |P|^2  &\:=\: \xi_1 \ , \\
        \sum_i \overline{Y}_{i,\alpha} Y_i^{\beta} - \Phi_{\alpha} \overline{\Phi}^\beta - \left( |P|^2 + |Q|^2\right)\delta_{\alpha}^\beta&\:=\: \xi_2 \delta_\alpha^\beta \ .
    \end{aligned}
\ee
In the above, we have denoted $P_{\alpha\beta} = P \epsilon_{\alpha\beta}$ and $Q_{\alpha\beta} = Q \epsilon_{\alpha\beta}$, $\alpha, \beta = 1,2$, where $\epsilon_{\alpha\beta}$ is the Levi-Civita symbol. To see this GLSM indeed realizes the space $\hat{X}$, one can do a similar analysis following \cite{Gu:2020oeb,Guo:2021dlz} and find that, in the phase $\xi_1,\xi_2 \gg 0$, the solutions to the $D$-term and $F$-term equations tell us that the vevs of $\Phi_\alpha$ and $Y_i^\alpha$ constitute the symplectic flag manifold $SF(1,2,4)$ (the F-term equations derived from \eqref{superpotential} impose the symplectic condition and force the $Q$ field to vanish), which is the base for $\hat{X}$, while $P_{\alpha\beta}$ contributes the fibers of $\hat{X}$.

The Coulomb branch equations are
\[
\begin{aligned}
    &\prod_{\alpha=1}^2\left(\sigma^{(1)} - \sigma^{(2)}_{\alpha}\right)  = q_1 (\sigma^{(1)} + \sigma^{(2)}_1+\sigma^{(2)}_2), \\
    &\left(\sigma^{(2)}_{\alpha}\right)^N = q_2 \left(\sigma^{(1)} - \sigma^{(2)}_{\alpha}\right)( \sigma^{(2)}_1+\sigma^{(2)}_2)(\sigma^{(1)} + \sigma^{(2)}_1+\sigma^{(2)}_2). 
\end{aligned}
\]
One can check that $\hat{X}$ is a Fano variety by computing the charges of any $U(1)\subset U(1)\times U(2)$. 

\subsubsection{The phases for $X_+$ and $X_-$}
The phase diagram of the GLSM for $\hat{X}$ can also be illustrated by Fig.~\ref{fig:phases}. The previous subsection has shown that phase I, namely, $\xi_1,\xi_2 \gg 0$, consists of a pure Higgs branch, realizing the manifold $\hat{X}$. In this subsection, we explore other phases and we will see that the Calabi-Yau fivefolds $X_\pm$ arise as Higgs branches in these phases.

\paragraph{Phase II: $\xi_1\ll 0 $ and $\xi_2\gg 0$.}
The $U(2)$ $D$-term equation shows that $Y_i^\alpha$ is non-degenerate, so they become homogeneous coordinates of $Gr(2,4)$ at low energy scale. The F-term equations impose the symplectic condition and force $Q$ to vanish. The $U(1)$ invariant fields $P \Phi_\alpha$ are in the $U(2)$ representation $\overline{{\yng(1)}} \otimes \wedge^2\overline{{\yng(1)}}$, which become the coordinates along the fiber. Therefore the vacuum manifold in this phase is 
\[
X_+ \:=\: {\rm Tot}\left(\mathcal{S}\oplus \wedge^2 \mathcal{S} \to SGr(2,4)\right)\ .
\]

\paragraph{Phase III: $\xi_1\gg 0$, $\xi_2\ll 0$ and $\xi_1 + 2 \xi_2 \gg 0$.}
If we take the trace of the second equation of \eqref{DeqEX4}, we get
\[
\sum_{\alpha,i} |Y^\alpha_i|^2 -\sum_\alpha |\Phi_\alpha|^2 - 2 \left( |P|^2 + |Q|^2 \right) = 2 \xi_2\, .
\]
Adding the first equation of \eqref{DeqEX4} to the equation above, we get
\[
\sum_{\alpha,i} |Y^\alpha_i|^2 - 3 |P|^2 -2 |Q|^2 = \xi_1 + 2 \xi_2\, ,
\]
therefore, $Y^\alpha_i$ cannot vanish simultaneously in this phase. Consequently, the F-term equations force the $Q$ field to vanish. Define the $SU(2)$ invariant fields $Z_i = \Phi_\alpha Y^\alpha_i$ and $B_{ij} = \epsilon_{\alpha \beta} Y^\alpha_i Y^\beta_j$, then the remaining degrees of freedom can be summarized as
\[
\centering
\begin{tabular}{c|ccc}
       & $Z_i$  & $B_{ij}$ & $P$    \\ \hline
 $U(1)$& 1  & 0  &$-1$ \\ 
 $\det {U(2})$ & $0$  & $1$   & $-1$ \\ 
\end{tabular}
\]
Moreover, from the D-term equations,
\[
\begin{aligned}
    \sum_{i} |Z_i|^2 &\:=\: \sum_{i=1} \sum_{\alpha,\beta} \overline{\Phi}^\beta \Phi_\alpha \overline{Y}_{\beta,i} Y^\alpha_i = \sum_{\alpha,\beta} \overline{\Phi}^\beta \Phi_\alpha (\overline{\Phi}^\alpha \Phi_\beta + (|P|^2 + |Q|^2 + \xi_2) \delta^\alpha_\beta)\, ,\\
    &\:=\: \sum_{\alpha} |\Phi_\alpha|^2 ( \sum_{\beta} |\Phi_\beta|^2 + |P|^2 + |Q|^2 + \xi_2) = (\xi_1 + \xi_2 + 2 |P|^2 + |Q|^2) (\xi_1 + |P|^2) > 0\, , 
\end{aligned}   
\]
so the $Z_i$'s cannot vanish simultaneously.
Then following the same analysis as that for phase III of section~\ref{phaseEX3}, we see $Z_i$ become the homogeneous coordinates of $\mathbb{P}^3$ and $B_{ij}$'s encode the subspace spanned by $Y^1$ and $Y^2$ in $\mathbb{C}^4$ containing $Z_i$. In addition, $Y^1$ and $Y^2$ must be orthogonal to each other with respect to the symplectic form on $\mathbb{C}^4$ due to the F-term equations, so the $B_{ij}$'s give us the vector bundle $(\cL^{\perp}/\cL)\otimes\cL^2$. As such, the vacuum manifold in this phase is
\[
    X_- \:=\: {\rm Tot}\left((\cL^{\perp}/\cL)\otimes\cL^2 \to \PP^3 \right)\, .
\]

\subsubsection{Effective theories on the phase boundaries}
The local model at the phase boundary of $\{ \xi_1=0, \xi_2 \gg 0\}$ is the $U(1)$ gauge theory with the following matter content
\begin{equation}\label{loc+EX4}
\centering
\begin{tabular}{c|cc}
  &  $\Phi_\alpha$ & $P$ \\ \hline
$U(1)$ & 1 & $-1$  
\end{tabular}
\end{equation}
The small window of this model consists of only one $U(1)$ charge. By a suitable choice of theta angle, this charge can be taken to be zero.

The local model at the phase boundary of $\{ \xi_2=0, \xi_1 \gg 0\}$ is the $U(2)$ gauge theory with the following matter content
\begin{equation}\label{loc-EX4}
\centering
\begin{tabular}{c|ccc}
  &  $Y_i^\alpha$ & Q & $P$ \\ \hline
$U(2)$ & ${\yng(1)}$ & $\det^{-1}$  & $\det^{-1}$
\end{tabular}
\end{equation}
The small window is given by Eq.~\eqref{smallwindowN4} with $m=2$. With a suitable choice of the theta angle, we can take the small window to be
\[
\left\langle { \yng(1)} \otimes {\det}^{-2} ,~ {\det}^{-1},~ { \yng(1)} \otimes {\det}^{-1},~ \cdot \right\rangle\, .
\]

\subsubsection{Derived equivalence induced by brane transport}
Now we can compute the functor for the equivalence of $D^b(X_+)$ and $D^b(X_-)$. Let $\pi_\pm$ be the projection of the vector bundle $X_\pm$ onto the corresponding base space. We choose the generators of $D^b(X_+)$ to be $\pi_+^*\cO_{SGr(2,4)}$, $\pi_+^* \cS_{SGr(2,4)}^\vee$, $\pi_+^* \wedge^2 \cS_{SGr(2,4)}^\vee$ and $\pi_+^* (\wedge^2 \cS_{SGr(2,4)}^\vee)^{\otimes 2}$. Since now we have a superpotential $W = Q \Omega(Y^1,Y^2)$, the lift of an IR brane is a matrix factorization of $W$.

$\pi_+^*\cO_{SGr(2,4)}$ can be lifted to the matrix factorization
\[
\xymatrix{
\mathcal{W}(0,\det^{-1})
\ar@<0.5ex>[rr]^-{\Omega(Y^1,Y^2)}
&& { \mathcal{W}(0,\cdot) }
\ar@<0.5ex>[ll]^-{Q}
}\,,
\]
which is in the small windows of both \eqref{loc+EX4} and \eqref{loc-EX4}, so it can be directly transported and projected to give the image $\pi_-^*\cO_{\mathbb{P}^3}$.

$\pi_+^* \wedge^2 \cS_{SGr(2,4)}^\vee$ can be lifted to the matrix factorization
\[
\xymatrix{
\mathcal{W}(0,\cdot)
\ar@<0.5ex>[rr]^-{\Omega(Y^1,Y^2)}
&& { \mathcal{W}(0,\det) }
\ar@<0.5ex>[ll]^-{Q}
}\, ,
\]
which fits in the small window of the local model \eqref{loc+EX4} at the first boundary. Because the vev of the $P$ field, which is in the $(-1,{\det}^{-1})$ representation, is nonzero in the IR limit, the matrix factorization above is IR equivalent to
\[
\xymatrix{
\mathcal{W}(-1,{\det}^{-1})
\ar@<0.5ex>[rr]^-{\Omega(Y^1,Y^2)}
&& { \mathcal{W}(-1,\cdot) }
\ar@<0.5ex>[ll]^-{Q}
}\, ,
\]
which is in the small window of the local model \eqref{loc-EX4} at the second boundary. Consequently, it can be projected to give the image $\pi_-^*\cO_{\mathbb{P}^3}(-1)$.

Similarly, $\pi_+^* (\wedge^2 \cS_{SGr(2,4)}^\vee)^{\otimes 2}$ can be transported to give the image $\pi_-^*\cO_{\mathbb{P}^3}(-2)$.

As in the previous example, for $\pi_+^* \cS_{SGr(2,4)}^\vee$, we need another empty brane. We can use the IR empty brane given by the exact sequence
\begin{equation}\label{emptyEX4}
\xymatrix{
\mathcal{W}(-1,\cdot)
\ar@<0.5ex>[rr]^-{\left( {\tiny \begin{array}{c} \Phi_1 \\ \Phi_2 \end{array} }\right)} \ar@<0.5ex>[d]^-{\Omega(Y^1,Y^2)}
&& { \mathcal{W}(0, { \overline{\yng(1)}} )}
\ar@<0.5ex>[rr]^-{(\Phi_2,-\Phi_1)} \ar@<0.5ex>[d]^-{\Omega(Y^1,Y^2)}
&& \mathcal{W}(1, {\det}^{-1}) \ar@<0.5ex>[d]^-{\Omega(Y^1,Y^2)} \\
\mathcal{W}(-1,\det)
\ar@<0.5ex>[rr]^-{\left( {\tiny \begin{array}{c} \Phi_1 \\ \Phi_2 \end{array} }\right)} \ar@<0.5ex>[u]^-{Q}
&& { \mathcal{W}(0,{ \yng(1)}) }
\ar@<0.5ex>[rr]^-{(\Phi_2,-\Phi_1)} \ar@<0.5ex>[u]^-{Q}
&& \mathcal{W}(1,\cdot ) \ar@<0.5ex>[u]^-{Q}
}
\end{equation}
to construct an empty brane. The lift of $\pi_+^* \cS_{SGr(2,4)}^\vee$ is
\begin{equation}\label{liftEX4}
\xymatrix{
\mathcal{W}(0, {\yng(1)} \otimes {\det}^{-1})
\ar@<0.5ex>[rr]^-{\Omega(Y^1,Y^2)}
&& { \mathcal{W}(0, {\yng(1)})}
\ar@<0.5ex>[ll]^-{Q}
}\, .
\end{equation}
From the empty brane \eqref{emptyEX4} and the equivalence
\[
\left(\xymatrix{
\mathcal{W}(-1, \cdot )
\ar@<0.5ex>[rr]^-{\Omega(Y^1,Y^2)}
&& { \mathcal{W}(-1, \det)}
\ar@<0.5ex>[ll]^-{Q}
}\right) \cong
\left(\xymatrix{
\mathcal{W}(-2, \det^{-1} )
\ar@<0.5ex>[rr]^-{\Omega(Y^1,Y^2)}
&& { \mathcal{W}(-2, \cdot)}
\ar@<0.5ex>[ll]^-{Q}
}\right)
\]
induced by the action of the $P$ field,
we see the IR image of \eqref{liftEX4} in the phase of $X_-$ is the complex $\text{Cone}(\pi_-^*\cO_{\mathbb{P}^3}(1)[-1] \rightarrow \pi_-^*\cO_{\mathbb{P}^3}(-2))$.

In summary, the brane transport with the choice of small windows mentioned above gives rise to the following map between the generators of $D^b(X_+)$ and $D^b(X_-)$:
\[
\begin{aligned}
\pi_+^*\cO_{SGr(2,4)} &\Longrightarrow \pi_-^*\cO_{\mathbb{P}^3}\, , \\
\pi_+^* \cS_{SGr(2,4)}^\vee &\Longrightarrow \text{Cone}\left(\pi_-^*\cO_{\mathbb{P}^3}(1)[-1] \rightarrow \pi_-^*\cO_{\mathbb{P}^3}(-2) \right)\, , \\
\pi_+^* \wedge^2 \cS_{SGr(2,4)}^\vee &\Longrightarrow \pi_-^*\cO_{\mathbb{P}^3}(-1)\,, \\
\pi_+^* (\wedge^2 \cS_{SGr(2,4)}^\vee)^{\otimes 2} &\Longrightarrow \pi_-^*\cO_{\mathbb{P}^3}(-2)\, .
\end{aligned}
\]

\section{Conclusion}
\label{sec:conclusion}
In this paper, we studied the brane transport in non-abelian GLSMs and proposed a realization of the derived equivalence between pairs of Calabi-Yau varieties, denoted as $(X_+,X_-)$, through a GLSM framework. Our approach involves embedding $X_+$ and $X_-$ into distinct phases of a GLSM for $\hat{X}$, as depicted in Fig.~\ref{fig:idea}. This embedding is designed to encode essential mixed branches $\mathcal{C}_{\pm}$ to ensure the well-defined nature of the model. The matching of Euler characteristics, or equivalently, the Witten indices in different phases, imposes the initial constraint on this construction.

Subsequently, we constructed the functor of the equivalence between $D^b(X_+)$ and $D^b(X_-)$ by tracing the brane transport across various phases. Given that $\hat{X}$ typically constitutes a Fano variety in our constructions, we computed the small window categories for some anomalous $U(2)$ gauge theories outlined in section~\ref{sec:small}. This computational analysis plays a pivotal role in our brane transport strategy. In section~\ref{sec:example}, we derived a necessary condition for the derived equivalence and illustrated our methodology through four concrete examples.

There are some limitations in the examples we have considered. For example, we have only considered the gauge group $U(2)$ in the non-abelian part. As two-parameter GLSMs, we have also only considered gauge groups $U(1)\times U(1)$ and $U(1)\times U(2)$.  But we expect our construction can be generalized to other two-parameter GLSMs with higher rank gauge groups straightforwardly. This extension hinges on computing small windows in the corresponding non-abelian gauge theories. Also, as a very first attempt in computing the small window of non-abelian GLSMs, it requires further explanation as well as examination on the small windows we found. Moreover, it is also interesting to apply a similar method as presented in this paper to comprehend the generalized flip induced derived equivalence as in \cite{2002math......6295B} from a GLSM point of view. We shall leave such considerations for future work. 

\section*{Acknowledgement}

We would like to thank J.~Knapp, M.~Romo, E.~Segal, L.~Smith and Y.~Wen for insightful conversations. We are grateful to J.~Knapp, M.~Romo and E.~Sharpe for their careful reading of the manuscript and useful suggestions and comments. We are very grateful to the anonymous referee for the detailed comments and valuable suggestions. We would also like to thank the workshop ``GLSM@30'' held at the Simons Center for Geometry and Physics, which stimulated many useful discussions. BL would like to thank the Birmingham University for its hospitality when part of the work was done. JG acknowledges support from the National Natural Science Foundation of China (Grant No.~12475005), the Natural Science Foundation of Shanghai Municipality (Grant No.~24ZR1468600), and the Fundamental Research Funds for the Central Universities. BL is supported by the KIAS Individual Grant (PG100701) at Korea Institute for Advanced Study and the Tsinghua scholarship for overseas graduate studies. HZ is supported in part by the China Postdoctoral Science Foundation (Grant No.~2022M720509), the National Natural Science Foundation of China (Grant No.~12405083,~12475005) and the Shanghai Pujiang Program (Grant No.~24PJA119).

\appendix

\section{Phases of the GLSM for $Fl(1,2,N)$}
\label{app:flagphase}

Gauge theories have been constructed to study the flag manifold in \cite{Donagi:2007hi,Ohmori:2018qza,Guo:2018iyr}, but their phase structures have been less explored. However, this would be helpful for our discussion in the main text. Therefore, in this appendix, we take a closer look at the two-step flag manifold $Fl(1,2,N)$. The GLSM for the flag manifold $Fl(1,2,N)$ is a $U(1)\times U(2)$ gauge theory with chiral matters $\Phi_\alpha$ and $X^\alpha_i$ with $\alpha = 1,2$ and $i=1,\dots,N$. 
\begin{equation}
\centering
\begin{tabular}{c|cc|c}
       & $\Phi_{\alpha}$  & $X^{\alpha}_{i}$   \\ \hline
 $U(1)$& 1  & 0   & $\xi_1$\\
 $U(2)$ & $\overline{{\yng(1)}}$  & ${\yng(1)}$ & $\xi_2$
\end{tabular} \label{eq:F124}
\end{equation}
where $\xi_1$ and $\xi_2$ are the FI-parameters associated with the gauge group $U(1)$ and $U(2)$ respectively. The supersymmetric vacuum is determined by the D-term equations given as
\begin{equation}\label{DeqApp}
\begin{aligned}
    D &\sim \sum_{\alpha=1}^2 |\Phi_\alpha|^2 - \xi_1\, , \\
    D^{\alpha}_\beta &\sim - \overline{\Phi}^\alpha \Phi_\beta + \sum_{i}\overline{X}_{\beta,i} X^{\alpha}_i  - \xi_2 \delta^\alpha_\beta\, . 
\end{aligned}
\end{equation}
The Coulomb vacua equations are 
\begin{equation}\label{coulombApp}
    \begin{aligned}
     &q_1 \:=\: (\sigma-\sigma_1)(\sigma -\sigma_2)\,, \\
    &\sigma_\alpha^N \:=\: - q_2 (\sigma - \sigma_\alpha)\,,  \\
    \end{aligned}
\end{equation}
for $\alpha =1,2$. Here, we denote by $\sigma$ the scalar component of the gauge field strength of $U(1)$, while $\sigma_\alpha$ the scalar components of the gauge field strength valued in the Cartan part of $U(2)$.

\begin{figure}[!h]
    \centering
    \tikzset{every picture/.style={line width=0.75pt}} 
    \begin{tikzpicture}[x=0.75pt,y=0.75pt,yscale=-0.6,xscale=0.6]
    \draw [line width=1.5]    (125,265) -- (260,265) -- (450,265) ;
    \draw [shift={(454,265)}, rotate = 180] [fill={rgb, 255:red, 0; green, 0; blue, 0 }  ][line width=0.08]  [draw opacity=0] (11.61,-5.58) -- (0,0) -- (11.61,5.58) -- cycle    ;
    \draw [line width=1.5]    (290,265) -- (290,120) ;
    \draw [shift={(290,114.83)}, rotate = 90] [fill={rgb, 255:red, 0; green, 0; blue, 0 }  ][line width=0.08]  [draw opacity=0] (11.61,-5.58) -- (0,0) -- (11.61,5.58) -- cycle    ;
    \draw [line width=1.5]    (290,265) -- (420,395) ;
    \draw (373,185) node [anchor=north west][inner sep=0.75pt]  [font=\normalsize]  {$I$};
    \draw (466,250) node [anchor=north west][inner sep=0.75pt]    {$\xi_{1}$};
    \draw (300,115) node [anchor=north west][inner sep=0.75pt]    {$\xi_{2}$};
    \draw (373,300) node [anchor=north west][inner sep=0.75pt]    {$III$};
    \draw  [draw opacity=0][fill={rgb, 255:red, 255; green, 255; blue, 255 }  ,fill opacity=1 ]  (277,315) -- (304,315) -- (304,350) -- (277,350) -- cycle  ;
    \draw (280,325) node [anchor=north west][inner sep=0.75pt]  [font=\normalsize]  {$IV$};
    \draw (190,185) node [anchor=north west][inner sep=0.75pt]  [font=\normalsize]  {$II$};
    \draw (435,370) node [anchor=north west][inner sep=0.75pt]    {$\xi_1 + \xi_2 = 0$};
    \end{tikzpicture}
    \caption{Phase diagram for $Fl(1,2,N)$.}
    \label{fig:phasesofflgs}
\end{figure}
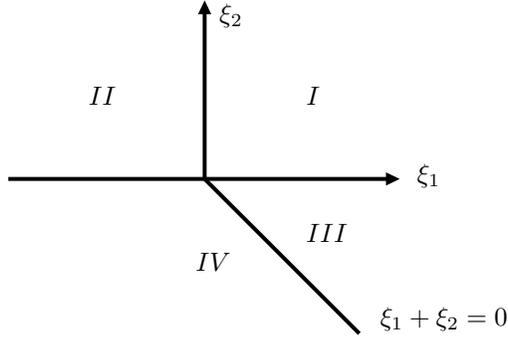

The classical configurations determined by the above $D$-term equations can be separated into four phases by different values of FI-parameters, as pictured in Fig.~\ref{fig:phasesofflgs}. A detailed semi-classical discussion of each phase is presented as follows:

\paragraph{Phase I: $\xi_1,\xi_2 \gg 0$.} This region consists of a pure Higgs branch, which realizes the flag manifold $Fl(1,2,N)$ as the symplectic quotient \cite{Donagi:2007hi}. One can also look at the solution to the Coulomb vacua equations in large $\xi$ limit. Here, we take $\xi_1 \to +\infty$ and $\xi_2 \to +\infty$, and the Coulomb vacua equations \eqref{coulombApp} become 
\[
    (\sigma - \sigma_1)(\sigma - \sigma_2) = 0\,, \quad \sigma_1^N = \sigma_2^N = 0\, .
\]
The only solution is $\sigma = \sigma_1 = \sigma_2 = 0$, indicating that there is only one pure Higgs branch. This pure Higgs branch can be obtained following \cite{Donagi:2007hi}, and it is the two-step flag manifold $Fl(1,2,N)$. Therefore, the Witten index in this phase is just the Euler characteristic of the flag manifold $Fl(1,2,N)$, which is
\[
    \chi(Fl(1,2,N))\:=\: \binom{N}{2}\binom{2}{1} = N (N-1)\, .
\]

\paragraph{Phase II: $\xi_1\ll 0,\xi_2\gg 0$.} Let us first look at the asymptotic behavior of the Coulomb vacua in the limit $\xi_1 \to - \infty$ and $\xi_2 \to + \infty$. Taking $\xi_2 \to +\infty$, we will have
\[
    \sigma_1^N = \sigma_2^N = 0\, .
\]
There is only a zero solution to the above equation $\sigma_1 =\sigma_2 = 0$. While the first equation becomes $\sigma^2 = q_1$, there are two non-zero solutions for $\sigma$: $\sigma = \pm \sqrt{q_1}$. The structure of vacua solutions suggests that there are only mixed Higgs-Coulomb branches in this phase.

Now let us find out what the mixed Higgs-Coulomb branches look like. When $\xi_1$ goes from positive to negative, there is no classical solution to the $U(1)$ D-term equation. In fact, the $U(1)$ part flows to the large $\sigma$ region as discussed in the previous paragraph. The field $\Phi_\alpha$ has mass proportional to $|\sigma|$ and should be integrated out. On the other hand, the fields $X^\alpha_{i}$ remain massless, and should satisfy the remaining $U(2)$ D-term equations which now become
\[
    \sum_{i=1}^N \overline{X}_{\beta,i} X^\alpha_{i} \:=\: \xi_2\delta^\alpha_\beta\, .
\]
This leads to nothing but the $Gr(2,N)$ \cite{Witten:1993yc,Witten:1993xi}. Therefore, the mixed Higgs-Coulomb branch in this phase consists of two copies of $Gr(2,N)$, each of which is located at a non-zero point in the $\sigma$-plane. Please see part (a) of Fig.~\ref{fig:flag124} for the case of $N=4$. Therefore, the total Witten index in this region is
\[
    2 \times \chi(Gr(2,N)) \:=\: 2 \times \binom{N}{2} \:=\: N(N-1)\, .
\]
This is consistent with the total Witten index obtained in Phase~I.

\begin{figure}[!h]
    \centering
    \begin{minipage}[b!]{0.45\textwidth}
    \begin{tikzpicture}
      \draw  [line width=1.5]  (3,1) -- (6,1) -- (8,2) -- (5,2) -- cycle ;
      \draw [line width=1.5]  [dash pattern={on 5.63pt off 4.5pt}]  (5,1.5) -- (5,2.8); 
      \draw [line width=1.5]  [dash pattern={on 5.63pt off 4.5pt}]  (6,1.5) -- (6,2.8);
      \fill (5,1.5) circle[radius=1.5pt];
      \fill (6,1.5) circle[radius=1.5pt];
      \draw (4.0,3.2) node [anchor=north west][inner sep=0.75pt]    {\footnotesize$Gr(2,4)$};
      \draw (5.6,3.2) node [anchor=north west][inner sep=0.75pt]    {\footnotesize$Gr(2,4)$};
      \draw (7.5,2.4) node [anchor=north west][inner sep=0.75pt]    {\footnotesize$\sigma$};
      \draw (5.5,0.5) node [anchor=north west][inner sep=0.75pt]    {$(a)$};
    \end{tikzpicture}
    \end{minipage}
    \begin{minipage}[b!]{0.45\textwidth}
    \begin{tikzpicture}[baseline=(current bounding box.south)]
      \draw  [line width=1.5]  (3,1) -- (6,1) -- (8,2) -- (5,2) -- cycle ;
      \draw [line width=1.5]  [dash pattern={on 5.63pt off 4.5pt}]  (4.5,1.5) -- (4.5,2.8); 
      \draw [line width=1.5]  [dash pattern={on 5.63pt off 4.5pt}]  (5.7,1.2) -- (5.7,2.5);
      \draw [line width=1.5]  [dash pattern={on 5.63pt off 4.5pt}]  (6.5,1.6) -- (6.5,2.8);
      \fill (4.5,1.5) circle[radius=1.5pt];
      \fill (5.7,1.2) circle[radius=1.5pt];
      \fill (6.5,1.6) circle[radius=1.5pt];
      \draw (4.4,3.2) node [anchor=north west][inner sep=0.75pt]    {\footnotesize$\mathbb{P}^3$};
      \draw (5.8,2.6) node [anchor=north west][inner sep=0.75pt]    {\footnotesize$\mathbb{P}^3$};
      \draw (6.6,3.2) node [anchor=north west][inner sep=0.75pt]    {\footnotesize$\mathbb{P}^3$};
      \draw (7.5,2.4) node [anchor=north west][inner sep=0.75pt]    {\footnotesize$\sigma_2$};
      \draw (5.5,0.5) node [anchor=north west][inner sep=0.75pt]    {$(b)$};
    \end{tikzpicture}
    \end{minipage}
    \caption{Structure of vacuum solutions in different phases when $N=4$.}
    \label{fig:flag124}
\end{figure}

\paragraph{Phase III: $\xi_1+\xi_2 \gg 0,\xi_2 \ll 0$.} Let us again first look at the asymptotic behavior of the Coulomb vacua equations \eqref{coulombApp}. In the limit $\xi_1\to +\infty$, the first equation becomes 
\[
    (\sigma - \sigma_1)(\sigma - \sigma_2) = 0\, ,
\]
and therefore $\sigma = \sigma_1$ or $\sigma = \sigma_2$. Without loss of generality, we can choose $\sigma = \sigma_1$ to fix the Weyl symmetry. Then, we have $\sigma_1^N = 0$, and so $\sigma = \sigma_1 = 0$. The remaining vacua equation is
\[
    \sigma_2^N = q_2 \sigma_2\, ,
\]
which has one zero solution $\sigma_2 = 0$ and $N-1$ nonzero solutions $\sigma_2 = e^{2\pi i \frac{\ell}{N-1} } q_2^{\frac{1}{N-1}}$ with $\ell=0,\dots,N-2$. Namely, in this phase, it seems that we have two sets of solutions:
\[
    \{\sigma = \sigma_1 = \sigma_2 = 0\} \quad \text{and} \quad \{ \sigma = \sigma_1 = 0, \sigma_2 = e^{2\pi i \frac{\ell}{N-1} } q_2^{\frac{1}{N-1}} | \ell =0,\dots,N-2 \}\, .
\]
However, the zero solution, $\{\sigma = \sigma_1 = \sigma_2 = 0\}$, is in fact not an honest solution, since there are no solutions to the $D$-term equation associated with $U(2)$ when $\xi_2 \ll 0$, by viewing $X^\alpha_i$ as entries of a hermitian matrix.

Another way to see this is to look at the IR effective theory. In fact, the GLSM flows to an effective gauged nonlinear sigma model defined by $Z_i := \Phi_{\alpha}X^{\alpha}_i$ and $B_{ij} := \epsilon_{\alpha\beta} X^\alpha_i X^\beta_j$ with gauge group and charges:
\[
\centering
\begin{tabular}{c|cc}
       & $Z_i$  & $B_{ij}$   \\ \hline
 $U(1)$& 1  & 0   \\
 $\det U(2)$ &0  & 1 
\end{tabular}
\]
The nonlinearity resides in the relations among coordinates $B_{ij}$ by definition, and this is the reason why we call this effective theory a {\it gauged nonlinear sigma model}. One should note that the coordinates $Z_i$ and $B_{ij}$ are also introduced in \cite[section~2.2]{Donagi:2007hi} for studying the Pl\"{u}cker embedding of the flag manifolds. Apparently, the second equation in \eqref{DeqApp} has no classical solution when $\xi_2$ is negative, which contradicts the pure Higgs branch suggested by the zero Coulomb vacua solution. 

While for each one of the remaining $N-1$ non-zero solutions, we have to integrate out $B_{ij}$'s as they are now massive and then the effective GLSM is just a $U(1)$ gauge theory with $N$ charge-1 chiral fields. Of course, we have to check the corresponding D-term equation has solutions in the present phase. Note that
\[
\begin{aligned}
    \sum_{i=1}^N |Z_i|^2 &= \sum_{i=1}^N \sum_{\alpha,\beta} \overline{\Phi}^\beta \Phi_\alpha \overline{X}_{\beta,i} X^\alpha_i = \sum_{\alpha,\beta} \overline{\Phi}^\beta \Phi_\alpha (\overline{\Phi}^\alpha \Phi_\beta + \xi_2 \delta^\alpha_\beta),\\
    &= \sum_{\alpha=1}^2 |\Phi_\alpha|^2 ( \sum_{\beta=1}^2 |\Phi_\beta|^2 + \xi_2) = (\xi_1 + \xi_2) \sum_{\alpha=1}^2 |\Phi_\alpha|^2 = \xi_1 (\xi_1+\xi_2) > 0
\end{aligned}   
\]
in the region $\xi_1 \gg 0$, $\xi_2 \ll 0$ and $\xi_1+\xi_2 \gg 0$, the D-term equation indeed has solutions, giving rise to the vacuum manifold inside the mixed Higgs-Coulomb branch, which is a projective space $\mathbb{P}^{N-1}$.

Thus, this phase only has mixed Higgs-Coulomb branches, and, more specifically, they are $N-1$ copies of $\mathbb{P}^{N-1}$ located at $N-1$ non-zero points on the Coulomb branch, see part~(b) of Fig.~\ref{fig:flag124} for the $N=4$ case. Based on this picture, one can calculate the total Witten index in this phase, which is
\[
    (N-1)\times \chi(\PP^{N-1}) \:=\: N(N-1)\,,
\]
matching the results from the other phases.

\paragraph{Phase IV: $\xi_1+ \xi_2 \ll 0, \xi_2 \ll 0$.} This is a pure Coulomb branch. First, it is clear that the region where $\xi_1 \ll 0$ and $\xi_2 \ll 0$ is of a pure Coulomb branch. Then, one can argue that there are no solutions to the D-term equations in the region $\xi_1 \gg 0$ and $\xi_1+\xi_2 \ll 0$, by following the previous discussions.

In this case, the total Witten index only has contributions from the Coulomb vacua, and the total number of solutions to the Coulomb vacua equations is indeed $N(N-1)$.

\bibliographystyle{utphys.bst}
\bibliography{ref}

@article{Segal_2016,
	doi = {10.1112/blms/bdw026},
	year = 2016,
	month = {apr},
	publisher = {Wiley},
	volume = {48},
	number = {3},
	pages = {533--538},
	author = {Ed Segal},
	title = {A new 5-fold flop and derived equivalence},
	journal = {Bulletin of the London Mathematical Society}
}

@article{Guo:2021aqj,
    author = "Guo, Jirui and Romo, Mauricio",
    title = "{Hybrid models for homological projective duals and noncommutative resolutions}",
    eprint = "2111.00025",
    archivePrefix = "arXiv",
    primaryClass = "hep-th",
    doi = "10.1007/s11005-022-01605-3",
    journal = "Lett. Math. Phys.",
    volume = "112",
    number = "6",
    pages = "117",
    year = "2022"
}

@article{Addington:2012zv,
    author = "Addington, Nicolas M. and Segal, Edward P. and Sharpe, Eric",
    title = "{D-brane probes, branched double covers, and noncommutative resolutions}",
    eprint = "1211.2446",
    archivePrefix = "arXiv",
    primaryClass = "hep-th",
    doi = "10.4310/ATMP.2014.v18.n6.a5",
    journal = "Adv. Theor. Math. Phys.",
    volume = "18",
    number = "6",
    pages = "1369--1436",
    year = "2014"
}

@article{halpern2020combinatorial,
  title={Combinatorial constructions of derived equivalences},
  author={Halpern-Leistner, Daniel and Sam, Steven},
  journal={Journal of the American Mathematical Society},
  volume={33},
  number={3},
  pages={735--773},
  year={2020}
}

@article{Segal:2010cz,
    author = "Segal, Ed",
    title = "{Equivalences between GIT quotients of Landau-Ginzburg B-models}",
    eprint = "0910.5534",
    archivePrefix = "arXiv",
    primaryClass = "math.AG",
    doi = "10.1007/s00220-011-1232-y",
    journal = "Commun. Math. Phys.",
    volume = "304",
    pages = "411--432",
    year = "2011"
}

@article{Donagi:2007hi,
    author = "Donagi, Ron and Sharpe, Eric",
    title = "{GLSM's for partial flag manifolds}",
    eprint = "0704.1761",
    archivePrefix = "arXiv",
    primaryClass = "hep-th",
    doi = "10.1016/j.geomphys.2008.07.010",
    journal = "J. Geom. Phys.",
    volume = "58",
    pages = "1662--1692",
    year = "2008"
}

@article{Ohmori:2018qza,
    author = "Ohmori, Kantaro and Seiberg, Nathan and Shao, Shu-Heng",
    title = "{Sigma Models on Flags}",
    eprint = "1809.10604",
    archivePrefix = "arXiv",
    primaryClass = "hep-th",
    doi = "10.21468/SciPostPhys.6.2.017",
    journal = "SciPost Phys.",
    volume = "6",
    number = "2",
    pages = "017",
    year = "2019"
}

@article{Guo:2018iyr,
    author = "Guo, Jirui",
    title = "{Quantum Sheaf Cohomology and Duality of Flag Manifolds}",
    eprint = "1808.00716",
    archivePrefix = "arXiv",
    primaryClass = "hep-th",
    doi = "10.1007/s00220-019-03462-z",
    journal = "Commun. Math. Phys.",
    volume = "374",
    number = "2",
    pages = "661--688",
    year = "2019"
}

@article{Guo:2021dlz,
    author = "Guo, Jirui and Zou, Hao",
    title = "{Quantum cohomology of symplectic flag manifolds}",
    eprint = "2107.09880",
    archivePrefix = "arXiv",
    primaryClass = "hep-th",
    doi = "10.1088/1751-8121/ac7487",
    journal = "J. Phys. A",
    volume = "55",
    number = "27",
    pages = "275401",
    year = "2022"
}

@book{weyman2003cohomology,
  title={Cohomology of vector bundles and syzygies},
  author={Weyman, Jerzy},
  number={149},
  year={2003},
  publisher={Cambridge University Press}
}

@article{Witten:1993yc,
    author = "Witten, Edward",
    editor = "Greene, B. and Yau, Shing-Tung",
    title = "{Phases of N=2 theories in two-dimensions}",
    eprint = "hep-th/9301042",
    archivePrefix = "arXiv",
    reportNumber = "IASSNS-HEP-93-3",
    doi = "10.1016/0550-3213(93)90033-L",
    journal = "Nucl. Phys. B",
    volume = "403",
    pages = "159--222",
    year = "1993"
}

@article{Witten:1993xi,
    author = "Witten, Edward",
    title = "{The Verlinde algebra and the cohomology of the Grassmannian}",
    eprint = "hep-th/9312104",
    archivePrefix = "arXiv",
    reportNumber = "IASSNS-HEP-93-41",
    month = "12",
    year = "1993"
}

@article{Hori:2006dk,
    author = "Hori, Kentaro and Tong, David",
    title = "{Aspects of Non-Abelian Gauge Dynamics in Two-Dimensional N=(2,2) Theories}",
    eprint = "hep-th/0609032",
    archivePrefix = "arXiv",
    doi = "10.1088/1126-6708/2007/05/079",
    journal = "JHEP",
    volume = "05",
    pages = "079",
    year = "2007"
}

@article{Chen:2020iyo,
    author = "Chen, Zhuo and Guo, Jirui and Romo, Mauricio",
    title = "{A GLSM View on Homological Projective Duality}",
    eprint = "2012.14109",
    archivePrefix = "arXiv",
    primaryClass = "hep-th",
    doi = "10.1007/s00220-022-04401-1",
    journal = "Commun. Math. Phys.",
    volume = "394",
    number = "1",
    pages = "355--407",
    year = "2022"
}

@article{halpern2016autoequivalences,
  title={Autoequivalences of derived categories via geometric invariant theory},
  author={Halpern-Leistner, Daniel and Shipman, Ian},
  journal={Advances in Mathematics},
  volume={303},
  pages={1264--1299},
  year={2016},
  publisher={Elsevier}
}

@article{Brunner:2021ulc,
    author = "Brunner, Ilka and Krumpeck, Lukas and Roggenkamp, Daniel",
    title = "{Defects and phase transitions to geometric phases of abelian GLSMs}",
    eprint = "2109.04124",
    archivePrefix = "arXiv",
    primaryClass = "hep-th",
    doi = "10.1007/s11005-024-01852-6",
    journal = "Lett. Math. Phys.",
    volume = "114",
    number = "5",
    pages = "114",
    year = "2024"
}

@article{Hori:2004zd,
    author = "Hori, Kentaro and Walcher, Johannes",
    title = "{D-branes from matrix factorizations}",
    eprint = "hep-th/0409204",
    archivePrefix = "arXiv",
    doi = "10.1016/j.crhy.2004.09.016",
    journal = "Comptes Rendus Physique",
    volume = "5",
    pages = "1061--1070",
    year = "2004"
}

@article{halpern2015derived,
  title={The derived category of a GIT quotient},
  author={Halpern-Leistner, Daniel},
  journal={Journal of the American Mathematical Society},
  volume={28},
  number={3},
  pages={871--912},
  year={2015}
}

@article{Gu:2020oeb,
    author = "Gu, W. and Sharpe, E. and Zou, H.",
    title = "{GLSMs for exotic Grassmannians}",
    eprint = "2008.02281",
    archivePrefix = "arXiv",
    primaryClass = "hep-th",
    doi = "10.1007/JHEP10(2020)200",
    journal = "JHEP",
    volume = "10",
    pages = "200",
    year = "2020"
}

@article{Clingempeel:2018iub,
    author = "Clingempeel, Joel and Le Floch, Bruno and Romo, Mauricio",
    title = "{Brane transport in anomalous (2,2) models and localization}",
    eprint = "1811.12385",
    archivePrefix = "arXiv",
    primaryClass = "hep-th",
    month = "11",
    year = "2018"
}

@article{Herbst:2008jq,
    author = "Herbst, Manfred and Hori, Kentaro and Page, David",
    title = "{Phases Of N=2 Theories In 1+1 Dimensions With Boundary}",
    eprint = "0803.2045",
    archivePrefix = "arXiv",
    primaryClass = "hep-th",
    reportNumber = "DESY-07-154, CERN-PH-TH-2008-048",
    month = "3",
    year = "2008"
}

@article{Green:1996dd,
    author = "Green, Michael B. and Harvey, Jeffrey A. and Moore, Gregory W.",
    title = "{I-brane inflow and anomalous couplings on d-branes}",
    eprint = "hep-th/9605033",
    archivePrefix = "arXiv",
    reportNumber = "DAMTP-96-40, EFI-96-13, YCTP-P8-96, RU-96-29",
    doi = "10.1088/0264-9381/14/1/008",
    journal = "Class. Quant. Grav.",
    volume = "14",
    pages = "47--52",
    year = "1997"
}

@article{Sharpe:2024dcd,
    author = "Sharpe, Eric",
    title = "{A survey of some recent developments in GLSMs}",
    eprint = "2401.11637",
    archivePrefix = "arXiv",
    primaryClass = "hep-th",
    doi = "10.1142/S0217751X24460011",
    journal = "Int. J. Mod. Phys. A",
    volume = "39",
    number = "33",
    pages = "2446001",
    year = "2024"
}

@article{Govindarajan:2005im,
    author = "Govindarajan, Suresh and Jockers, Hans and Lerche, Wolfgang and Warner, Nicholas P.",
    title = "{Tachyon condensation on the elliptic curve}",
    eprint = "hep-th/0512208",
    archivePrefix = "arXiv",
    reportNumber = "CERN-PH-TH-2005-259",
    doi = "10.1016/j.nuclphysb.2006.12.009",
    journal = "Nucl. Phys. B",
    volume = "765",
    pages = "240--286",
    year = "2007"
}

@inproceedings{Aspinwall:2004jr,
    author = "Aspinwall, Paul S.",
    title = "{D-branes on Calabi-Yau manifolds}",
    booktitle = "{Theoretical Advanced Study Institute in Elementary Particle Physics (TASI 2003): Recent Trends in String Theory}",
    eprint = "hep-th/0403166",
    archivePrefix = "arXiv",
    reportNumber = "DUKE-CGTP-04-04",
    doi = "10.1142/9789812775108_0001",
    pages = "1--152",
    month = "3",
    year = "2004"
}

@article{Hellerman:2001bu,
    author = "Hellerman, Simeon and Kachru, Shamit and Lawrence, Albion E. and McGreevy, John",
    title = "{Linear sigma models for open strings}",
    eprint = "hep-th/0109069",
    archivePrefix = "arXiv",
    reportNumber = "SLAC-PUB-8984, SU-ITP-00-24",
    doi = "10.1088/1126-6708/2002/07/002",
    journal = "JHEP",
    volume = "07",
    pages = "002",
    year = "2002"
}

@article{Brunner:2006tc,
    author = "Brunner, Ilka and Gaberdiel, Matthias R. and Keller, Christoph A.",
    title = "{Matrix factorisations and D-branes on K3}",
    eprint = "hep-th/0603196",
    archivePrefix = "arXiv",
    doi = "10.1088/1126-6708/2006/06/015",
    journal = "JHEP",
    volume = "06",
    pages = "015",
    year = "2006"
}

@inproceedings{Hori:2000ic,
    author = "Hori, Kentaro",
    title = "{Linear models of supersymmetric D-branes}",
    booktitle = "{KIAS Annual International Conference on Symplectic Geometry and Mirror Symmetry}",
    eprint = "hep-th/0012179",
    archivePrefix = "arXiv",
    reportNumber = "HUTP-00-A051",
    pages = "111--186",
    month = "12",
    year = "2000"
}

@incollection{van2004non,
  title={Non-commutative crepant resolutions},
  author={Van den Bergh, Michel},
  booktitle={The Legacy of Niels Henrik Abel: The Abel Bicentennial, Oslo, 2002},
  pages={749--770},
  year={2004},
  publisher={Springer}
}

@inproceedings{Hori:2019vkm,
    author = "Hori, Kentaro and Romo, Mauricio",
    booktitle = "{Notes on the hemisphere}",
    title = "{Notes on the hemisphere}",
    doi = "10.2969/aspm/08310127",
    year = "2019"
}

@article{Aspinwall:2001zq,
    author = "Aspinwall, Paul S.",
    title = "{Some navigation rules for D-brane monodromy}",
    eprint = "hep-th/0102198",
    archivePrefix = "arXiv",
    reportNumber = "DUKE-CGTP-01-01, NSF-ITP-01-11",
    doi = "10.1063/1.1409963",
    journal = "J. Math. Phys.",
    volume = "42",
    pages = "5534--5552",
    year = "2001"
}

@article{Morrison:1994fr,
    author = "Morrison, David R. and Plesser, M. Ronen",
    title = "{Summing the instantons: Quantum cohomology and mirror symmetry in toric varieties}",
    eprint = "hep-th/9412236",
    archivePrefix = "arXiv",
    reportNumber = "DUKE-TH-94-78, IASSNS-HEP-94-82",
    doi = "10.1016/0550-3213(95)00061-V",
    journal = "Nucl. Phys. B",
    volume = "440",
    pages = "279--354",
    year = "1995"
}

@article{Hori:2013ika,
    author = "Hori, Kentaro and Romo, Mauricio",
    title = "{Exact Results In Two-Dimensional (2,2) Supersymmetric Gauge Theories With Boundary}",
    eprint = "1308.2438",
    archivePrefix = "arXiv",
    primaryClass = "hep-th",
    month = "8",
    year = "2013"
}

@article{Honda:2013uca,
    author = "Honda, Daigo and Okuda, Takuya",
    title = "{Exact results for boundaries and domain walls in 2d supersymmetric theories}",
    eprint = "1308.2217",
    archivePrefix = "arXiv",
    primaryClass = "hep-th",
    reportNumber = "UT-KOMABA-13-8",
    doi = "10.1007/JHEP09(2015)140",
    journal = "JHEP",
    volume = "09",
    pages = "140",
    year = "2015"
}

@article{Sugishita:2013jca,
    author = "Sugishita, Sotaro and Terashima, Seiji",
    title = "{Exact Results in Supersymmetric Field Theories on Manifolds with 
Boundaries}",
    eprint = "1308.1973",
    archivePrefix = "arXiv",
    primaryClass = "hep-th",
    reportNumber = "KUNS-2458, YITP-13-72",
    doi = "10.1007/JHEP11(2013)021",
    journal = "JHEP",
    volume = "11",
    pages = "021",
    year = "2013"
}

@article{Ballard:2016ncw,
    author = "Ballard, Matthew and Favero, David and Katzarkov, Ludmil",
    title = "{Variation of geometric invariant theory quotients and derived categories}",
    doi = "10.1515/crelle-2015-0096",
    journal = "J. Reine Angew. Math.",
    volume = "2019",
    number = "746",
    pages = "235--303",
    year = "2019"
}

@article{Minasian:1997mm,
    author = "Minasian, Ruben and Moore, Gregory W.",
    title = "{K theory and Ramond-Ramond charge}",
    eprint = "hep-th/9710230",
    archivePrefix = "arXiv",
    reportNumber = "YCTP-P21-97",
    doi = "10.1088/1126-6708/1997/11/002",
    journal = "JHEP",
    volume = "11",
    pages = "002",
    year = "1997"
}

@article{Lin:2024fpz,
    author = "Lin, Ban and Romo, Mauricio",
    title = "{B-brane Transport and Grade Restriction Rule for Determinantal Varieties}",
    eprint = "2402.07109",
    archivePrefix = "arXiv",
    primaryClass = "hep-th",
    doi = "10.1007/s00220-024-05153-w",
    journal = "Commun. Math. Phys.",
    volume = "405",
    number = "11",
    pages = "268",
    year = "2024"
}

@article{Sharpe:1999qz,
    author = "Sharpe, Eric R.",
    title = "{D-branes, derived categories, and Grothendieck groups}",
    eprint = "hep-th/9902116",
    archivePrefix = "arXiv",
    reportNumber = "DUKE-CGTP-99-03",
    doi = "10.1016/S0550-3213(99)00535-0",
    journal = "Nucl. Phys. B",
    volume = "561",
    pages = "433--450",
    year = "1999"
}

@article{Douglas:2000gi,
    author = "Douglas, Michael R.",
    title = "{D-branes, categories and N=1 supersymmetry}",
    eprint = "hep-th/0011017",
    archivePrefix = "arXiv",
    reportNumber = "RUNHETC-2000-42",
    doi = "10.1063/1.1374448",
    journal = "J. Math. Phys.",
    volume = "42",
    pages = "2818--2843",
    year = "2001"
}

@article{Hori:2011pd,
    author = "Hori, Kentaro",
    title = "{Duality In Two-Dimensional (2,2) Supersymmetric Non-Abelian Gauge Theories}",
    eprint = "1104.2853",
    archivePrefix = "arXiv",
    primaryClass = "hep-th",
    doi = "10.1007/JHEP10(2013)121",
    journal = "JHEP",
    volume = "10",
    pages = "121",
    year = "2013"
}

@article{Aspinwall:2006ib,
    author = "Aspinwall, Paul S.",
    title = "{The Landau-Ginzburg to Calabi-Yau dictionary for D-branes}",
    eprint = "hep-th/0610209",
    archivePrefix = "arXiv",
    reportNumber = "DUKE-CGTP-06-04",
    doi = "10.1063/1.2768185",
    journal = "J. Math. Phys.",
    volume = "48",
    pages = "082304",
    year = "2007"
}

@article{Galkin2016Gamma,
  author    = {Sergey Galkin and Vasily Golyshev and Hiroshi Iritani},
  title     = {Gamma classes and quantum cohomology of Fano manifolds: Gamma conjectures},
  journal   = {Duke Mathematical Journal},
  volume    = {165},
  number    = {11},
  pages     = {2005--2077},
  year      = {2016},
  month     = aug,
  doi       = {10.1215/00127094-3476593},
  publisher = {Duke University Press},
     eprint = {1404.6407}
}

@ARTICLE{2002math......6295B,
       author = {{Bondal}, Alexei and {Orlov}, Dmitri},
        title = "{Derived categories of coherent sheaves}",
      journal = {arXiv Mathematics e-prints},
         year = 2002,
        month = jun,
          eid = {math/0206295},
        pages = {math/0206295},
          doi = {10.48550/arXiv.math/0206295},
archivePrefix = {arXiv},
       eprint = {math/0206295},
 primaryClass = {math.AG},
       adsurl = {https://ui.adsabs.harvard.edu/abs/2002math......6295B}
}

@article{Donovan:2013gia,
    author = "Donovan, Will",
    editor = "Donagi, Ron and Katz, Sheldon and Klemm, Albrecht and Morrison, David R.",
    title = "{Grassmannian twists, derived equivalences and brane transport}",
    eprint = "1304.2913",
    archivePrefix = "arXiv",
    primaryClass = "math.AG",
    journal = "Proc. Symp. Pure Math.",
    volume = "90",
    pages = "251--264",
    year = "2015"
}

@article{Gerhardus:2015sla,
    author = "Gerhardus, Andreas and Jockers, Hans",
    title = "{Dual pairs of gauged linear sigma models and derived equivalences of Calabi\textendash{}Yau threefolds}",
    eprint = "1505.00099",
    archivePrefix = "arXiv",
    primaryClass = "hep-th",
    reportNumber = "BONN-TH-2015-05",
    doi = "10.1016/j.geomphys.2016.12.005",
    journal = "J. Geom. Phys.",
    volume = "114",
    pages = "223--259",
    year = "2017"
}

@article{Knapp:2023izn,
    author = "Knapp, Johanna",
    title = "{Grade restriction and D-brane transport for a nonabelian GLSM of an elliptic curve}",
    eprint = "2312.07639",
    archivePrefix = "arXiv",
    primaryClass = "hep-th",
    doi = "10.1142/S0217751X24460047",
    journal = "Int. J. Mod. Phys. A",
    volume = "39",
    number = "33",
    pages = "2446004",
    year = "2024"
}

@article{thomas2005notes,
  title={Notes on GIT and symplectic reduction for bundles and varieties},
  author={Thomas, Richard P},
  journal={arXiv preprint math/0512411},
  year={2005}
}

@article{Donovan:2020jfh,
    author = "Donovan, Will",
    title = {{Stringy K{\"a}hler Moduli for the Pfaffian-Grassmannian Correspondence}},
    eprint = "2009.12630",
    archivePrefix = "arXiv",
    primaryClass = "math.AG",
    doi = "10.3842/SIGMA.2021.028",
    journal = "SIGMA",
    volume = "17",
    pages = "028",
    year = "2021"
}

@Article{Eager2017,
author={Eager, Richard
and Hori, Kentaro
and Knapp, Johanna
and Romo, Mauricio},
title={Beijing lectures on the grade restriction rule},
journal={Chinese Annals of Mathematics, Series B},
year={2017},
month={Jul},
day={01},
volume={38},
number={4},
pages={901-912},
issn={1860-6261},
doi={10.1007/s11401-017-1103-8}
}

@article{Hori:2016txh,
    author = "Hori, Kentaro and Knapp, Johanna",
    title = "{A pair of Calabi-Yau manifolds from a two parameter non-Abelian gauged linear sigma model}",
    eprint = "1612.06214",
    archivePrefix = "arXiv",
    primaryClass = "hep-th",
    month = "12",
    year = "2016"
}

@article{Acosta:2014cma,
    author = "Acosta, Pedro",
    title = "{Asymptotic Expansion and the LG/(Fano, General Type) Correspondence}",
    eprint = "1411.4162",
    archivePrefix = "arXiv",
    primaryClass = "math.AG",
    month = "11",
    year = "2014"
}

@article{Cecotti:1991me,
    author = "Cecotti, Sergio and Vafa, Cumrun",
    title = "{Topological antitopological fusion}",
    reportNumber = "HUTP-91-A031, SISSA-69-91-EP",
    doi = "10.1016/0550-3213(91)90021-O",
    journal = "Nucl. Phys. B",
    volume = "367",
    pages = "359--461",
    year = "1991"
}

@article{hori2000d,
    author = "Hori, Kentaro and Iqbal, Amer and Vafa, Cumrun",
    title = "{D-branes and mirror symmetry}",
    eprint = "hep-th/0005247",
    archivePrefix = "arXiv",
    reportNumber = "HUTP-00-A017, MIT-CTP-2941",
    month = "5",
    year = "2000"
}

@article{eagon1962ideals,
  title={Ideals defined by matrices and a certain complex associated with them},
  author={Eagon, John A and Northcott, Douglas Geoffrey},
  journal={Proceedings of the Royal Society of London. Series A. Mathematical and Physical Sciences},
  volume={269},
  number={1337},
  pages={188--204},
  year={1962},
  publisher={The Royal Society London},
  doi = {10.1098/rspa.1962.0170}
}

@article{Hori:202x,
    author = "Eager, Richard and Hori, Kentaro and Knapp, Johanna and Romo, Mauricio",
    journal = "To appear"
}

@article{Bertolini:2013xga,
    author = "Bertolini, Marco and Melnikov, Ilarion V. and Plesser, M. Ronen",
    title = "{Hybrid conformal field theories}",
    eprint = "1307.7063",
    archivePrefix = "arXiv",
    primaryClass = "hep-th",
    doi = "10.1007/JHEP05(2014)043",
    journal = "JHEP",
    volume = "05",
    pages = "043",
    year = "2014"
}

@article{Erkinger:2022sqs,
    author = "Erkinger, David and Knapp, Johanna",
    title = "{On genus-0 invariants of Calabi-Yau hybrid models}",
    eprint = "2210.01226",
    archivePrefix = "arXiv",
    primaryClass = "hep-th",
    doi = "10.1007/JHEP05(2023)071",
    journal = "JHEP",
    volume = "05",
    pages = "071",
    year = "2023"
}

@article{Guo:2025yed,
    author = "Guo, Jirui and Romo, Mauricio and Smith, Lucy",
    title = "{B-brane transport in nonabelian GLSMs for $K_{Gr(2,N)}$}",
    eprint = "2503.06293",
    archivePrefix = "arXiv",
    primaryClass = "hep-th",
    doi = "10.1007/s00023-025-01643-2",
    journal = "accepted by Ann. Henri Poincar\'e",
    month = "11",
    year = "2025",
}

\end{document}